\documentclass[rmp,aps,twocolumn,nofootinbib,endfloats,eqsecnum]{revtex4}

\usepackage{graphics,graphicx}
\usepackage{epsfig}
\usepackage{bm}
\usepackage{amsmath}
\usepackage{amssymb}

\newcommand{\bS}{{\bf S}}

\newcommand{\bk}{{\bf k}}

\newcommand{\br}{{\bf r}}

\newcommand{\bsig}{\mbox{\boldmath{$\sigma$}}}

\newcommand{\Tr}{\,{\rm{Tr}\,}}

\newcommand{\DD}{\Delta_0}

\newcommand{\beqa}{\begin{eqnarray}}
\newcommand{\eeqa}{\end{eqnarray}}

\begin{document}
\title{Impurity-induced states in conventional and unconventional superconductors}

\author{A. V. Balatsky}
\email{avb@viking.lanl.gov,  http://theory.lanl.gov}
\affiliation{Theoretical Division, Los Alamos National Laboratory,
Los Alamos, New Mexico {\rm  }87545}

\author{I. Vekhter}
\email{vekhter@phys.lsu.edu}
\affiliation{Department of Physics
and Astronomy, Louisiana State University, Baton Rouge, Louisiana
70803}

\author{Jian-Xin Zhu}
\email{jxzhu@viking.lanl.gov}
\affiliation{Theoretical Division,
Los Alamos National Laboratory, Los Alamos, New Mexico 87545}

\begin{abstract}
We review recent developments in our understanding of how impurities
influence the electronic states in the bulk of superconductors. Our
focus is on the quasi-localized states in the vicinity of impurity
sites in conventional and unconventional superconductors and our
goal is to provide a unified framework for their description. The
non-magnetic impurity resonances in unconventional superconductors
are directly related to the Yu-Shiba-Rusinov states around magnetic
impurities in conventional $s$-wave systems. We  review the physics
behind these states, including quantum phase transition between
screened and unscreened impurity, and emphasize recent work on
$d$-wave superconductors. The bound states are most spectacularly
seen in scanning tunneling spectroscopy measurements on high-$T_c$
cuprates, which we describe in detail. We also discuss very recent
progress on the states coupled to impurity sites which have their
own dynamics, and impurity resonances in the presence of an order
competing with superconductivity. Last part of the review is devoted
to influence of local deviations of the impurity concentration from
its average value on the density of states in $s$-wave
superconductors. We review how these fluctuations affect the density
of states and show that $s$-wave superconductors are, strictly
speaking, gapless in the presence of an arbitrarily small
concentration of magnetic impurities.
\end{abstract}

\date{\today}
\maketitle

\tableofcontents

\section{Introduction}
\label{sec:INTRODUCTION}

\subsection{Aim and scope of this article}

Real materials are not pure. Sometimes  excessive impurities hinder
observations of beautiful physics that would be there in cleaner
systems. Magnetic disorder destroys the coherence of the
superconducting state. At the very least, in conventional metals,
impurities lead to higher resistivity. It is therefore very tempting
to treat impurities as unfortunate obstacles to our understanding of
the true underlying physics of the systems we study, strive to make
cleaner and better materials, and ignore imperfections whenever
possible.

Yet sometimes impurities directly lead to the desired physical
properties. They are crucial in achieving functionality of doped
semiconductors: undoped semiconductors are just band insulators and
not useful for applications in electronics. The entire multi-billion
dollar semiconducting electronics industry is based on the precise
control and manipulation of electronic states due to dopant
(impurity) states.

Consequently, sensitivity of a physical system to disorder can be a
blessing in disguise. It can lead not only to achieving new
applications but also to uncovering the nature of exotic ground
states, elucidating properties of electronic correlations, and
producing electronic states that are impossible in the bulk of a
clean system. Until recently this idea has not been emphasized
enough in the study of correlated electron systems, but in recent
years more and more efforts are focused on understanding the changes
produced by disorder in a wide variety of strongly interacting
electronic matter. One of the most promising directions is the study
of disorder near quantum critical points, where several types of
order compete and exist in delicate balance that impurities have the
power to tip in favor of one of the orders \cite{AJMillis:2003}.

This is a review of the impurity effects on the electronic states in
superconductors. The main purpose of our article is to give a reader
an appreciation of recent developments, review the current
understanding and outline further questions on impurity effects in
conventional, and especially unconventional superconductors.
Superconductors present probably the first example of a nontrivial
many-electron system where effects of disorder on the electronic
states were studied experimentally and theoretically, and this
review focuses on these effects.

The subject of impurity effects in superconductors is well
established and well covered, see, for example,  excellent textbooks
and reviews \cite{AAAbrikosov:1963,JRSchrieffer:1964,MTinkham:1996,
PGdeGennes:1989,ALFetter:1971,MSigrist:1991,JFAnnett:1990}. The main
classical results, such as Abrikosov-Gor'kov theory of pairbreaking
by magnetic impurities \cite{AAAbrikosov:1960}, and Anderson
theorem, that explains why non-magnetic impurities do not destroy
conventional superconductivity, \cite{PWAnderson:1959} are well
known from the 60s, and are now taught in graduate school. The need
to review the subject arose since a) there are many new results; b)
the analyses of the classical papers have been substantially
modified in applications to novel materials; c) the emphasis of the
study of the impurity effects shifted from macroscopic to
microscopic length scales.

From the early days of superconductivity, impurity doping was one of
the most important tools to identify the nature of the pairing state
and microscopic properties. A classical study of the role of
magnetic impurities in conventional superconductors was carried out
by ~\onlinecite{MAWoolf:1965}, and followed by many detailed
investigations.  Both magnetic and nonmagnetic impurities are
pairbreakers in unconventional superconductors, and often impurity
suppression of superconductivity is a strong early hint of the
unconventional pairing state. For example, the rapid suppression of
the transition temperature, $T_c$,  in Al doped SrRuO$_4$
superconductor was the first and strong indication that it is a
p-wave superconductor \cite{APMackenzie:1999,APMackenzie:2003}.

In the past two decades we have witnessed a tremendous growth of the
number of novel superconductors. Many of them belong to the general
class of  strongly correlated electron systems, and, as a result of
Coulomb interaction, the superconductivity is unconventional, see
below.  Study of the effect of impurities on unconventional
superconductors is a still developing field, yet it is mature enough
to warrant an overview.

Sometimes superconducting state emerges from competition between
different phases, such as magnetically ordered and paramagnetic in
high-temperature cuprates, organic materials and heavy fermion
systems. Experimentally, superconductivity often is the strongest
when the two competing states are nearly degenerate, near quantum
critical points. This is the case for example for Ce based heavy
fermion materials \cite{VASidorov:2002} and  UGe$_2$
\cite{SSSaxena:2000}. Study of impurity effects in these materials
allows (at least, in principle) to determine the nature of the
superconducting state and reveal competing electronic correlations.

This has driven in part the study of impurity effects in high-$T_c$
superconductors. At present, despite much progress, there is no
complete microscopic description and certainly no consensus in the
community on the mechanism of superconductivity. Study of competing
orders in the neighborhood of impurity atoms has the potential to
reveal the nature and origin of the superconducting state.

The new states and structures that appear due to disorder often are
confined to micro- or mesoscopic length scales. They would remain in
the realm of academic discussion were it not for the development of
new techniques and probes of disorder. At the time of classical
work, experimental interest lied solely in macroscopic properties of
materials: transition temperature, $T_c$, specific heat, and the
average density of states (obtained from planar junction tunneling
measurements) were the experimentally measured quantities. With
perfection of more local probes such as nuclear magnetic resonance
(NMR), and especially with development of scanning tunneling
microscopy and spectroscopy (STM/STS), it became possible to
experimentally determine the structures on the atomic scales around
the impurity sites. Therefore the emphasis of theoretical work also
shifted to the study of these local properties.

It is therefore timely and useful to review new results and ideas
about impurity-generated states in superconductors.

We had to be selective about the topics that are included in this
article. In the spirit of new approaches, our review primarily
discusses the physics of the single impurity bound or quasi-bound
states and the local electronic effects in the vicinity of defects.
We also discuss the physics on the mesoscopic scales, and the
behavior of impurities in the presence of competing orders. In the
specific case of high-$T_c$ materials we discuss possible
competition between superconducting state and some competing orders
in the so called pseudogap state of these superconductors.

In all our discussions we restrict ourselves to the study of the
behavior of the density of states. A more comprehensive review of
all the effects that were studied experimentally and discussed
theoretically is a much more difficult task and would take
substantially more space. We do not discuss the behavior of
transport coefficients: while this is a subject of intense current
interest and many important results have been obtained there, it is
beyond the scope of this article.

To keep this review useful and relevant for people entering the
field, we start with a simple Bardeen-Copper-Schrieffer (BCS) model
for superconductivity, and use a modified version of this model
throughout the article. We do not consider the corrections due to
strong coupling that appear in the  Eliashberg analysis; in the
known cases of electron-phonon interaction these corrections are
quantitative rather than qualitative
\cite{JPCarbotte:1990,ESchachinger:1982,ESchachinger:1980,Eschachinger:1984}.
In many unconventional materials dynamical glue in the
self-consistent theory is not known. For example, there is an
ongoing debate on the very nature of the normal state in the
high-$T_c$ cuprate superconductors. Yet most people agree that the
superconducting state of cuprates is less anomalous then the normal
state, and has the superconducting gap of $d$-wave symmetry. We take
a view that at low energies it could be described {\it for the
purposes of our article} by BCS with d-wave pairing state.

At the same time, while this is a review of recent work on impurity
effects in unconventional superconductors, it is emphatically not a
comprehensive review of impurity effects in high-T$_c$ cuprates.
Nature of superconducting state, detailed microscopic description
and competing orders in the cuprates are still a subject of intense
debate at present. There is a number of excellent reviews of physics
of cuprates, including scanning tunneling microscopy
(STM)~\cite{OFischer:2004}, angle-resolved photoemission
spectroscopy (ARPES) \cite{ADamascelli:2003,JCCampuzano:2004} and on
nature of pseudogap state \cite{TTimusk:1999}. Reader is referred to
these reviews for the in depth discussion of the issues specific to
high-$T_c$ materials.

\subsection{Unconventional superconductivity}

Examples of exotic superconductors discovered in the last two
decades include high-$T_c$, heavy fermion superconductors, organic
superconductors, SrRuO$_4$. The common feature of all of them is
that they are unconventional, i.e. the pairing symmetry is not
s-wave, in contrast to conventional materials, such as lead.

Here any superconductor with the gap function that transforms
according to a trivial representation of the point group of the
crystal will be called an s-wave superconductor.  We call a
superconducting order parameter unconventional if it transforms as a
nontrivial representation of the symmetry group. To be more precise,
superconductivity is characterized by an order parameter, that
describes pairing of fermions with time-reversed momenta, $\bk$ and
$-\bk$,
    \beqa \Psi(\bk)_{\alpha,
\beta} = \langle \psi_{\bk, \alpha} \psi_{-\bk, \beta}\rangle
 \label{EQ:Intro1},
    \eeqa
where $\alpha, \beta$ are spin indices of the paired fermionic
states. We distinguish between the spin singlet pairing (total spin
of the pair $S=0$), for which $\Psi(\bk)_{\alpha, \beta} = \Psi(\bk)
(i\sigma^{(y)}_{\alpha \beta}$,  where $\sigma^{(y)}$ is the Pauli
matrix in spin space, and spin triplet state ($S=1$), when
$\Psi_{\alpha\beta}$ is a symmetric spinor in $\alpha, \beta$. Since
the order parameter has to be antisymmetric with respect of
permutation of fermion operators in Eq.~(\ref{EQ:Intro1}), the {\em
spatial} part of $\Psi(\bk)_{\alpha, \beta}$ is even for spin
singlet superconductors and odd in the spin-triplet case. Expanding
in eigenfunctions of orbital momentum, it follows that spin singlet
pairing corresponds to {\em even} orbital function of momentum $\bk$
and hence we call it $s$- (for $l=0$), $d$-wave (for $l=2$), etc.
superconductor in analogy with the notation for the atomic states.
For spin triplet superconductor, the orbital part is an {\em odd}
function of $\bk$, and hence spin triplet superconductor can be
$p$-wave ($l=1$), $f$-wave ($l=3$) etc.  More rigorously one would
characterize pairing states by the irreducible representation of the
symmetry of the crystal lattice, including the spin-orbit
interaction \cite{GEVolovik:1984,EIBlount:1985,MSigrist:1991}.
Characterization in terms of orbital moment is an
oversimplification, and we will use this terminology with
understanding that the correct symmetries are used for a given
crystal structure.  The above classification is given for BCS-like
or {\em even frequency} superconductors. This classification is
opposite for {\em odd-frequency} pairing, where, for example, spin
singlet state has {\em odd} parity because pairing wave function is
odd function of time \cite{VLBerezinskii:1974,AVBalatsky:1992}. We
will focus on BCS like or even-frequency superconductors here.

A reasonable definition of unconventional pairing state, that we
adopt here, is that the order parameter average over the Fermi
surface vanishes :
    \beqa
    \sum_{\bk} \Psi(\bk)_{\alpha \beta} = 0.
    \label{EQ:Intro2}
    \eeqa
Hence superconductors with the constant or nearly constant order
parameter on the Fermi surface are s-wave, while $p$-, $d$- or
higher wave states, where Eq.~(\ref{EQ:Intro2}) holds, are
signatures of an unconventional superconductor. There are several
excellent recent reviews that address the unconventional nature of
superconducting pairing states in specific compounds, such as
$p$-wave superconductivity in SrRuO$_4$ \cite{APMackenzie:2003} and
$d$- wave state in high-$T_c$ materials
\cite{JFAnnett:1990,DJvanHarlingen:1995,CCTsuei:2000}.

\subsection{Outline}

We start with the general overview of BCS-like superconductivity. To
review the effects of impurities we need to discuss the properties
of superconductors in general.  In cuprates, as well as in some
heavy fermion systems and other novel superconductors, there is some
evidence for the existence of an order competing with
superconductivity on all or parts of the Fermi surface. The exact
nature of the competing order parameter is only conjectured. A
general feature of all such models is the enhancement of the
competing order once superconductivity is destroyed, for example in
the vicinity of a scattering center. It has been suggested that the
reaction of the system to the introduction of impurities can be an
important test of the order, or even growing correlations towards
such an order, in the superconducting state.

The prerequisite for such a test is the detailed understanding of
the behavior of ``simple'' superconductors with impurities. Work
aimed at developing this understanding spans a period of more than
40 years, and some of the very recent results continue to be fresh
and unexpected. Therefore we devote a large fraction of this review
to the discussion of the properties of superconductors with
impurities in the absence of any competing order. In this case, from
a theoretical standpoint, before discussing the impurity effects we
need to agree upon methods to describe the very phenomenon that
makes the impurity effects so interesting: superconductivity. Even
in the most exotic compounds investigated so far the superconducting
state itself is not anomalous, in that it results from pairing of
fermionic quasiparticles, and in that these Cooper pairs may be
broken by interaction with impurities or external fields.

Impurity effects in conventional superconductors were subject of the
very early studies by Anderson, so called ``Anderson theorem",
\cite{PWAnderson:1959} and by Abrikosov and Gor'kov
\cite{AAAbrikosov:1960}, hereafter AG. This pioneering work laid the
foundation for our understanding of impurity effects in conventional
and unconventional superconductors, described in terms of electron
lifetime due to scattering on an ensemble of impurities. AG
predicted the existence of the gapless superconductivity that was
subsequently observed in experiments \cite{MAWoolf:1965}. The brief
summary of the AG theory and its extensions to non-$s$-wave
superconductivity is given Table~\ref{TABLE:ANDERSON} where effect
of impurities on the superconducting state {\em on average, or
globally} is listed.

\begin{table}[H]
  \begin{tabular}{|c|c|c|c|}
     \hline
      & $S$-WAVE & $P$-WAVE & $D$-WAVE \\
     \hline
     POTENTIAL SCATTERING& $-$ & $+$ & $+$ \\
     \hline
     MAGNETIC SCATTERING& $+$ & $+$ & $+$ \\
     \hline
   \end{tabular}
   \caption{Effects of potential and magnetic
   impurity scattering on the $s$-, $p$- and $d$-wave superconductors is shown qualitatively.
   ``$+$'' indicate that impurity
   scattering is a pairbreaker and ``$-$'' is that impurity scattering is not a pairbreaker.
   There is a qualitative difference between the potential scattering in $s$-wave superconductors and any other
   case. Potential scattering impurities are not pairbreakers in $s$-wave case due to Anderson theorem.
   This is an exceptional case. For any other case any impurity scattering will suppress superconductivity. Obviously
   the details depend on scattering strength and other details.  At high enough concentrations both magnetic and
   nonmagnetic
   impurities will suppress superconductivity regardless of
   symmetry.
   \label{TABLE:ANDERSON}}
    \end{table}

After intense interest in the early days of the BCS theory, the
subject was considered ``closed'' in mid-60s, with most
experimentally relevant problems solved. However, as often happens,
recently there has been a revival of the interest in the studies of
``traditional'' low-temperature $s$-wave superconductors with
magnetic and non-magnetic impurities, with many new theoretical and
experimental results changing our perspective on this classical
problem.

A special place in this review is devoted to the study of impurity
induced {\em local} bound states or resonances. This is an old
subject, going back to the 60's when the  bound states near magnetic
impurities in $s$-wave superconductors were predicted in a
pioneering work of Yu, Shiba and Rusinov
\cite{LYu:1965,HShiba:1968,AIRusinov:1969}. They  considered
pairbreaking by a {\em single magnetic impurity} in a
superconductor, and found that there are quasiparticle states inside
the energy gap that are localized in the vicinity of the impurity
atom. The corresponding gap suppression occurs locally and the
concept of lifetime broadening is inapplicable. In general, in this
situation it is more useful to focus on local quantities, such as
local density of states (LDOS), local gap etc., rather than on
average impurity effects (which vanish for the single impurity in
the thermodynamic limit). Yet it is clear that this local physics at
some finite concentration of impurities suppresses superconductivity
completely. This connection was discussed in
\cite{LYu:1965,HShiba:1968,AIRusinov:1969}. In particular, formation
of the intragap bound state and impurity bands due to magnetic
impurity leads to filling of the superconducting gap, and therefore
connects to the AG theory \cite{AAAbrikosov:1960}.

At the time there were no experimental techniques to directly
observe single impurity  states. As a result the entire subject was
largely forgotten until the STM was applied to study the impurity
states by Yazdani et al.~\cite{AYazdani:1997}. This reinvigorated
the field and lead to a firm shift in the interest from global to
local properties. Soon afterwards STM was used to observe local
impurity states near vacancies and impurities in the high-$T_c$
cuprates~\cite{SHPan:2000b,AYazdani:1999,EWHudson:1999,EWHudson:2001}.
These discoveries opened a new field of research where impurities
open a window into the study of electronic properties of exotic
materials with atomic spatial resolution. As a first test of
theories this allowed a direct comparison of the local electronic
features in tunneling characteristics  with the theoretical
predictions for the density of states.

We start  by briefly reviewing the BCS theory in Sec.~\ref{sec:BCS}.
Our main goal there is to review three approaches that will be used
to analyze the impurity effects: direct diagonalization of the
hamiltonian via Bogoliubov-Valatin transformation, variational wave
function of the original BCS paper, and the Green's function method
which is well suited to the analysis of multiple impurity problems.
Then we define different types of impurity scattering in
Sec.~\ref{sec:Imp}. We pay special attention to distinguishing
between magnetic and non-magnetic impurities, and differentiating
between static and dynamic scatterers. The basic features of
non-magnetic scattering in $s$-wave superconductors are outlined in
Sec.~\ref{sec:Anderson}.

To keep in tune with our intention to make the review readable by
graduate students and researchers entering the field, we begin the
discussion of the localized states by considering an example of an
impurity bound state in a two-dimensional (2D) metal in
Sec.~\ref{sec:imp2Dmetal}. Then we  discuss the low-energy bound
state in $s$- and $d$-wave superconductors in Sec.~\ref{sec:Shiba}
and Sec.~\ref{sec:Dwave} respectively. Changes in the ground state
of a superconductor containing a classical spin as a function of the
coupling strength between the spin and conduction electrons are
discussed in Sec.~\ref{sec:QPT}.

We proceed to consider the situations when the impurities have their
own dynamics, so that their effect on the electrons is complicated,
see Sec.~\ref{sec:DynamicalImp}, and the combined influence of the
collective modes and impurities, Sec.~\ref{sec:Interplay}. We
briefly touch upon possible existence of impurity resonances in
different models of the pseudogap state of the cuprates in
Sec.~\ref{sec:PG}, and discuss recent STM measurements on both
conventional and unconventional superconductors in
Sec.~\ref{sec:STM}. The final two parts of our review are devoted to
the discussion of the effects on impurities on meso- and macroscopic
scale. For completeness, we briefly review the basics ideas of
computing the average density of states for a macroscopic sample in
Sec.~\ref{sec:AverDOS}. For lack of space we cannot do justice to
this very rich subject and use it largely to discuss new results on
the impurity effect on the scales small compared to the sample size,
but large relative to the superconducting coherence length. In that
situation there are dramatic consequences of local impurity
realizations that may be different from the average, and we overview
the results for the density of states in Sec.~\ref{sec:OF}. We
conclude with the summary in Sec.~\ref{sec:Conclusion}.

\subsection{Other related work}

In focusing largely on the properties of impurities on atomic or
mesoscopic scales, we cannot give due attention within the confines
of this review to several other questions that have been important
in the studies of impurities. One of these is how exactly does the
impurity band grow out of bound states on individual impurity sites,
i.e. what is the effect of interference between such sites is in
real space. We briefly review some of recent work in
Sec.~\ref{sec:AverDOS}, but do not discuss the subject in depth. The
answer to this question is still somewhat unsettled even in general:
while the usual finite lifetime approach
\cite{LPGorkov:1985,SSchmitt-Rink:1986,PJHirschfeld:1986} gives a
constant density of states at the Fermi level in a $d$-wave
superconductor, and even though the same result has been obtained in
field theoretical models of Dirac fermions in two dimensions,
mimicking the $d$-wave superconductor
\cite{KZiegler:1996a,KZiegler:1996b}, it has also been argued that
this DOS diverges \cite{CPepin:1998,CPepin:2001}, or vanishes.
Vanishing DOS can occur with different power laws in energy
depending on the approach one uses to treat disorder
\cite{TSenthil:1999,AANersesyan:1995,AANersesyan:1997} (see also
\cite{MJBhaseen:2001}). The vanishing itself can be traced to level
repulsion when the system is treated within random matrix theory
\cite{AAltland:1997}. Detailed self-consistent numerical studies
indicate, however, the the behavior of the DOS depends on the
details of the impurity scattering and electronic structure
\cite{WAAtkinson:2000,JXZhu:2000c}. In particular, the divergence
only occurs in perfectly particle-hole symmetric systems, and
generically Atkinson et al. find that there is a non-universal
suppression of the density of states over a small energy scale close
to the Fermi level.

The interference between many impurities have been investigated
recently \cite{LZhu:2004,WAAtkinson:2003,LZhu:2003} with the eye on
the importance of these effects for the interpretation of the
features in the STM data on the high-$T_c$ cuprates collected over a
large area of the sample. The interference is also responsible for
the formation of the impurity bands and therefore is crucial for
determining the transport properties, which we do not address in
this review.  Within the framework of the $t$-matrix approximation
transport properties of unconventional superconductors in general
\cite{PJHirschfeld:1986,SSchmitt-Rink:1986,CPethick:1986,PJHirschfeld:1988,BArfi:1988,PJHirschfeld:1989,MJGraf:1996},
and high-$T_c$ cuprates in particular
\cite{PJHirschfeld:1993,PJHirschfeld:1994,SMQuinlan:1994,MJGraf:1995,SMQuinlan:1996,PJHirschfeld:1997,DDuffy:2001}
have been extensively discussed, and the experiments on both
microwave, optical, and thermal conductivity are used to extract
properties of impurity scattering, see \cite{TTimusk:1999} for a
review as well as very recent results in both experiment
\cite{AHosseini:1999,GLCarr:2000,MChiao:2000,JCorson:2000,JJTu:2002,GPSegre:2002,PJTurner:2003,RWHill:2004,YSLee:2004}
and theory
\cite{MHHettler:1999,AJBerlinsky:2000,AVChubukov:2003,EJNicol:2003,PCHowell:2004}.
The question of localization in both $s$-wave \cite{MMa:1985} and
$d$-wave
\cite{PALee:1993,TSenthil:1998,SVishveshwara:2000,TSenthil:2000,AGYashenkin:2001,WAAtkinson:2002}
continues to be investigated. Some of these results have been
summarized in recent reviews on high-T$_c$ systems
\cite{TTimusk:1999}. We also do not touch upon the rich phenomena
related to the surfaces playing the role of extended impurities that
can also lead to the formation of the bound states
\cite{CRHu:1994,GEBlonder:1982,LJBuchholtz:1981,MCovington:1997,MFogelstrom:1997,MAprili:1998,SKashiwaya:2000}.

By now there are also few reviews available on the subject of
impurity states. Joynt \cite{RJoynt:1997} reviewed early work on the
impurity states within the $t$-matrix theory focusing on anomalous
transport due to finite lifetime of the quasibound states around
impurities. Byers, Flatte and Scalapino , contributed substantially
to studies of the detailed electronic structure of the resonance
state and interference patterns
\cite{JMByers:1993,MEFlatte:1997a,MEFlatte:1997b,MEFlatte:1998}, and
reviewed their and related work \cite{MEFlatte:1999}. An excellent
review of thermal and transport properties of low-energy
quasiparticles in nodal superconductors was recently given by Hussey
\cite{NEHussey:2002}.

The subject is so rich and well developed that it does not seem
possible to do justice to both local quasiparticle properties around
a single impurity site and the questions of interference and
transport within the confines of a single paper.  With this in mind
we now are ready for a main discussion.


\section{A BCS theory primer}

\label{sec:BCS}

We begin by reviewing the Bardeen-Cooper-Schrieffer (BCS) theory.
This section only briefly summarizes the results pertinent to our
discussion; many excellent textbooks provide an in-depth view on the
theory
\cite{JRSchrieffer:1964,PGdeGennes:1989,MTinkham:1996,JBKetterson:1999}.
Consider a general hamiltonian ${\cal H}_{BCS}=\widehat H_0({\bf r})
+ H_{int}$, where
\begin{equation}
  \widehat H_0({\bf r})=\sum_\alpha \int d^d{\bf r}
    \psi^\dagger_\alpha({\bf r}) [\epsilon(-i \bm{\nabla}_{\bf
    r})-\mu]
    \psi_\alpha({\bf r})
\end{equation}
is the band hamiltonian of quasiparticles with dispersion
$\epsilon({\bf k})$, $\mu$ is the chemical potential, and the
interaction part
    \begin{equation}
    H_{int}=-\frac{1}{2}\sum_{\alpha,\beta \atop \gamma,\delta}\int d^d{\bf r}
    d^d{\bf r}^\prime
    \psi^\dagger_\alpha({\bf r})\psi^\dagger_\beta({\bf r}^\prime)
    V_{\alpha\beta\gamma\delta} ({\bf r},{\bf r}^\prime)
    \psi_\gamma({\bf r}^\prime)\psi_\delta({\bf r}).
    \end{equation}
Here {\bf r} is the real space coordinate, $\alpha$ and $\beta$ are
the spin indices, and $\psi^\dagger$ and $\psi$ are the fermionic
creation and annihilation operators respectively. The mean field
approximation consists of decoupling the four-fermion interaction
into a sum of all possible bilinear terms, so that
    \begin{eqnarray}
    \label{EffectiveH:general}
    H_{int}&=&\sum_{\alpha,\beta}\int d^d{\bf r} d^d{\bf r}^\prime
    \biggl\{
    \widetilde V_{\alpha\beta} ({\bf r},{\bf r}^\prime)
    \psi^\dagger_\alpha({\bf r}) \psi_\beta({\bf r}^\prime)
    \\
    \nonumber
    &&
    +
    \Delta_{\alpha\beta} ({\bf r}, {\bf r}^\prime)
    \psi^\dagger_\alpha({\bf r})\psi^\dagger_\beta({\bf
    r}^\prime)
    + \Delta^\star_{\alpha\beta} ({\bf r}, {\bf r}^\prime)
    \psi_\beta({\bf r})\psi_\alpha({\bf
    r}^\prime)
    \biggr\}.
    \end{eqnarray}
The effective potential, $\widetilde V_{\alpha\beta} ({\bf r},{\bf
r}^\prime)$ is the sum of the Hartree and Fock (exchange) terms, and
the last two terms account for superconducting pairing. The pairing
field, $\Delta$, is determined self-consistently from
    \begin{equation}
    \label{Delta:BCS}
    \Delta_{\alpha\beta} ({\bf r}, {\bf r}^\prime)=
    \frac{1}{2} V_{\alpha\beta \gamma\delta} ({\bf r},{\bf r}^\prime)
    \langle \psi_\gamma({\bf r}^\prime)\psi_\delta({\bf
    r})\rangle.
    \end{equation}

The pairing occurs only below the transition temperature, $T_c$;
above $T_c$ the average of the two annihilation operators in
Eq.~(\ref{Delta:BCS}) vanishes, and therefore
$\Delta_{\alpha\beta}=0$. In contrast, Hartree and Fock terms are
finite at all temperatures, and can be incorporated in the
quasiparticle dispersion, $\epsilon({\bf k})$. These terms do change
below $T_c$, upon entering the superconducting state. Their relative
change, however, is of the order of the fraction of electrons
participating in superconductivity, and therefore is small for weak
coupling superconductors ($\sim \Delta/W\ll 1$, where $W$ is the
electron bandwidth). Therefore the effective potential, $\widetilde
V$, is not explicitly included in the following discussion except
where specified.

Therefore we start with a reduced mean field BCS hamiltonian,
    \begin{eqnarray}
    \label{Hamiltonian:BCS}
    {\cal H}_{BCS}&=&\sum_\alpha \int d^d{\bf r}
    \psi^\dagger_\alpha({\bf r}) \widehat H_0({\bf r})
    \psi_\alpha({\bf r})
    \\
    \nonumber
    &+&
    \sum_{\alpha,\beta}\int d^d{\bf r} d^d{\bf r}^\prime
    \biggl\{\Delta_{\alpha\beta} ({\bf r}, {\bf r}^\prime)
    \psi^\dagger_\alpha({\bf r})\psi^\dagger_\beta({\bf
    r}^\prime)
    + h. c.
    \biggr\}.
    \end{eqnarray}

The spatial and spin structure of $\Delta_{\alpha\beta} ({\bf r},
{\bf r}^\prime)$ determines the type of superconducting pairing. In
most of this review we consider singlet pairing, when $\Delta$ has
only the off-diagonal matrix elements in spin space, and it is
common to write $\Delta_{\alpha\beta} ({\bf r}, {\bf r}^\prime)=
(i{\sigma}^y)_{\alpha\beta}\Delta ({\bf r}, {\bf r}^\prime)$, where
$\Delta$ is now a scalar function, see previous section.

In a uniform superconductor the interaction depends only on the
relative position of the electrons, so that $V({\bf r},{\bf
r}^\prime)=V({\bm \rho}\equiv{\bf r}-{\bf r}^\prime)$. Therefore in
the absence of impurities, the structure of the order parameter in
real space depends on the symmetry properties of $V({\bm\rho})$.
These are easier to consider in momentum, rather than coordinate,
space. In models with local attraction, when
$V({\bm\rho})=V_0\delta({\bm\rho})$, the Fourier transform of the
interaction is featureless, and $\Delta({\bf k})=\Delta_0$; an
example of an isotropic, or $s$-wave superconductor.

In the remainder of this section we overview the main methods
solving the BCS hamiltonian since the same methods are commonly
applied to the studies of impurity effects in superconductors. The
approaches that we consider are: a) direct diagonalization via
Bogoliubov-Valatin transformation; b) variational determination of
the ground state energy from the trial wave function; and c) Green's
function formalism.

\subsection{Bogoliubov transformation}
\label{sec:BdG}

Since the effective hamiltonian of Eq.~(\ref{Hamiltonian:BCS}) is
bilinear in fermion operators, $\psi$ and $\psi^\dagger$, it can be
diagonalized by a canonical transformation of the form
    \begin{equation}
    \psi_\alpha ({\bf r})=\sum_n
    \biggl[
    u_{n\alpha}({\bf r})\gamma_{n} +
    v_{n\alpha}({\bf r})\gamma^\dagger_{n}
    \biggr],
    \end{equation}
subject to condition $|u_{n\alpha}({\bf r})|^2+
    |v_{n\alpha}({\bf r})|^2=1$. The resulting equations on the coefficients
$u$ and $v$ are
    \begin{eqnarray}
    \label{BogoliubovEqU:general}
    E u_\alpha({\bf r})&=& H_0 ({\bf r})u_\alpha({\bf r})
        +
    \int d^d{\bf r}^\prime
    \Delta_{\alpha\beta}({\bf r},{\bf r}^\prime)
            v_\beta({\bf r}^\prime),
    \\
    \label{BogoliubovEqV:general}
    - E v_\alpha({\bf r})&=& H^\star_0 ({\bf r})v_\alpha({\bf r})
        +
    \int d^d{\bf r}^\prime
    \Delta^\star_{\alpha\beta}({\bf r},{\bf r}^\prime)
            u_\beta({\bf r}^\prime).
    \end{eqnarray}
Here we suppressed the label $n$ for brevity. Clearly, when
$\Delta=0$, coefficients $u$ and $v$ do not couple, and there is no
particle-hole mixing.

For each $n$ there are four functions, $u_\uparrow({\bf
r}),u_\downarrow({\bf r}),v_\uparrow({\bf r}),u_\downarrow({\bf r})$
that need to be determined. However, for a singlet superconductor,
for example, the matrix $\Delta_{\alpha\beta}$ is off-diagonal in
the spin indices, so that $u_\uparrow$ ($u_\downarrow$) couples only
to $v_\downarrow$ ($v_\uparrow$), so that in practice only two of
the equations are coupled. In the presence of the impurity
potential, however, in general all four components become
interdependent.

Equations~(\ref{BogoliubovEqU:general})-(\ref{BogoliubovEqV:general}),
are coupled integro-differential equations for the functions
$u_{n\alpha}({\bf r})$ and $v_{n\alpha}({\bf r})$. They have to be
complemented by the self-consistency equations on
$\Delta_{\alpha\beta}$, which can be obtained directly from
Eq.~(\ref{Delta:BCS}) to be
    \begin{eqnarray}
    \nonumber
    \Delta_{\alpha\beta} ({\bf r}, {\bf r}^\prime)&=&
    -\frac{1}{2}
    V_{\alpha\beta \gamma\delta}({\bf r},{\bf r}^{\prime})
    \sum_n
    \biggl[
    u_{n\gamma}({\bf r}^\prime) v^\star_{n\delta} ({\bf r}) f(E_n)
    \\
    \label{Eq:DeltaBCSSelf-Cons}
    &&
    \qquad
    +
    v^\star_{n\gamma}({\bf r}^{\prime})
    u_{n\delta}({\bf r})
    (1-f(E_n))
    \biggr].
    \end{eqnarray}
Here the Fermi function $f(E)=[\exp(E/T)+1]^{-1}$.

In a uniform superconductor the Fourier transform of the Bogoliubov
equations,
Eqs.~(\ref{BogoliubovEqU:general})-(\ref{BogoliubovEqV:general}),
into the momentum space gives
   \begin{eqnarray}
    \label{BogoliubovEqK:U}
    (\xi_{\bf k}-E_{\bf k})u_{{\bf k}\alpha}
    +\Delta_{\alpha\beta}({\bf k})
    v_{{\bf k}\beta}
    =0,
    \\
    \label{BogoliubovEqK:V}
    (\xi_{\bf k}+E_{\bf k})v_{{\bf k}\alpha}
    +\Delta^\star_{\alpha\beta}(-{\bf k})
    u_{{\bf k}\beta}
    =0,
    \end{eqnarray}
where $\xi_{\bf k}$ is the bare quasiparticle energy, measured with
respect to the chemical potential, $\xi_{\bf k}=\epsilon({\bf
k})-\mu$. In a singlet superconductor
   \begin{eqnarray}
    \label{BogoliubovEqK:singletU}
    (\xi_{\bf k}-E_{\bf k})u_{{\bf k}\uparrow}
    +\Delta ({\bf k})
    v_{{\bf k}\downarrow}
    =0,
    \\
    \label{BogoliubovEqK:singletV}
    (\xi_{\bf k}+E_{\bf k})v_{{\bf k}\downarrow}
    -\Delta^\star({\bf k})
    u_{{\bf k}\uparrow}
    =0,
    \end{eqnarray}
and recover the familiar energy spectrum $E_{\bf k}=\sqrt{\xi_{\bf
k}^2+|\Delta ({\bf k})|^2}$, with the coefficients $u$ and $v$ given
by
\begin{equation}
    \label{uv}
  \Biggl(  \begin{matrix}
    u_{\bf k}^2 \\ v_{\bf k}^2
    \end{matrix} \Biggr)=\frac{1}{2}\Biggl[1
    \pm\frac{\xi_{\bf k}}{E_{\bf k}}\Biggr].
\end{equation}

\subsection{BCS variational wave function}
\label{sec:BCSWaveFunction}

Superconductivity originates from the instability of the Fermi sea
towards pairing of time-reversed quasiparticle states. Therefore a
variational wave function approach, originating with the classic BCS
paper, is to restrict the trial wave function to the subspace of
either empty or doubly occupied states,
    \begin{equation}
    \label{PsiBCS}
    |\Psi({\bf r})\rangle= \prod_n
    (a_n+b_n c^\dagger_{n\uparrow} c^\dagger_{\bar n \downarrow})
    |0\rangle,
    \end{equation}
and to minimize the energy, $E_{BCS}=\langle\Psi|H|\Psi\rangle$.
This is a legitimate approximation at $T=0$, and is a very good
approach at low temperatures. In Eq.~(\ref{PsiBCS}) the vacuum state
$|0\rangle$ denotes the filled Fermi sea, and
$c^\dagger_{n\uparrow}$ ($c^\dagger_{\bar n\downarrow}$) creates a
quasiparticle with spin up (down) and with the wave function
$\phi_n({\bf r})$ ($\phi^\star_n({\bf r})$) that is the
eigenfunction of the single particle Hamiltonian. Normalization
requires that $|a_n|^2+|b_n|^2=1$.

In the absence of impurities these eigenfunctions can be labeled by
the same indices, {\bf k} and $\alpha$, as in the previous section.
Consequently, the variational approach is completely equivalent to
the Bogoliubov analysis with the choice $u_n ({\bf r})= a_n
\phi_n({\bf r})$, and $v_n ({\bf r})= b_n \phi_n({\bf r})$. In
general, however, interaction with impurities may lead to the
appearance of the single particle states in the ground state wave
function, see Sec.~\ref{sec:QPT}. Moreover, it is worth remembering
that energy of the state described by the BCS wave function is
greater or equal to that of the exact ground state obtained by
solving the Bogoliubov equations.

\subsection{Green's functions}
\label{sec:GreensFunction}

The third approach that we will use in this work is the Green's
function method, which originates with the work of Gor'kov.
Following Nambu we introduce a 4-vector that is a spinor
representation of the particle and hole states,
\begin{equation}
  \Psi^\dagger({\bf r})=(\psi^\dagger_\uparrow,
  \psi^\dagger_\downarrow, \psi_\uparrow, \psi_\downarrow).
\end{equation}
The matrix Green's function is defined as the imaginary-time ordered
average
\begin{equation}
  \widehat G (x, x^\prime)=-\langle T_{\tau} \Psi(x)
   \Psi^{\dagger}(x')\rangle,
\end{equation}
where the four-vector $x=({\bf r},\tau)$ combines the real space
coordinate, {\bf r}, and the imaginary time, $\tau$. The time
evolution of the creation and annihilation operators in the
Heisenberg approach is given by $\partial\psi/\partial\tau=[{\cal
H}_{BCS}-\mu N, \psi]$.

For a singlet homogeneous superconductor the Hamiltonian of
Eq.~(\ref{Hamiltonian:BCS}) in the Nambu notation takes the form,
\begin{equation}
  H_{BCS}=\int d{\bf r} \Psi^\dagger({\bf r})
  (\xi(-i\nabla)\tau_3 + \Delta\tau_1\sigma_2)\Psi({\bf r}),
\end{equation}
and we find~\cite{KMaki:1969}
\begin{equation}
    \label{Eq:G-BCS}
  \widehat G_0^{-1}({\bf k}, \omega)=i\omega_n -\varepsilon({\bf k})\tau_3-
  \Delta ({\bf k}) \sigma_2\tau_1.
\end{equation}
Here $\omega_n=\pi T (2n+1)$ is the Matsubara frequency, $\sigma_i$
are the Pauli matrices acting in spin space, $\tau_i$ are the Pauli
matrices in the particle-hole space, and $\tau_i\sigma_j$ denotes a
direct product of the matrices operating in the 4-dimensional Nambu
space. The self-consistency equation for a single superconductor
takes the form
\begin{equation}
  \Delta({\bf k})=-T\sum_{\omega_n}\int d{\bf k}^\prime V({\bf k},
  {\bf k}^\prime) \Tr [\tau_1\sigma_2 G_0].
\end{equation}
In BCS the interaction is restricted to a thin shell of electrons
near the Fermi surface, and therefore
\begin{equation}
    \label{Eq:DeltaSelfConGen}
  \Delta(\widehat\Omega)=-TN_0\sum_{\omega_n}\int d\widehat\Omega^\prime
  V(\widehat\Omega,\widehat\Omega^\prime)
  \Tr \biggl[\tau_1\sigma_2 \int d\xi_{\bf k} G_0\biggr],
\end{equation}
where $\widehat\Omega$ denotes a direction on the Fermi surface, and
$N_0$ is the normal state density of states.

The off-diagonal component of $\widehat G_0$, is often called the
Gor'kov's anomalous $F$, (Gor'kov) Green's functions since it
describes the pairing average
\begin{eqnarray}
   F_{\alpha\beta}(x, x')=
     -\langle T_{\tau}\psi_{\alpha}(x)\psi_{\beta}(x')\rangle .
\end{eqnarray}
In general $F_{\alpha\beta}(x, x^\prime)={\mathrm g}_{\alpha\beta}F
(x, x^\prime)$, where $\mathrm{g}$ is the matrix describing the spin
structure of the superconducting order. For the singlet pairing
${\mathrm g}=i\sigma^{(y)}$, where $\sigma^{(y)}$ is the Pauli
matrix. Therefore in a singlet spatially uniform superconductor
normal and anomalous components of $\widehat G_0$ are
\begin{eqnarray}
  G(\omega_n, {\bf k})&=&\frac{i\omega_n+\xi_{\bf k}}
  {(i\omega_n)^2-\xi^2_{\bf k}-|\Delta({\bf k})|^2},
  \\
    F(\omega_n, {\bf k})&=&\frac{\Delta({\bf k})}
  {(i\omega_n)^2-\xi^2_{\bf k}-|\Delta({\bf k})|^2}.
\end{eqnarray}

The connection with the Bogoliubov's transformation is provided by
rewriting the Green's functions as
\begin{eqnarray}
  G(\omega_n, {\bf k})&=&\frac{u_{\bf k}^2}{i\omega_n-E_{\bf k}}+
  \frac{v_{\bf k}^2}{i\omega_n+E_{\bf k}},
  \\
    F(\omega_n, {\bf k})&=&u_{\bf k} v_{\bf k}^\star
  \biggl(\frac{1}{i\omega_n-E_{\bf k}}-
  \frac{1}{i\omega_n+E_{\bf k}}\biggr),
\end{eqnarray}
where $u_{\bf k}$ and $v_{\bf k}$ are given by Eq.~(\ref{uv}).

The three approaches discussed above are complementary and
equivalent in the case of homogeneous superconductors. However, some
of them are better suited for addressing specific questions in the
presence of impurities. In particular, we will see that the Green's
function method is sometimes advantageous for determining the
thermodynamic properties of a material and averaging over many
impurity configurations. For inhomogeneous problems, where we are
interested in the spatial variations of the superconducting order
and electron density, both Bogoliubov equations and Green's
functions are often used. In general, the choice of a specific
methods depends on the type of question asked in the presence of
impurities, and we briefly describe the basic models and issues
related to impurity scattering in superconductors below.

\section{Impurities in superconductors}

\label{sec:Imp}

\subsection{Single impurity potential}

If we are to address theoretically the question of what defects do
to superconductivity, we must describe the defects and
superconductivity in the same framework. Grain and surface
boundaries, twinning planes, and other structural inhomogeneities
scatter conduction electrons and therefore affect the resulting
order parameters. However, here  we focus on only one type of
imperfection: impurity atoms.

\paragraph{Potential scattering.}  First and foremost an impurity atom has
a different electronic configuration than the host solid, and
therefore interacts with the density of conduction electrons via a
Coulomb potential.
    \begin{equation}
    \label{Eq:HPotExt}
    H_{imp}=\sum_{\alpha} \int d{\bf r} \psi^\dagger_\alpha({\bf r})
    U_{pot} ({\bf r})
    \psi_\alpha({\bf r}).
    \end{equation}
In good metals the Coulomb interaction is screened at the length
scales comparable to the lattice spacing, and therefore the
resulting scattering potential is often assumed to be completely
local, $U_{pot} ({\bf r})=U_0 \delta({\bf r}-{\bf r}_0)$, with the
impurity at ${\bf r}_0$. The resulting scattering occurs only in the
isotropic, $s$-wave, angular momentum channel. If finite range of
the interaction is relevant, scattering in $l\neq 0$ channels needs
to be considered. In that case the treatment is similar to that of
magnetic scattering in conventional superconductors, see
Sec.~\ref{sec:Shiba}, and was applied to unconventional
superconductors in, for example \cite{AVBalatsky:1994,APKampf:1997}.

In the 4-vector notation of the previous section the potential
scattering has to have the same matrix structure as the chemical
potential, or $\varepsilon({\bf k})$ in Eq.~(\ref{Eq:G-BCS}), so
that
\begin{equation}
    \label{Eq:HPot4}
    H_{imp}=\int d{\bf r} \Psi^\dagger({\bf r})
    U_{pot}({\bf r}) \tau_3
    \Psi({\bf r}),
    \end{equation}
or, in Nambu notation,
\begin{equation}
  \widehat U_{pot}=U_0 \tau_3\delta({\bf r}-{\bf r}_0)
\end{equation}

\paragraph{Magnetic scattering.}
In addition to the electrostatic interactions, if the impurity atom
has a magnetic moment, there is an exchange interaction between the
local spin on the impurity site and the conduction electrons,
\begin{equation}
\label{Eq:HMagExt}
    H_{imp}=\sum_{\alpha\beta}\int d{\bf r} J({\bf r})
    \psi^\dagger_\alpha({\bf r})
    {\bf S}\cdot {\bm{\sigma}}_{\alpha\beta}
    \psi_\beta({\bf r}).
\end{equation}
The range of interaction here is determined by the quantum
mechanical structure of the electron cloud associated with the
localized spin. Again, in reality we often consider a simplified
exchange hamiltonian with $J({\bf r})=J_0\delta({\bf r}- {\bf
r}_0)$, which captures the essential physics of the problem. In the
4-vector notations of the previous section the electron spin
operator becomes
\begin{equation}
\label{MatrixAlpha}
  {\bm\alpha}=\frac{1}{2}\biggl[
  (1+\tau_3){\bm\sigma}+(1-\tau_3)\sigma_3{\bm\sigma}\sigma_3\biggr].
\end{equation}
Therefore
\begin{equation}
    \label{Eq:HMag4}
    H_{imp}=\int d{\bf r} \Psi^\dagger({\bf r})
    J({\bf r}){\bf S}\cdot {\bm{\alpha}}
    \Psi({\bf r}),
\end{equation}
or, in Nambu notation,
\begin{equation}
  \widehat U_{mag}=J({\bf r}){\bf S}\cdot {\bm{\alpha}}.
\end{equation}

\paragraph{Anderson impurity.}
However, even if the ground state of an isolated impurity has an
electron spin, the result of putting such an impurity into a host
matrix may modify the spin configuration as the impurity electrons
couple to the conduction band.  Therefore a realistic model for an
impurity site is based on the Anderson model, with the Hamiltonian
\begin{eqnarray}
    \label{AndersonImp1}
  H_{A}&=&\sum_\alpha E_0 d^\dagger_\alpha d_\alpha +
  Un_{d\uparrow}n_{d\downarrow} +H_{sd},
  \\
  \label{AndersonImp2}
  H_{sd}&=&\sum_{\bf k,\alpha} V_{sd} c^\dagger_{{\bf k},\alpha}
  d_\alpha + h.c.
\end{eqnarray}
Here $E_0$ is the position of the impurity level relative to the
Fermi energy, $d^\dagger$ and $d$ operate on the impurity site, $U$
is the Coulomb repulsion for the electrons localized on the impurity
site, and $c^\dagger_{\bf k},c_{\bf k}$ create and annihilate the
conduction electrons. This Hamiltonian allows the electrons to hop
on and off the impurity site, resulting in a finite width of the
impurity level, $\Gamma=\pi |V_{sd}|^2N_0$. The model describes the
potential scattering, when $U\ll \Gamma$. On the other hand, when
$E_0\ll E_F$, $E_0+U\gg E_F$, and $U\gg \Gamma$, we expect the local
levels to remain split, so that the impurity state is singly
occupied and has a local spin. Therefore it allows a natural
interpolation between potential and magnetic scattering, as well as
the study of the mixed valence regime. The price to pay for such a
rich behavior of the Anderson impurities is the difficulty of
studying them analytically, and therefore in practice many results
have been obtained in the simplified models above, although a number
of very thorough numerical renormalization group studies of Anderson
impurities in superconductors exist. We will review some of them for
completeness, but will not focus on those extensively.

\subsection{Many impurities}

\label{ImpAvBasic}

In all of our discussions we assume noninteracting impurities, so
that the net impurity potential is simply
\begin{eqnarray}
  \widehat U_{imp}({\bf r})&=&
  \sum_i \widehat U_{imp}({\bf r}-{\bf
  r}_i)
  \\
  &=&\int d{\bf
  r}^\prime\rho_{imp}({\bf r}^\prime)
  \widehat U_{imp} ({\bf r}-{\bf r}^\prime).
\end{eqnarray}
Here $\widehat U$ denotes the matrix structure of the potential in
both spin and particle-hole space, and we introduced the impurity
density,
\begin{equation}
  \rho({\bf r})=\sum_i \delta({\bf r}-{\bf r}_i).
\end{equation}
 We also assume the dilute impurity limit of the average
impurity concentration $n_i\ll 1$, where
\begin{equation}
  n_i=\int \frac{d{\bf r}}{V}\rho({\bf r}).
\end{equation}
For magnetic scatterers it was explicitly shown that the effect of
the RKKY interaction between scattering centers on the
superconducting properties is small
\cite{AILarkin:1971,VMGalitskii:2002}.

If we now compute a local physical quantity, such as the density of
states measured at the position {\bf r} by the STM, it will depend
on the distance from the nearby impurities, and therefore will be
different for different realization of impurity distributions. In
contrast, thermodynamic quantities, such as the density of states
measured in planar junctions, or the specific heat, average the
density of states over many random local configurations of
impurities. Therefore in computing their values we average over {\it
all}  impurity configurations \cite{AAAbrikosov:1963}, so that, for
example,
\begin{equation}
\label{Gimp}
  \bar G(\omega_n, {\bf k})=
  \prod_{i=1}^{N_i}\biggl[\frac{1}{V}\int d{\bf r}_i G(\omega_n, {\bf k},
  {\bf r}_1,
  \ldots, {\bf r}_{N_i})\biggr].
\end{equation}
Here a bar denotes such an impurity average.

By definition $\bar\rho_{imp}=n_i$. We also assume an {\it
uncorrelated}, or random, impurity distribution, which means
\begin{eqnarray}
    \nonumber
  \overline{\rho({\bf r})\rho ({\bf r}^\prime)}&\equiv&
  \prod_{i=1}^{N_i}\biggl[\frac{1}{V}\int d{\bf r}_i
    \rho({\bf r},{\bf r}_1,
  \ldots, {\bf r}_{N_i})\rho ({\bf r}^\prime,{\bf r}_1,
  \ldots, {\bf r}_{N_i})\biggr]
  \\
  &=&n_i\delta({\bf r}-{\bf r}^\prime) +
  n_i^2.
\end{eqnarray}
Since the impurities are dilute, $n_i^2\ll n_i$, and we neglect the
second term compared to the first. In Sec.~\ref{sec:AverDOS} we
implement this impurity averaging procedure to determine the average
density of states.

\subsection{The self-energy and the $T$-matrix approximation}

In practice to compute the Green's function in the presence of
impurities we will often employ the $T$-matrix approximation. This
method is described in detail in many original articles and reviews
\cite{PJHirschfeld:1986,PJHirschfeld:1988,PJHirschfeld:1993,THotta:1993,NEHussey:2002,GDMahan:2000},
and we only briefly summarize it.

For a single impurity with the scattering potential $\widehat
U_{{\bf k},{\bf k}^\prime}$ in the momentum space (given by one of
the models discussed at the beginning of this chapter), the
$T$-matrix accounts exactly for multiple scattering off of one
impurity. In the language of Feynman diagrams, the corresponding
process is shown in Fig.~\ref{fig:TMatrix}. Here, and throughout the
review, the hat over a letter means that it denotes a matrix in
Nambu space. Therefore the full Green's function is
\begin{eqnarray}
  \label{GTmom}
  \widehat G({\bf k}, {\bf k}^\prime)&=& \widehat G_0({\bf k})
  + \widehat G_0({\bf k})\widehat U_{{\bf k},{\bf k}^\prime}
  \widehat G_0({\bf k}^\prime)
  \\
  \nonumber
  &&
  + \sum_{{\bf k}^{\prime\prime}}\widehat G_0({\bf k})\widehat U_{{\bf k},{\bf
  k}^{\prime\prime}}
  \widehat G_0({\bf k}^{\prime\prime})\widehat U_{{\bf k}^{\prime\prime},{\bf k}^\prime}
  \widehat G_0({\bf k}^\prime)+\ldots.
\end{eqnarray}
Here we suppressed the frequency index in the Green's function as
the scattering is elastic. The series can be summed to write (see
Fig.~\ref{fig:TMatrix})
\begin{equation}
    \label{GSingleImp}
  \widehat G({\bf k}, {\bf k}^\prime)= \widehat G_0({\bf k})+
    \widehat G_0({\bf k})\widehat T_{{\bf k},{\bf k}^\prime}
  \widehat G_0({\bf k}^\prime),
\end{equation}
where the $T$-matrix is given by
\begin{eqnarray}
  \widehat T_{{\bf k},{\bf k}^\prime}&=&\widehat U_{{\bf k},{\bf k}^\prime}
  +\sum_{{\bf k}^{\prime\prime}}\widehat U_{{\bf k},{\bf
  k}^{\prime\prime}}
  \widehat G_0({\bf k}^{\prime\prime})\widehat U_{{\bf k}^{\prime\prime},{\bf k}^\prime}
  +\ldots
    \\
    &=&\widehat U_{{\bf k},{\bf k}^\prime}
  +\sum_{{\bf k}^{\prime\prime}}\widehat U_{{\bf k},{\bf
  k}^{\prime\prime}}
  \widehat G_0({\bf k}^{\prime\prime})\widehat T_{{\bf k}^{\prime\prime},{\bf
  k}^\prime}.
\label{TmatrixSingle}
\end{eqnarray}
This equation needs to be solved for $\widehat T$. If the impurity
scattering is purely local, $\widehat U({\bf r}-{\bf r}^\prime)$,
the scattering is isotropic, $\widehat U_{{\bf k},{\bf
k}^\prime}=\widehat U$, greatly simplifying the process of solving
the equation for the $T$-matrix, as $\widehat T$ depends only on
frequency.

Notice that we could draw the set of diagrams in
Fig.~\ref{fig:TMatrix} in real space, and write the corresponding
set of equations for the $T$-matrix and Green's function $\widehat
G({\bf r},{\bf r}^\prime)$ in complete analogy with
Eq.~(\ref{TmatrixSingle}). The main observation here is that, in the
vicinity of the impurity, the translational invariance is broken,
and the Green's function depends on two momenta, {\bf k} and {\bf
k}$^\prime$.
\begin{equation}
    \widehat G(\br,\br';\omega)  = \widehat G_0(\br,\br';\omega) +
    \widehat G_0(\br, {\bf r}_0;\omega)
    \widehat T(\omega)
    \widehat G_0({\bf r}_0,\br^\prime;\omega)
    \label{GTreal}
\end{equation}
\begin{figure}
  \caption{Multiple scattering on a single impurity. Thick (thin)
  line denotes full (bare) Green's function, and the dashed line
  denotes scattering process. The second line defines the $T$-matrix
  according to Eq.~(\ref{TmatrixSingle}).}
    \label{fig:TMatrix}
  \end{figure}

The $T$-matrix lends itself easily to describe the effect of an
ensemble of impurities. The so called self-consistent $T$-matrix
approach (SCTM) considers multiple scattering on a single site of an
electron that has already been scattered on all other impurity sites
\cite{PJHirschfeld:1988,PJHirschfeld:1986}. This results in
replacing the bare Green's function in Eq.~(\ref{TmatrixSingle}) by
its impurity-averaged counterpart, $\widehat G({\bf k},\omega)$.
Notice that after averaging over the random impurity distribution
the translational invariance is restored, so that the Green's
function depends on a single momentum {\bf k}. The combined effect
of impurities is given by the self energy, $\widehat\Sigma({\bf
k},\omega)=n_{imp}\widehat T_{{\bf k},{\bf k}}$, so that
\begin{equation}
    \label{G-SCTM}
    \widehat G^{-1}({\bf k},\omega)=\widehat G_0^{-1}({\bf
    k},\omega) - \widehat\Sigma({\bf k},\omega).
\end{equation}
In contrast to the single impurity case where Eq.~(\ref{GSingleImp})
with the $T$-matrix given by Eq.~(\ref{TmatrixSingle}) is the exact
solution of the problem, the Green's function given above is an
approximation, and much recent research is motivated by questions
about how accurately it describes the properties of nodal
superconductors with impurities.

\subsection{Static and dynamic impurities}

So far we only discussed the {\it static} impurities, and most of
our review addressed such a situation. However, even for purely
potential scattering a situation is possible when a vibrational mode
leads to a time-dependent modulation of the charge on an impurity
site, and, as a result, $U_{pot}$ acquires a characteristic
frequency. Such a mode can be extended, as a phonon, or local.
Influence of the dynamical impurity on the {\it local} properties of
a superconductor is a relatively new subject of research and we
summarize recent results in Sec.~\ref{sec:DynamicalImp}

For magnetic scattering the situation is more complex even in a
normal metal. The degeneracy between the spin-up and spin-down
states on the impurity site and the non-trivial commutation
relations between different spin components ensure that quantum
dynamics of the impurity is generated even if the exchange constant
is purely static. In simple words, if the scattering process, which
flips both the spin of the conduction electron and the impurity
spin, is relevant, the dynamics of the local spin flips becomes
essential. This dynamics leads to Kondo screening of the impurity
spin in a metal, and in Sec.~\ref{sec:DynamicalImp} we briefly
discuss the current status of the yet not fully understood problem
of Kondo effect in a superconductor.

In the limit of large impurity spin, however, the change of the
impurity spin by 1 during the spin flip scattering is not relevant,
and its dynamics does not play a major role. In this limit of {\it
classical} spin the static local and global density of states is
discussed in Sec.~\ref{sec:Shiba} and Sec.~\ref{sec:AverDOS}
respectively. Such a spin acquires dynamics only when placed in an
external magnetic field, which also affects the superconducting
state.

\section{Non-magnetic impurities and Anderson's theorem}

\label{sec:Anderson}

One of the most important early results was the robustness of the
conventional superconductivity to small concentrations of
non-magnetic impurities. Theoretical underpinning of this result is
known as Anderson's theorem~\cite{PWAnderson:1959}. Anderson's
observation was that, since superconductivity is due to the
instability of the Fermi surface to pairing of time-reversed
quasiparticle states, any perturbation that does not lift the
Kramers degeneracy of these states does not affect the mean field
superconducting transition temperature.

This is most clearly seen from the BCS analysis, which we carry out
following \onlinecite{MMa:1985}. We consider an isotropic pairing
potential, $V_{\alpha\beta\gamma\delta}({\bf r}, {\bf
r}^\prime)=V\delta({\bf r}-{\bf r}^\prime)$. In the absence of a
magnetic field the coefficients $a_n=\sin\theta_n$ and
$b_n=\cos\theta_n$ can be taken real without loss of generality, so
that the self-consistency condition,
Eq.~(\ref{Eq:DeltaBCSSelf-Cons}) reads
    \begin{equation}
    \label{AndersonGap}
    \Delta_n=V\sum_{m\neq n}\frac{\Delta_m}{(\xi_m^2+\Delta_m^2)}
    \int d^{d}{\bf r}\phi_n^2({\bf r}) \phi_m^2({\bf r}).
    \end{equation}
Here
    \begin{equation}
    \Delta_n=\int d^{d}{\bf r}\Delta({\bf r}) \phi_n^2({\bf r}).
    \end{equation}
As noted above, in the BCS approach $\phi$'s are the eigenfunctions
of the single particle hamiltonian. In the absence of impurities the
system is translationally invariant, so that $\Delta({\bf
r})=\Delta_n=\Delta_0$. The most important assumption underlying
Anderson's theorem is that the superconducting order parameter can
be taken to be uniform, $\Delta({\bf r})=\Delta_1$, even in the
presence of impurities. In that case the individual eigenfunctions
of the single particle hamiltonian {\it including impurities} are
rather complicated. However, the gap equation,
Eq.~(\ref{AndersonGap}), takes the form
    \begin{equation}
    \label{AndersonGap2}
    \frac{1}{V}=\int d\epsilon
    \frac{N(\epsilon,{\bf
    r})}{\sqrt{\epsilon^2+\Delta_1^2}},
    \end{equation}
equivalent to that of a pure superconductor provided the density of
states
    \begin{equation}
    N(\epsilon,{\bf r})=\sum_n \phi_m^2({\bf
    r})\delta(\epsilon-\epsilon_m),
    \end{equation}
is unchanged compared to the pure metal, $N(\epsilon,{\bf
r})\approx\rho_0$. If this condition is satisfied, the solution
$\Delta_1$ of the gap equation Eq.~(\ref{AndersonGap2}) must be
identical to that of the BCS equation in the absence of impurities,
and therefore the transition temperature and the gap are insensitive
to the impurity scattering at the mean field level.

Anderson's theorem helped explain why superconductivity was robust
to impurities in many early experiments. It is important to realize
however that it is an {\it approximate} statement about the
thermodynamic averages of the system. Beginning with the next
section we will analyze in more detail the changes that impurities
create in superconductors in their immediate surrounding as well as
on average. We will see that even purely potential scattering does
induce changes in the local properties of superconductors, albeit
the corresponding change in the transition temperature remains
minimal. Anderson's theorem brings to the fore the need to separate
the study of impurity effects on different lengths scales, from
lattice spacing to the coherence length, to sample size.

Before we proceed to study the local properties we discuss the
extensions of the Anderson's treatment of impurities. In weakly
disordered systems the density of states remains nearly constant as
a function of disorder. Ma and Lee~\cite{MMa:1985} argued that
Anderson's theorem remains valid in the form above even in a
strongly disordered system provided the localization length, $L\gg
(\rho_0\Delta_0)^{1/d}$. In that case there is a large number of
states localized within energy $\Delta_0$ of the Fermi surface, and
these state form a local superconducting patch. The Josephson
interaction between the patches then leads to global phase coherence
at $T=0$. Moreover, they argued that the theorem holds all the way
to the limit of site localization.

It is important to note that the superfluid stiffness, i.e. the
ability of the superconductor to screen out the magnetic field, is
affected by disorder. In particular, when the quasiparticle
lifetime, $\tau$, becomes sufficiently short, $\Delta_0\tau\ll 1$,
the superfluid density $\rho_s\approx \Delta_0\tau$. Consequently
the superconductor is sensitive to the local phase fluctuations of
the order parameter, and the experimentally observed transition
temperature may be severely suppressed compared to the mean field
$T_c$, as it is, for example, in granular superconductors.
Approaches extending beyond the mean field picture are largely
outside the scope of this review.

Therefore for dilute impurities Anderson's theorem is valid provided
the superconducting order parameter can be taken to be nearly
uniform. Since the ``healing length'' of $\Delta({\bf r})$ over
which it can change appreciably is the coherence length,
$\xi_0\simeq \hbar v_F/\Delta_0$, where $v_F$ is the Fermi velocity,
while the Coulomb screening length for the charged impurities in
metals is of the order of the lattice spacing, $a$, for $\xi_0\gg a$
the order parameter remains essentially uniform, and Anderson's
theorem holds. Much work has been done recently on the effect of the
ultrashort coherence length on the impurity scattering in
superconductors. In particular, it has been shown that when the
superconducting pairing is of the order of the electron bandwidth,
Anderson's theorem is violated
\cite{RMoradian:2001,KTanaka:2000,AGhosal:1998}.

Ghosal et al.~\cite{AGhosal:1998}, and Xiang and
Wheatley~\cite{TXiang:1995} have explored in detail the discrepancy
between the single particle excitation gap and the superconducting
order parameter as a function of disorder in these circumstances.
Beyond Anderson's regime of the constant density of states, both
quantities decrease at first, since the disorder depletes the
density of states. Then, however, the spectral gap persists while
the superconducting order vanishes. As pointed out by Ma and
Lee~\cite{MMa:1985} in the limit of strong disorder the models with
on-site pairing, such as those studied by Tanaka and Marsiglio, and
Ghosal et al., show the on-site spectral gap (so-called Anderson
negative-$U$ glass) without the off-diagonal long range order and
without symmetry breaking.

In most experimentally relevant situations, however, the corrections
to the main statement of Anderson's theorem are quantitative rather
than qualitative. This is generally true of most results pertaining
to the impurity scattering in superconductors, and therefore it is
very instructive to consider this problem in BCS-like systems.


\section{Single impurity bound state in two-dimensional metals}
\label{sec:imp2Dmetal}

Before we proceed to calculate the effect of impurity in a d-wave
superconductor it is instructive to review a simpler problem of a
single impurity in a metal. We show here a $T$-matrix calculation
for finding the bound states due to a single impurity in $d$
dimensions with an attractive potential $U_0 \leq 0$. The
Hamiltonian for the problem is
    \beqa H = \sum_\bk [\epsilon(\bk) - \mu] c^{\dag}_{\bk, \sigma}
    c_{\bk, \sigma} + \sum_{\bk,\bk'} U_0 c^{\dag}_{\bk,\sigma}c_{\bk',
    \sigma} \label{impmetal1} \eeqa
the $U_0$ term describes the on-site  energy change of  electron
density $n(\br)$ in external potential $U(\br) = U_0 \delta(\br)$.
We consider a single particle ($\mu = 0$), although the calculation
for the normal metal with a finite density of states follows simply
by replacing $\epsilon({\bf k})\rightarrow \xi({|bf k})$ in the
following.

The bare Green's function for a free particle is
    \beqa
    G_0(\omega, \bk) = [\omega - \epsilon(\bk)]^{-1}.
    \label{impmetalG1}
    \eeqa
Since the vertex of the impurity interaction, $U_0$ is momentum
independent, the equation for the $T$-matrix is particularly simple
and follows from Eq.~(\ref{TmatrixSingle}),
    \beqa
    T(\omega) = U_0 +  U_0\sum_\bk G_0(\omega, \bk)T(\omega) \nonumber\\
    T(\omega) = \frac{U_0}{1 - U_0\sum_\bk G_0(\omega,\bk)}
    \label{impmetalT1}
    \eeqa
Summation over momentum in $\sum_\bk$ is easily performed using the
density of states
 \beqa
 N(\epsilon) = \sum_\bk \delta (\epsilon -
 \epsilon(\bk)) =  \Gamma _d
\epsilon^{\frac{d}{2} - 1},
 \label{impmetalT1'}
 \eeqa
where $\Gamma _d$ is a constant dependent on dimension. Therefore
    \beqa
    g_0(\omega)=\sum_\bk G_0(\omega,\bk) = \int_0^{W} \frac{d \epsilon
    N(\epsilon)}{\omega - \epsilon} \simeq  - \Gamma_d
    \omega^{\frac{d-2}{2}},
    \label{impmetalT2}\eeqa
for $d\neq 2$, where $W$ is the electron half bandwidth. In two
dimensions $g_0\simeq -\Gamma_2\ln (W/|\omega|)$. Consequently, the
$T$-matrix for $d\neq 2$ is given by
    \beqa
    \label{TmatrixMetal}
        T = \frac{U_0}{1 - g_d  \omega^{\frac{d-2}{2}}}
        \label{impmetalT3}
        \eeqa
where $ g_d= - U_0 \Gamma_d$ is the effective coupling constant, and
by the same expression with the obvious substitution of $\ln
(W/\omega)$ for $d=2$.

Since the Green's functions in the presence of impurity scattering
is $G = G_0 + G_0 T G_0$, see Eq.~(\ref{GSingleImp}), poles of the
$T$-matrix are the new poles of $G$ that are not poles of $G_0$,
signifying the appearance of new states. We can find this pole,
$\omega_0$, from Eq.~(\ref{TmatrixMetal}) for different dimension
$d$. The bound state ($\omega_0<0$, see Fig.~\ref{FIG:ImpStNmetal})
is formed for an arbitrarily small potential $|U_0|$ in d = $1,2$,
but requires a critical coupling for $d=3$. The energy of this state
is given by
    \beqa
    &&\omega_0 \sim (g_1)^2, \qquad\qquad\qquad\qquad \mbox{if $d=1$};\\
    &&\omega_0 = W \exp(-\frac{1}{g_2}),\qquad\qquad \ \ \ \mbox{if $d=2$};
    \label{2DBound} \\
    &&\omega_0 \sim [g_3^{\frac{1}{2}} - g_{3c}^{\frac{1}{2}}], \ g_3
    \geq g_{3c},
    \qquad \mbox{if $d=3$},
    \label{impmetalT4}
    \eeqa
where the $d=3$ critical coupling $g_{3c}\sim W^{-1/2}$.

\begin{figure}
\caption{Impurity bound state in a metal at energy $\omega_0$ is
formed as a result of a multiple scattering.}
\label{FIG:ImpStNmetal}
\end{figure}

We focus in more detail on the two-dimensional case, when $g_2 =
\Gamma_2 |U_0|$ and $\Gamma_2 = \frac{m}{2\pi}$ is the electron
density of states. The bandwidth, $W \simeq \frac{\hbar^2}{2ma^2}$
is the ultraviolet cutoff corresponding to the lattice parameter $a$
for free particle. This result can be compared to the solution of
the Schr\"odinger's  equation  for the particle in  the 2D
attractive potential $U_0$ \cite{LDLandau:2000}, Ch. 45. For an
arbitrary potential $U(\br)$ the solution obtained using the
T-matrix is asymptotically correct if the scattering length is
greater than $a$. For shallow potential the bound state energy
$-\omega_0$ is small, and the characteristic extent of the bound
state wave function is $l_0 = (\frac{\hbar ^2}{2 m \omega_0})^{1/2}
\gg a$  Therefore in this limit we can safely approximate $U(\br) =
U_0\delta(\br)$, where $U_0 = \int U(\br) d\br$.

Finding the energy of the bound state, Eq.~(\ref{2DBound}), is only
one part to the solution. We also want to determine the corrections
to the {\em local Density of States} due to bound state. We write
the equation for the Green's function in real space,
Eq.~(\ref{GTreal}),
    \beqa
    G(\br,\br';\omega)  = G_0(\br,\br';\omega) +
    G_0(\br, 0;\omega)T(\omega) G_0(0,\br';\omega)
    \nonumber
    \eeqa
and read off the position dependent Density of states (DOS)
    \begin{eqnarray}
    N(\br, \omega) &=& -\frac{1}{\pi}Im G(\br,\br;\omega)\nonumber \\
    & = &N_0(\br, \omega)
    + \delta N(\br, \omega).
    \label{impmetalDOS1}
    \end{eqnarray}
Here the first term is the DOS of a clean system, and the second is
the correction due to the bound state. We focus on the energy range
close to the bound state energy, $\omega \simeq \omega_0$. Since the
bound state is below the bottom of the band, the unperturbed Green's
function $G^0$ has no imaginary part in this range ($N_0=-\mbox{Im}
g_0 (\omega=0)/\pi$). Therefore the only contribution to the
imaginary part of the full Green's function,
Eq.~(\ref{impmetalDOS1}) comes from the $T$-matrix :
    \begin{eqnarray}
     \mbox{Im} T(\omega) &=& \mbox{Im} \frac{1}{1/g_2 -
    \log[W/(-\omega)]} \nonumber \\
    &=& \mbox{Im} \ln^{-1}[\frac{\omega + i\delta}{\omega_0}]\nonumber \\
    &= &
     \pi \delta (\omega - \omega_0),
     \label{impmetalT6}
     \end{eqnarray}
and the correction to the DOS of a clean system is :
    \beqa
    \delta N(\br,\omega) = |G_0(\br, \omega_0)|^2 \delta(\omega -
    \omega_0)
    \label{impmetalDOS2}
    \eeqa
with $G_0(\br, \omega) = N_0 J_0(k_F r)\ln[\frac{W}{\omega}]$ is the
real part of Green's finction in 2D systems.
Equations~(\ref{impmetalT4}) and (\ref{impmetalDOS2}) are the main
results of this section. They establish the logic we will adhere to
in finding impurity induced bound states: a) find the poles of the T
matrix in the and  the poles of the dressed Green's function
Eq.~(\ref{impmetalT4}), b) find the inhomogeneous DOS due to
impurity induced state, Eq.~(\ref{impmetalDOS2}). One should keep in
mind that this approach is just one of many one can implement in a
search for scattering induced bound states. Alternatively one can
use for example the exact numerical solution of a finite system. As
we will argue for superconducting case the self-consistency
condition can not be implemented analytically and the numerical
solution remains the only method available.

\section{Low-energy states in $s$-wave superconductors}

\label{sec:Shiba}

\subsection{Potential scattering}

Even though the potential scattering does not change the bulk
properties of isotropic superconductors, it does affect the local
density of states \cite{KMachida:1972,HShiba:1973}.  Let us consider
the Anderson impurity model,
Eqs.~(\ref{AndersonImp1})-(\ref{AndersonImp2}) in the limit $U=0$
(resonance scattering). As discussed above the localized state
acquires a finite width, $\Gamma=\pi |V_{sd}|^2N_0$, due to
hybridization with the conduction band. The Green's function of the
conduction electrons can be evaluated in the $T$-matrix approach,
with the result at real frequencies \cite{KMachida:1972,HShiba:1973}
\begin{equation}
  \widehat T(\omega)=|V_{sd}|^2 \tau_3
  \biggr[
  \omega-E_0\tau_3-|V_{sd}|^2\tau_3\sum_{\bf k} \widehat G_0({\bf
  k},\omega)\tau_3\biggl]^{-1} \tau_3.
\end{equation}
The poles of the $T$-matrix determine the location of the bound
states
\begin{equation}
  \omega^2\biggl[1+\frac{2\Gamma}{\sqrt{\Delta^2-\omega^2}}\biggr]
  =E_0^2+\Gamma^2.
\end{equation}
In most physical situations $\Gamma\gg\Delta$, so that the bound
states are located at
\begin{equation}
  \omega_0=\pm\Delta (1-2\pi^2 (\Delta N_d(0))^2),
\end{equation}
where $N_d(0)=\pi^{-1}\Gamma/(\Gamma^2+E_0^2)$ is the density of
states of the resonant impurity level.  For typical densities of
states $\Delta N_d(0)\sim 10^{-3}$, so that the bound states lies
essentially at the gap edge. Shiba considered a finite but small
value of the Coulomb repulsion and allowed for the induced pairing
on the impurity site \cite{HShiba:1973}. He concluded that, even
though there may be a shift of the bound state to lower energies, it
still lies within $10^{-3}\Delta$ of the mean field gap edge, and
therefore can be neglected in the discussions of physical
properties.

\subsection{Classical spins}

If the substitution atoms have a magnetic moment, the time-reversal
symmetry is violated, and therefore superconductivity will be
suppressed. We consider the magnetic scattering,
Eq.~(\ref{Eq:HMag4}), which we rewrite in the momentum space as
\begin{equation}
    \label{Eq:ShibaImpSimple}
  H_{ex}=\frac{1}{2N}\sum_{{\bf k,k}^\prime \atop\alpha\beta}
  J({\bf k}-{\bf k}^\prime) c^\dagger_{{\bf
  k},\alpha}{\bm\sigma}_{\alpha\beta}\cdot {\bf S} c_{{\bf
  k}^\prime \beta}.
\end{equation}
We first review a simplified version of this problem, where we do
not need to consider Kondo screening. We review scattering on {\it
classical spins} first studied independently at about the same time
by Shiba, Rusinov, and Yu
\cite{HShiba:1968,AIRusinov:1968,AIRusinov:1969,LYu:1965}. Quantum
mechanical properties of spin can be neglected when
$S\rightarrow\infty$, and we simultaneously take $J\rightarrow 0$ so
that the product $JS=const$. In this limit the localized spin acts
as a local magnetic field.

Therefore we study the effect of the impurity with the potential
$U({\bf r})= U_0+U_{ex}$, or $H_{imp}=H_{imp}+H_{ex}$, on a BCS
$s$-wave superconductor with the unperturbed hamiltonian of the form
\begin{equation}
  H_0=\sum_{{\bf k}\alpha} \varepsilon_{\bf k}c^\dagger_{{\bf
  k},\alpha}c_{{\bf k} \alpha} + \Delta_0 \sum_{\bf k}
  \{ c^\dagger_{{\bf k}\uparrow }c^\dagger_{-{\bf k} \downarrow}
  +c_{-{\bf k} \downarrow} c_{{\bf k}\uparrow } \}.
\end{equation}
This problem serves as a starting point for all subsequent analysis
of the resonance states in superconductors.

To find a localized state with energy $0<E<\Delta_0$ near a single
paramagnetic impurity we perform a Bogoliubov transformation
\cite{AIRusinov:1968,LYu:1965} to find
\begin{eqnarray}
  Eu_\alpha({\bf r})=\varepsilon({\bf k}) u_\alpha({\bf r})
  +i\Delta\sigma^{y}_{\alpha\beta}v_\beta({\bf r}) +
  U_{\alpha\beta}({\bf r}) u_\beta({\bf r}),
  \\
  Ev_\alpha({\bf r})=-\varepsilon({\bf k}) v_\alpha({\bf r})
  -i\Delta\sigma^{y}_{\alpha\beta}u_\beta({\bf r}) -
  U_{\alpha\beta}({\bf r}) v_\beta({\bf r}).
\end{eqnarray}
This system is solved by Fourier transforming the equations and
expanding the impurity potential in spherical harmonics, and has
solutions with energies
\begin{equation}
  \frac{E_l}{\Delta_0}=\frac{1+(\pi N_0V_l)^2-(\pi N_0J_lS/2)^2}
  {\sqrt{[1+(\pi N_0V_l)^2 - (\pi N_0J_l S/2)^2]^2+4(\pi N_0J_l
  S/2)^2}},
\end{equation}
where $N_0$ is, again, the density of states at the Fermi energy in
the normal state. This result can be written in a more elegant form
if we introduce the phase shifts, $\delta_l$, of scattering for up
(+) and down (-) electrons, in each angular channel,
\begin{equation}
  \tan \delta_l^{\pm}=(\pi N_0)(V_l\pm J_l S/2).
\end{equation}
Then the energies of the states in the gap become
\begin{equation}
    \label{Eq:EnergyOfShibaStates}
  \epsilon_l=\frac{E_l}{\Delta_0}=\cos(\delta_l^+ - \delta_l^-).
\end{equation}
Clearly, for purely potential scattering ($\delta_l^+=\delta_l^-$)
the spectrum begins at the gap edge, and there are no intragap
states. However, as the magnetic scattering increases, a series of
low-energy states below the gap edge appear. Purely magnetic
scattering corresponds to $\delta_l^+=-\delta_l^-$, and strong
scattering (large phase shift) yields a localized state deep in the
gap, while weak scattering (small phase shift) results in the bound
state very close to the gap edge.

The same result can be obtained using the Green's function
formulation \cite{HShiba:1968,AIRusinov:1969} and solving the single
impurity problem using the $T$-matrix method described above. With
the impurity hamiltonian of Eq.~(\ref{Eq:HMag4}) in the Nambu
notations the matrix Green's function for the system is
\begin{equation}
    \label{Eq:GSingleImpTMatrix}
 \widehat G({\bf k}, {\bf k}^\prime;\omega)=
    \widehat G_0({\bf k}, \omega)\delta({\bf k}-{\bf k}^\prime) +
    \widehat G_0({\bf k}, \omega) \widehat T({\bf k},{\bf k}^\prime)
    \widehat G_0({\bf k}^\prime, \omega).
\end{equation}
Here the $T$-matrix is computed as in Sec.~\ref{sec:Imp}, and we sum
over the indices of the matrix ${\bm\alpha}$ in each vertex. The
$l$-th angular component of the $T$-matrix satisfies the matrix
equation (for a spherical Fermi surface and isotropic gap)
\begin{equation}
  {\widehat T}_l(\omega)={\widehat U}_l + {\widehat U_l}
  \int d\varepsilon \widehat G_0({\bf k}, \omega) {\widehat
  T}_l(\omega).
\end{equation}

The full expressions for the $T$-matrix for both potential and
magnetic scattering in all angular channels is straightforward to
obtain \cite{AIRusinov:1969} but is rather cumbersome, so that we
don't give it here. Even the case of only spherically symmetric
scattering ($l=0$) with both $U_0\neq 0$ and $J\neq 0$ the
$T$-matrix is simple yet lengthy \cite{YOkabe:1983}. The main
results for the energy of the Shiba states remains the same, of
course as Eq.~(\ref{Eq:EnergyOfShibaStates}).

In the particular case of purely magnetic spherically symmetric
exchange, $J({\bf k}-{\bf k}^\prime)=J$, only $l=0$ components are
non-vanishing and the $T$-matrix has a particularly simple
form~\cite{HShiba:1968}. The diagonal in spin indices component is,
\begin{equation}
  T^{(1)}(\omega)=\frac{1}{N}\frac{(JS/2)^2
  \widehat g_0(\omega)}{I-(JS\widehat g_0(\omega)/2)^2}.
\end{equation}
Here $\widehat g_0$ is the local matrix Green's function,
\begin{equation}
  \widehat g_0(\omega)=\frac{1}{N}\sum_{\bf k}\widehat G_0({\bf k}, \omega)
  =-\pi N_0\frac{\omega +\Delta_0\sigma_2\tau_2}{\sqrt{\Delta_0^2-\omega^2}}.
\end{equation}
The bound state energy
\begin{equation}
  \epsilon_0=\frac{E_0}{\Delta_0}=
  \frac{1-(JS\pi N_0/2)^2}{1+(JS\pi
  N_0/2)^2}.
\end{equation}

The wave functions of the bound states at $E_l$ can be computed
using the Bogoliubov equations above. In the simplest case of
isotropic scattering at distances $r\gg p_F^{-1}$, both $u({\bf r})$
and $v({\bf r})$ vary as \cite{AIRusinov:1969,ALFetter:1965}
\begin{equation}
  \frac{\sin (p_F r -\delta_0^\pm)}{p_F r}
  \exp (-r/\xi_0 |\sin(\delta_0^+-\delta_0^-)|,
\end{equation}
that is, the state is localized near the impurity site at distances
\begin{equation}
r_0\sim \frac{\xi_0}{|\sin(\delta_0^+-\delta_0^-)|}
=\frac{\xi_0}{\sqrt{1-\epsilon_0^2}}.
\end{equation}
The square of these coefficients gives the spatial dependence of the
amplitude of the particle and hole components of the density of
states at a given position {\bf r} \cite{AYazdani:1997}.

The analysis above was carried out under the assumption that the
variation of the superconducting order parameter, $\Delta$, around
the impurity site does not change the position of the resonance low
energy state. There are several characteristic length scales for
this variation, $\delta\Delta({\bf r})$. Far away from the impurity,
$r\gg\xi_0$, at temperatures close to $T_c$, where this variation
can be determined perturbatively, $\delta\Delta({\bf
r})/\Delta_0\simeq 1/(p_F r)$ \cite{JHeinrichs:1968,AIRusinov:1968}.
This power law is insensitive to the phase shifts of scattering on
the impurity. At low temperatures a fully self-consistent treatment
is required, which leads to $\delta\Delta({\bf r})$ decaying as
$(p_F r)^{-3}$ and oscillating on the scale of $\xi_0
\Delta_0/\omega_D$, where the Debye temperature $\omega_D$ sets the
scale for the interaction between
electrons~\cite{PSchlottmann:1976}.

In the immediate vicinity of impurity, $v_F/\omega_D\ll r\ll\xi_0$,
the variation of the order parameter is $\delta\Delta({\bf
r})/\Delta_0\simeq 1/(p_F r)^2$ in the linear response approximation
\cite{AIRusinov:1968}. In the fully self-consistent treatment at
distances $r\ll \xi_0 \omega_D/E_F$, this dependence was found to
acquire an oscillating factor $\sin^2 p_F
r$~\cite{PSchlottmann:1976}.

In all these cases, since the suppression of the order parameter is
determined by the Fermi wavelength, the effect is negligible in
determination of the position of the bound state.

\section{Impurity-induced virtual bound states in $d$-wave
 superconductors}

\label{sec:Dwave}

We are now  ready  to extend our discussion to impurity induced
states in $d$-wave superconductors. Scalar (non-magnetic) impurities
are pair-breakers for ``higher-orbital-momentum'' states, such as a
$d$-wave pairing state. This occurs because change of the momentum
of particles in the Cooper pair upon scattering disrupts the phase
assignment for particular momenta in a nontrivial pairing
\cite{PWAnderson:1959,TTsuneto:1962,DMarkowitz:1963}. More
rigorously this follows from the analysis of the normal and
anomalous self-energies due to scattering within the
Abrikosov-Gorkov theory \cite{AAAbrikosov:1963}. We also note that
one of the first arguments about pairbreaking effects of potential
scattering was given by Larkin \cite{AILarkin:1965}.

As we have emphasized, for pairbreaking impurities the local
properties of the superconductor near an impurity site, such as the
local density of states and the gap amplitude, will be modified
dramatically. To capture these modifications, we use a variation of
the Yu-Shiba-Rusinov approach
\cite{LYu:1965,HShiba:1968,AIRusinov:1968}, which treats magnetic
impurities in the strong scattering limit, see Sec.~\ref{sec:Shiba}.
We restrict our consideration to the $s$-wave scatterers with the
phase shift close to the unitarity limit, $\delta_0 \simeq \pi/2$,
when the bound state has energy away from the gap edge. In contrast
to the $s$-wave superconductors, in $d$-wave systems the density of
states below the gap maximum is non-zero, and varies linearly with
energy in a pure system. Consequently, the overlap with the
particle-hole continuum only allows the formation of {\em virtual
bound states} with a finite lifetime.

We focus in this section on point-like defects, and use the
$T$-matrix approach. A closely related method uses quasiclassical
approximation and picture of Andreev scattering ideas to reproduce
the results of $T$-matrix calculation
\cite{CHChoi:1990,DCChen:1996,ASchnirman:1999}. Even more
interesting  results  are obtained within the quasiclassical
formalism for extended defects. For example, it has been shown that
index theorem dictates the existence of the low energy quasi-bound
state \cite{IAdagideli:1999}.

Zn substitutions in cuprates are one example of nonmagnetic atoms
that are predominantly potential scatterers in high-$T_c$
superconductors. Although Zn ions are nominally non-magnetic, $T_c$
is strongly suppressed by Zn substitution of Cu in the planes
\cite{KIshida:1991,THotta:1993}. Therefore, it is reasonable to
assume that Zn ions are non-magnetic unitary scatterers, see below.

We analyze virtual impurity-bound states in a $d$-wave
superconductor and, within this framework, explore possible
implications of the assumption that the pairing in cuprates is in
the $d_{x^2-y^2}$ channel. We model cuprates as a 2D $d$-wave
superconductor, based on strong anisotropy of electronic transport.
Our results, can be easily extended for any nontrivial pairing state
and may be relevant, e.g. for heavy-fermion superconductors with
impurities. Here we closely follow the references
\cite{PCStamp:1987,AVBalatsky:1995,MISalkola:1996,MISalkola:1997,LJBuchholtz:1981}.

Main results of this section are as follows: (${\bf i}$) A
strongly-scattering scalar impurity is a requirement for a
localized, virtual or virtually bound state ( or resonance) to exist
in a $d$-wave superconductor. It is intuitively obvious that any
strong enough pair-breaking impurity --- magnetic or
 non-magnetic --- will induce such a state.  Indeed, the low-lying
quasiparticle states close to the nodes in the energy gap will be
strongly influenced even by a non-magnetic impurity potential,
resulting in a virtual bound state in the unitary limit. (${\bf
ii}$)This should be compared to the fact that, in $s$-wave
superconductors, both magnetic and resonant non-magnetic impurities
produce bound states inside the energy gap  \cite{KMachida:1972}.
The energy $\Omega'$ and the decay rate $\Omega''$ of this state are
given by
 \beqa \Omega \equiv \Omega'+i\Omega''= - \Delta_0
 \frac{\pi c/2}{\log(8/\pi c)} \left[ 1 +
 \frac{i\pi}{2} \frac{1}{\log (8/\pi c)}
\right] \label{eq:impdwave1} \eeqa where $c=\cot \delta_0$. These
results are computed assuming  logarithmic accuracy is sufficient,
with $\log (8/\pi c) \gg 1$.   In the unitary limit, defined as
$\delta_0 \rightarrow \pi/2$ ($c\rightarrow 0$), the virtual bound
state becomes a resonance  at $ \Omega\rightarrow 0$ with $
\Omega''/ \Omega' \rightarrow 0$. In the opposite case of weak
scattering with $c\lesssim 1$, the energy of the virtual bound state
formally approaches $\Omega' \sim \Delta_0$ and the state is
ill-defined because $\Omega'' \sim \Omega'$ (see
Fig.~\ref{FIG:density}.
The wave function of the bound state is found to decay as a power
law: $\Psi(r) \sim 1/r$ and is not normalizable. This is consistent
with the virtually bound state being  not really a bound state. Wave
function is localized along the directions of the vanishing gap, so
called nodal directions.

\subsection{Single potential impurity problem}

Consider the single scalar impurity problem with \beqa H_{\rm int} =
\sum_{\bk\bk'\sigma} U_0 c^\dagger_{\bk\sigma}c_{\bk'\sigma} \eeqa
where $U_0$ is the strength of the scalar impurity potential at
$\br=0$, resulting in $s$-wave phase shift $\delta_0$.

The scattering is described by a $T$-matrix, $\hat{T}(\omega)$,
which is independent of wavevector. The Green's function in the
presence of an impurity is \beqa \widehat{G}({\bk,\bk'};\omega)=
\widehat{G}_0 ({\bk},\omega)\delta_{\bk\bk'} + \widehat{G}_0({\bk},
\omega) \widehat{T}(\omega) \widehat{G}_0 ({\bk'},\omega)\;, \eeqa
where   $\Delta_{\bk}= \Delta_0 \cos 2\varphi$ is the gap function
of $d_{x^2-y^2}$-symmetry,

>From the previous analyses~\cite{CJPethick:1986,
SSchmitt-Rink:1986,PCStamp:1987,PJHirschfeld:1988,PALee:1993,PJHirschfeld:1993,
AVBalatsky:1994,HShiba:1968}, it is known that $\hat{T} = T_0
\hat{\tau}_0 +T_3 \hat{\tau}_3$ for $s$-wave scattering. The
$T$-matrix takes the form \beqa
 T(\omega)_{11} = 1/[c -
g_{11}(\omega)], \label{eq:7} \eeqa
  where $g_{11}(\omega) = \frac{1}{2\pi N_0} \sum_{\bk} {\rm Tr} \,
\widehat{G}^{(0)}({\bk},\omega)( \widehat{\tau}_0 +
\widehat{\tau}_3)_{11} $. The quasi bound states in the
single-impurity problem are given by the poles of the $T$-matrix
\beqa c= g_{11}(\Omega), \label{eq:impdwave2} \eeqa which is an
implicit equation for impurity resonance $\Omega_0$ as a function of
$c$, the strength of impurity scattering. Choosing the gap function
at the Fermi surface so that $\Delta(\varphi)= \Delta_0 \cos
2\varphi$, one finds for particle-hole symmetric case $g_{11} =
g_0(\omega)=\left\langle \omega/ \sqrt{\Delta(\varphi)^2
-\omega^2}\right\rangle _{FS}$, where the angular brackets denote
averaging over the Fermi surface; for simplicity, we take $\langle
\bullet  \rangle _{FS} = \int \bullet \ d\varphi/2\pi$ \footnote{We
assume that the energy gap has line nodes in three dimensions with
weak quasiparticle dispersion along the $z$ axis; an extension to a
general three-dimensional case is straightforward.}. For
$|\omega|\ll \Delta_0$, one finds \beqa g_0(\omega) = -
\frac{2\omega}{\pi \Delta_0} \left( \log \frac{4\Delta_0}{\omega} -
\frac{i\pi}{2}\right). \label{eq:impdwave3} \eeqa In
Fig.~\ref{FIG:ImpGraphic} we illustrate a solution of the
Eq.~(\ref{eq:impdwave2}).

\begin{figure}[tbp]
\caption{
Graphic solution of the Eq.~(\ref{eq:impdwave2}) for large $U_0$ is
shown. Only  physically relevant intersections  with small imaginary
part $\Omega_0" \ll \Omega'_0$ are shown. In the region where the
imaginary part of the local Green's function exceeds the real part,
the resonance is broadened and merges with continuum. Resonances
below (or above, for a different sign of $U_0$) the fermionic band
are the sharpest, with most of spectral weight, and the virtual
bound state inside the gap is well resolved for large $U_0$ (small
$c$). } \label{FIG:ImpGraphic}
\end{figure}

In principle, the solution of Eq.~(\ref{eq:impdwave2}) is complex,
indicating a resonant nature of the quasiparticle state, better
described as a virtual state. This is easily seen form
Eq.~(\ref{eq:impdwave1}), which solves Eq.~(\ref{eq:impdwave2}) to
logarithmic accuracy. However, as $c\rightarrow0$, the resonance can
be made arbitrarily sharp. For $c=0$, the virtual state becomes a
sharp resonance  state bound to the impurity \cite{AVBalatsky:1995}.
As $c\rightarrow 1^-$, $\Omega'$ and $\Omega''$ increase without
bound so that $\Omega''/\Omega' \rightarrow 1^-$, and the solution
becomes unphysical. For $c>1$, no solution has been found for
$\Omega$. \footnote{The related model of the Anderson impurity in an
unconventional superconductor has been considered by L. Borkowskii
and P. Hirschfeld,  Phys. Rev. B {\bf 46}, 9274 (1992). The results
found here for pure potential scattering require the generalization
of the Anderson model to include the impurity potential phase shift,
independent of the Kondo temperature. This aspect of impurity
scattering has not been addressed previously.}

To properly solve a single impurity state one has to retain both
components of the $T$-matrix. Assumptions about particle-hole
symmetry alone are not sufficient to leave the $T_3$ contribution
out while computing the density of states around the impurity site.
Full solution, Eq.~(\ref{eq:7}), leads to the definite sign of the
resonance energy depending on the sign of $c$. Indeed, if we assume
that $T_3$ is zero we find that \beqa T_0 = \frac{g_0(\omega)}{c^2 -
g^2_0(\omega)}\;,\;\; T_3 = \frac{c}{c^2 - g^2_0(\omega)}\;,
\label{EQ:T3} \eeqa and $T_0$ has now {\em two} poles at $c = \pm
g_0(\omega)$. This means that there would be two solutions to the
"poles" of the $T$-matrix for each $c$, one on positive and one on
negative frequency. This would clearly contradict the obvious
particle hole asymmetry introduced by impurity. One can not get a
symmetric density of states if we have only repulsive ($U_0 > 0$) or
attractive ($U_0 <0$) impurity potential. This argument shows that
one is not allowed to ignore $T_3$ contributions, because $T_3$ is
not a smooth function of energy near the pole.

Now we turn to the physical implications of these virtual bound
states in a $d$-wave superconductor. Consider the most interesting
case of unitary impurities in the dilute limit, separated by a
distance greater than the coherence length $\xi$. Before averaging
over impurities, these bound states are {\it nearly localized} close
to the impurity sites (see below) and can substantially modify the
local characteristics of the superconductor: for example, the local
density of states, observed in STM  and the local NMR relaxation
rates of atoms close to the impurities.

Consider a local density of states, defined as \beqa N(\br,\omega) =
-\frac{1}{\pi}{\rm Im}\,g_{11}(\br,\br;\omega+i0^+) \eeqa with the
total Green's function in the presence of the impurity
\begin{eqnarray}
\widehat{G}(\br,\br';\omega)&=& \widehat{G}_0 (\br-\br',\omega)
\nonumber
\\ &&+ \widehat{G}_0(\br,\omega) \widehat{T}(\omega)
\widehat{G}_0 (\br',\omega).
\end{eqnarray}
We find two terms in the local density of states \beqa
 N(\br,\omega) = N(\omega) + N_{{\rm
imp}}(\br,\omega)\;.
 \eeqa
The first term originates from the bulk quasiparticles, which are
described by plane-wave eigenstates with $E_{{\bk}} =
\sqrt{\xi^2_{{\bk}} + \Delta^2_{{\bk}}}, \ \ g^{(0)}(0,\omega) =
\sum_{{\bk}} [u_{\bk}^2/(\omega - E_{{\bk}}) + v_{\bk}^2/(\omega +
E_{{\bk}}) ]$, where $u_{\bk}$ and $v_{\bk}$ are the standard
Bogoliubov factors. The bulk density of states is constant in the
system with $N(\omega)/N_0 = \omega/\Delta_0$, for $\omega \ll
\Delta_0$. The second term, \beqa N_{{\rm imp}}(\br,\omega) =
-\frac{1}{\pi} {\rm Im} \, [{\widehat{G}_0({\br},\omega)
\hat{T}(\omega) \widehat{G}_0 ({\br},\omega)}]_{11} \eeqa originates
from the virtual bound  state created at the impurity.
This impurity state will have a form of a cross with long tails
extended along the gap nodes see Fig.~\ref{FIG:SalkolaImpRes}. As an
clarifying
 example, consider the limit of unitary scattering for which the
resonant state is formed at $E_{{\rm imp},n}\equiv\Omega\rightarrow
0$, see also previous  Sec.~\ref{sec:imp2Dmetal}. Because ${\rm
Im}\, G^{(0)}({\br},\omega = 0) =-\pi N(\omega = 0) = 0$, only the
imaginary part of the T-matrix contributes to $N_{{\rm imp}}$ and
the bound-state probability density is found to decay as the inverse
second power of the distance from the impurity, mostly localized
along the nodes of the gap function, \beqa N_{{\rm imp}}(\br,\omega
= 0) = {\rm Re}\,[\hat{G}^{(0)}({\br},\omega = 0)]^{2} \propto
r^{-2}, \label{eq:impdwave20} \eeqa and similarly, but with smaller
amplitude,  in the vicinity of the extrema of the gap function,
\beqa N_{{\rm imp}}({\vec {r}},\omega = 0) \propto
\frac{\Delta^2_0}{E_F^2} r^{-2}, \eeqa

In addition to the  power law decaying large distance asymptotics
there is also an additional exponentially decaying piece that decays
with $\xi(\varphi)$, the angle-dependent coherence length of the
superconductor, defined as $\xi(\varphi) = \hbar
v_F/|\Delta(\varphi)|$. Exponentially decaying part is  important to
compare the induced DOS to the observed in STM near impurity site
although it does not change asymptotic behavior at large distances.
In practice the intensity near impurity is mapped out only within
few lattice sites.

Gap nodes lead to the power law decay of the wave function along all
directions at large distances $r >> \xi$. This follows from the
power counting of the d-wave propagator we estimate: $G(\br, \omega
\rightarrow 0) \sim \int d^2 k \exp(i\bk \cdot \br)G(\bk, \omega
\rightarrow 0) \sim \int k dk \exp(i\bk \cdot \br)\frac{v_F k}{k^2}
\sim 1/r$. The fact that the impurity state is virtually  bound is
reflected in the logarithmically divergent normalization. This
divergence should be cut off at an average distance between
impurities at any finite concentration. More generally, for an
arbitrary position of the resonance, taking into account that only
one state has been produced with $ E_{{\rm imp},n} = \Omega'
+i\Omega''$, we find
\begin{eqnarray}
N_{{\rm imp}} (\br,\omega) &=& \frac{\Omega_i''}{\pi}\sum_i
\biggl{[} \frac{\vert u(\br - \br_i)\vert^{2}}{(\omega-\Omega_i')^2
+ \Omega_i''^2} \nonumber \\
&&- \frac{\vert v(\br - \br_i)\vert^{2}}{(\omega+\Omega_i')^2 +
\Omega_i''^2}\biggr{]}\;, \label{eq:impdwave5}
\end{eqnarray} where we have introduced the sum
over different impurities, located at $\br_i$, and $u(\br - \br_i),
v(\br-\br_i)$ are the eigenfunction of the Bogoliubov-de Gennes
equation at the impurity level.

The local effects of impurities are  best revealed  by local probes.
NMR experiments on Cu in Zn-doped cuprates are quite useful in this
regard. From Eq.~(\ref{eq:impdwave20}) and below, one concludes
immediately that the local NMR signal would show two distinct
relaxation rates (or even the hierarchy of rates): one coming from
the Cu sites, far away from the impurities, and another from the
sites, close to the impurities. The Cu sites near the impurities
will be sensitive to the higher local  density of states and will
have a higher relaxation rate at low temperatures. At finite
impurity concentration ($\sim 2\%$), the volume-averaged density of
states will have a finite limit at $\omega\rightarrow 0$, as follows
from Eq.~(\ref{eq:impdwave20}). The relaxation rates of Cu atoms
close to and away from an impurity will, therefore, have the same
temperature dependence $(T_1T)^{-1} = const$, but will be of a
different magnitude. Precisely this behavior has been observed
experimentally:  Ishida {\it et al.}~\cite{KIshida:1991} have
measured two NMR relaxation rates for Cu in Zn-doped
YBa$_2$Cu$_3$O$_{7-\delta}$. The second NMR signal with higher
relaxation was inferred arising from the near-impurity Cu sites.
Alloul and collaborators have pointed out that the NMR signal coming
from  the sites close to impurties shows a distribution of
relaxation times and reflect local electronic and magnetic
distortions produced by impurities, see~\cite{JBobroff:2001} and
references therein.

More direct evidence for the impurity induced resonances in
high-$T_c$ is coming from STM experiments. Local variations of the
density of states can be probed using scanning-tunneling microscopy.
These experiments were crucial in establishing the existence of the
impurity induced resonances in cuprates and their anisotropic nature
\cite{SHPan:2000b,EWHudson:1999}, see section~\ref{sec:STM}.

We would like to contrast our picture of the dilute limit of
strongly scattering centers to the usual approach of averaging over
impurities at finite concentration. If one considers averaging over
impurities, two NMR relaxation rates, arising from unequivalent
sites, cannot  be resolved; similarly   local inhomogeneous aspect
of the localized states will be  lost after averaging over impurity
positions.

For practical purposes the distinction between the a true  bound
states and continuum in our case is not  well defined,  as it  is in
$s$-wave superconductors. Any finite temperature will produce a
finite lifetime for these bound states, and they will be hybridized
with the continuum of low-energy quasiparticles as they are not
separated by a  well defined gap.

\subsection{Single magnetic impurity problem}

The similar analysis for the magnetic impurity is more involved. For
a quantum spin $s = 1/2$ in d-wave superconductor one needs to
address the Kondo effect. It is discussed in more details in
Sec.~\ref{sec:DynamicalImp}. For a classical spin $S \gg 1$ the
analysis within the mean field is  similar to the one in the
previous section~\cite{MISalkola:1997}.

\begin{figure}[tbp]
\caption{Illustration of the cross shaped nature of the impurity
state. Shown is the spectral density $A_{\sigma}(\br, \pm \Omega_0)$
as a function of position and spin ~in units of $N_0\Delta_0$ for a)
$\mu = 0$ and b) $\mu = - W$, $2W$ is the bandwidth,   in a
two-dimensional d-wave superconductor as a function of position
around a classical magnetic moment ( $N_0J_0 = 10$ and $U_0 = 0$)
located at $\br = 0$; a is the lattice spacing. These results are
computed self-consistently with $\xi = 10a$. At half filling, the
spectral density obeys particle-hole symmetry: $A_{\uparrow}(\br,
\Omega_0) = A_{\downarrow}(\br, -\Omega_0)$. The energies of the
shown virtual-bound states are a) $\Omega_0 =0.05 \Delta_0$ and b)
$\Omega_0 = 0.5 \Delta_0$. From~\cite{MISalkola:1997}}
\label{FIG:SalkolaImpRes}
\end{figure}

The main effect is that  exchange coupling between the local spin
$S$ and electron spin leads to the renormalization of the effective
scattering potential. Namely for electrons of two  spin polarization
the net impurity potential is $
  U_0 \pm J$, where $U_0$ is the potential scattering strength and $J$
  is the exchange coupling to impurity spin, see Sec.~\ref{sec:STM}.
  There are two virtual bound
  states,
   one for each electron spin orientation.
  STM data on Ni-doped Bi2212 are fit well using this simple
  formula.
  This mean field approach does not address the dynamics of the
  large spin $S$. More analysis is required to address this
  problem.

\subsection{Self-consistent gap solution near impurity}

Impurity scattering will produce local modifications of the order
parameter. We already addressed some of the effects in s-wave
superconductors in Sec.~\ref{sec:QPT}. Here we will discuss the
self-consistently determined gap in d-wave superconductors.

To address these effects at small distances one would need to use an
numerically determined exact spectra near impurity and solve
self-consistent gap equation. We define: \begin{eqnarray} \Delta(i,
i+\delta) &=& \frac{V_{i, i+\delta}}{2}\sum_n [u_n(i
+\delta)v^*_n(i) + u^*_n(i)v_n(i+\delta)]\nonumber
\\
&&\times \tanh(\frac{E_n}{2k_B T})\;. \label{EQ:impdwave selfcons1}
\end{eqnarray} Numerical solution of this problem was presented in
~\cite{MFranz:1996,MISalkola:1997,HTsuchiura:2000,JXZhu:2000a}. The
main result is that impurity scattering suppresses the gap
magnitude. Suppression is strongest on the impurity site and quickly
gap recovers a bulk value, although there are always oscillating
tails at far distance due to $2k_F$ oscillations,
Fig.~\ref{FIG:Franz1}.

\begin{figure}[tbp]
\caption{
Self-consistently determined gap function near scalar impurity in a
2D $d$-wave superconductor. Gap suppression is strongly localized
near impurity site aside from weak oscillating tails.  From
\cite{MFranz:1996}} \label{FIG:Franz1}
\end{figure}

In practice, the difference between self-consistent and non
self-consistent solution is not important at distance beyond few
lattice sites away from impurity. Local gap suppression is clearly
seen in STM data, see Sec.~\ref{sec:STM}.

\subsection{Spin-orbit scattering  impurities}

Spin-orbit coupling in impurity scattering in superconductors is the
least discussed among all other kinds of impurity scattering.
Standard spin orbit scattering is of the form: \beqa H_{SO imp} =
\sum_{\bk, \bk'} \lambda_{SO}  c^{\dag}_{\bk,\alpha}
\vec{\sigma}_{\alpha \beta}\cdot(\bk \times \bk') c_{\bk' \beta}
\label{EQ:ImpSO1} \eeqa where $\lambda_{SO}$ is the strength of Spin
Orbit scattering. This kind of SO  scattering would be present even
for nonmagnetic impurities, it will be a pairbreaker and it  will
produce the the quasi-bound states inside the gap, although the
detailed calculation has not been done to our knowledge.

Another kind of SO scattering impurities scattering in d-wave from
the magnetic impurity can  be considered as well. It  was initially
motivated by experiments on Ni doped
Bi2212~\cite{RMovshovich:1998,WKNeils:2002}. In this approach
impurity spin is coupled to the orbital motion of the conduction
electrons~\cite{CGrimaldi:1999,YSBarash:1997,MJGraf:2000,AVBalatsky:1998}:
\beqa H_{SO, imp} = \sum_{\bk, \bk'} \gamma_{SO}
c^{\dag}_{\bk,\sigma} \bS \cdot (\bk \times \bk') c_{\bk' \sigma}
\label{EQ:ImpSO2} \eeqa where $\gamma_{SO}$ is the strength of
coupling and $\bS$ is the impurity spin. This term is the SO
coupling  $H_{SO, imp} = \gamma_{SO} \hat{{\bf L}} \cdot \bS$
written in second quantized notation. Predominantly in plane motion
of electrons as is the case in Bi2212 will couple  $L_z$  to $S_z$,
$\hat{{\bf L}}_z = i \hbar
\partial_{\phi}$ is the angular momentum operator $L_z$ with respect to the impurity
site.
 The net effect of this term is twofold. This scattering term is
a pairbreaker,  locally gap is suppressed and resonance is formed.
However the more interesting and nontrivial is the distortion of the
$d$-wave order parameter in the vicinity of impurity, which results
from  the nontrivial orbital structure of the $d$-wave order. The
$d$-wave state is a linear combination of the state with $l = 2$ and
$l = -2$, $\Delta(\phi) = \Delta_0 \cos(2\phi) \propto \exp(2i\phi)
+ \exp(-2i\phi) \sim x^2 - y^2$. The orbital angular momentum
components will be affected differently as a result of scattering.
In the {\em first order} perturbation theory in $H_{SO imp}$ one
generates the correction to the order parameter $\Delta^{'} = i
\Delta_0 \gamma_{SO} \sin(2\phi) \sim xy$. There is a finite
amplitude for incoming $d$-wave pair $|in \rangle \propto |x^2 - y^2
\rangle$ to scatter into $|out\rangle \propto i|xy \rangle$ channel:
 \beqa
  |out\rangle =
i\gamma_{SO}\Delta_0 \hat{{\bf L}}\cdot \bS |in \rangle = i \hbar
\gamma_{SO} \Delta_0 \sin(2\phi)\;.
 \label{EQ:SOImp3}
 \eeqa Therefore
as is the case for SO scattering, sometimes  impurity scattering can
produce nontrivial distortions of the initial order parameter, aside
from trivial suppression. For more details,
see~\cite{MJGraf:2000,AVBalatsky:1998,JXZhu:2002e}. Similarly,
magnetic field, that acts similarly to the $S_z$ term in
Eq.~(\ref{EQ:SOImp3}), not only suppresses the d-wave order
parameter but also produces the secondary $d_{xy}$ component,
see~\cite{RBLaughlin:1998,AVBalatsky:2000,MFranz:1998,YTanuma:1998,KKuboki:1998}.

\subsection{Effect of doppler shift and magnetic field}

Here we are focusing on the orbital effect of magnetic field and
thus the problem is closely related to the effect of Doppler shift
on impurity resonance.

In the presence of a superflow with velocity ${\bf v}_S$ propagators
are modified: $G(\bk, \omega) \rightarrow G(\bk, \omega - \bk \cdot
{\bf v}_S)$  for a planar wave state at vector $\bk$ and similar
change for $F$ function. Hence the rest of the calculation for the
impurity state goes through as before.  Since only local propagators
enter into solution for impurity resonance Eq.~(\ref{eq:impdwave1}),
the modifications will arise as a result of changes in density of
states due to Doppler shift.

Interesting effect of the superflow produced by the screening
currents on the impurity state was studied by Samokhin and
Walker~\cite{KVSamokhin:2001}. They pointed out that   Doppler shift
will result in the broadening of the impurity induced resonance.
This is a consequence of the local scattering nature of impurity
that means summing over all momenta to obtain local Green's function
$G_0(\omega)$. One would need to compare the typical value of the
Doppler shift $v_S k_F$ and the energy of the resonance $\Omega'$.
In the case when Doppler shift is small effect is negligible. In the
opposite limit of $v_S k_f \gg \Omega'$ superflow produces
broadening of the resonance but not the energy shift of the
resonance.

\subsection{Sensitivity of impurity state to details of band structure}

In the above discussion we were using the single band model with
particle-hole symmetric structure as a  simplest example to prove
the existence of the impurity induced resonance. The effect of
asymmetric band about the gap midpoint was considered by
Joynt~\cite{RJoynt:1997}, by assuming a constant DOS with different
energy ranges outside the gap edge. To make a comparison with the
real experimental data on impurity resonances, see
Sec.~\ref{sec:STM}, one has  to understand the details of realistic
band structure. Microscopically, relevant bands in Cu-O planes are
Cu $d_{x^2-y^2}$ and O $p_{x,y}$ bands. In the above analysis we
have {\em assumed} that upon the reduction of the complicated band
structure of high-$T_c$ to a single band model, one can still
describe nonmagnetic impurity by a single parameter, i.e. the
on-site potential $U_0$. We do not have a proof for this and assume
that the major physical effects, such as that the impurity induced
resonance will be properly captured in a simplified model.

Within the framework of this simplified one band model  one  can
still investigate some effects beyond the simplest assumptions. One
of the most obvious questions  is the position of the
impurity-induced resonance with respect to the Fermi energy. We find
the resonance depends on the sign of the impurity potential,  the
electron occupation, and the band structure. We have performed
numerical exact diagonalization for the $t$-$t^{\prime}$-$V$ model
with nearest-neighbor hopping $t$, next nearest neighbor hopping
$t^{\prime}$, and a negative $V$, that describes the nearest
neighbor attraction to produce effectively a $d$-wave pairing. The
single particle energy dispersion for the normal state is given by:
\begin{equation}
\xi_{k}=-2t(\cos k_x + \cos k_y) -4t^{\prime} \cos k_x \cos k_y
-\mu\;, \label{EQ:ENERGYDISP}
\end{equation}
with $\mu$ the chemical potential. Impurity was modeled as an
on-site potential $U_0$. We have looked at three possibilities: (i)
$t=1,t'=0, \mu=0$ (the filling factor $n=1.0$), with band
particle-hole symmetry present, Fig.~\ref{FIG:PHfig1}; (ii) $t=1,
t'=-0.2, \mu=-0.784 (n=0.84)$, with no band particle-hole symmetry,
Fig.~\ref{FIG:PHfig2}; (iii) $t=1,t'=-0.3,\mu=-1.0 (n=0.85)$, again
with band particle-hole symmetry absent, Fig.~\ref{FIG:PHfig3}. Here
we talk about the band particle-hole symmetry because as long as an
impurity is introduced, the local particle-hole symmetry is always
broken.

\begin{figure}[tbp]
\caption{The LDOS as
a function of energy at the impurity site (left panels) and at one
of its nearest neighbors (right panels) in a 2D lattice. The upper
panels are for various values of repulsive potential, $U_0=0$ (black
line), 2 (blue line), 5 (green line), and 10 (red line) while the
lower panels are for various values of attractive potential, $U_0=0$
(black line), -2 (blue line), -5 (green line), and -10 (red line).
The band structure parameter values are $t=1$, $t^{\prime}=0$, and
the chemical potential $\mu = 0$. } \label{FIG:PHfig1}
\end{figure}

\begin{figure}[tbp]
\caption{Same as
Fig.~\ref{FIG:PHfig1} except that the band structure parameter
values are $t=1$, $t^{\prime}=-0.2$, and the chemical potential $\mu
= -0.784$.  } \label{FIG:PHfig2}
\end{figure}

\begin{figure}[tbp]
\caption{Same as
Fig.~\ref{FIG:PHfig1} except that the band structure parameter
values $t=1$, $t^{\prime}=-0.3$ and the chemical potential
$\mu=-1.0$. } \label{FIG:PHfig3}
\end{figure}

As shown in these figures,  for the cases (i) and (ii), the band DOS
has two coherent peaks. Also for the case (ii), the DOS is
asymmetric with respect to the zero energy point. In these two
cases, a repulsive potential $U_0>0$ leads to a negative energy
impurity state $\Omega_0' < 0$. This position is manifested as a
resonance peak appearing below the Fermi energy in the LDOS directly
impurity site but appearing above the Fermi energy in the LDOS at
its four nearest neighbors. An attractive impurity potential $U_0<0$
induces a positive energy impurity state $\Omega_0'>0$, as reflected
as a resonance peak above the Fermi energy in the LDOS directly on
the impurity site but below the Fermi energy in the LDOS at its
nearest neighbors.

For the case (iii), in addition to the two coherent peaks, there are
also two van Hove singularity peaks (the one on the negative energy
side being more  pronounced and the other on the positive energy
side being faint). Then for a repulsive impurity, the on-site
resonance peak does shift from the negative energy side across the
zero energy. This phenomenon is absent for the cases of (i) and
(ii). For $U_0<0$, the result is similar to the cases (i) and (ii).
We point that in the STM data, the van Hove singularity peaks are
absent. Here we have chosen for (ii) and (iii) only in the optimal
doping regime. In other doping regime, all possibilities uncovered
above could appear. More detailed analysis, especially realistic
band structure calculations will allow us better address the details
of impurity states in high-$T_c$ materials.

As for the sign of the impurity potential from the Zn and Ni atoms
in cuprates, it is still an unsettled issue. It is believed that
these atoms substitute Cu in the Cu-O plane, and do not change the
hole doping. In case of Zn that has $3d^{10}4s^{2}$ electrons,
Zn$^{++}$ is in $d^{10}$ configuration. The third ionization energy
should be a rough measure of the energy level for $d$-orbital of Cu
and Zn/Ni ions though the electrons from the $d$-orbital of Cu form
a band. By comparing the energies of Cu atom $E_{Cu^{++}} = -36.83
eV$, and Zn atom $E_{Zn^{++}} = -39.722eV$, we estimate energy to be
on the order of $U_0 \simeq -2.89 eV$. Therefore, Zn atom plays the
role of a strong attractive potential.

In case of Ni doping, Ni$^{++}$ has a $3d^8$ configuration and a
spin $S=1$ ground state is formed. Therefore, Ni impurity will
produce both potential scattering $U_0$ and magnetic scattering $J$.
We can estimate the energy $U_0$ again by taking the difference
between atomic energies for $E_{Cu^{++}} = -36.83 eV$ and
$E_{Ni^{++}} = -35.17 eV$. We estimate $U_0 \simeq 1.66 eV$ for Ni.
It provides a weaker repulsive potential for the non-magnetic
scattering part.

\section{Single impurity bound state in a pseudogap state of two-dimensional metals}
\label{sec:PG}

In the high-temperature cuprates, many
experiments~\cite{ChRenner:1998a,MRNorman:1998a,JWLoram:2000} show
that the electronic density of states near the Fermi surface is
suppressed within the range of $\Delta_{PG}$ above the
superconducting phase transition temperature $T_c$ but below a
characteristic temperature $T^{*}$. So far, the mechanism for the PG
state is still hotly debated. For a review
see~\cite{TTimusk:1999,TTimusk:2003}. The typical competing
scenarios for this anomalous phenomenon, including mainly the
pre-formed pair with phase-fluctuation model~\cite{VJEmery:1995},
the Bose-Einstein condensation of Cooper pairs~\cite{QChen:1998},
the time-reversal-symmetry-breaking circulating current
model~\cite{CMVarma:1999}, and the $d$-density-wave model
(DDW)~\cite{SChakravarty:2001}, which is typical kind of staggered
flux state~\cite{IAffleck:1988,JBMarston:1989,TCHsu:1991}. In the
first scenario, the normal state contains preformed Cooper pairs,
and the phase fluctuation of the pairing field destroys the long
range order, that is, the superconductivity. Since the pairing field
has at the onset a $d$-wave symmetry in the momentum space for the
relative motion degrees of freedom, a $d$-wave-like pesudogap
follows naturally. In the second category of scenarios, it is
speculated that the PG state comes as a result of a new order not
inheriting from the superconducting pairing.

In this Section, we are not going to discuss the origin for the PG.
Instead we are particularly interested in the consequences of the
electronic property around a single impurity in the PG state. One
can consider the temperature evolution of the impurity state as the
temperature increases and eventually becomes larger than $T_c$.
 Then there are two possibilities for  the  evolution of impurity
 resonance at $T>T_c$: a) the impurity resonance
 gradually broadens until the superconducting gap
  vanishes, at which point the impurity resonance totally disappears and b)
the resonance gets broader but survives above $T_c$. Which of the
possibilities is realized
 depends on the normal state phase the
superconductor evolves into. It has been
argued~\cite{JWLoram:2000,VMKrasnov:2000} that in the underdoped
regime the superconducting gap opens up in addition to the PG
present well above $T_c$.  Hence, we find that the impurity
resonance survives above $T_c$ in the PG state of high-$T_c$
materials.
 The position and the
width of the resonance are determined by the impurity scattering
strength and PG scale. In the absence of PG above $T_c$ the impurity
state disappears.

Specifically we calculate the resonant state generated by the
substitution of one Cu atom with a Zn atom using  the $T$-matrix
approach. We rely on the fact that the density of states (DOS) is
depleted at the Fermi energy in the PG regime. We argue that the
mere fact that the  DOS is depleted at the Fermi energy is
sufficient  to produce a resonance near the nonmagnetic impurity,
such as Zn.
Before we consider the impurity effect within specific scenarios, we
give a general analysis, which is valid in the PG state {\em with no
superconducting phase or amplitude fluctuations above $T_c$}, as
long as there are interactions that lead to the PG state.
 The approach  we take is similar to the previous analysis of
the nonmagnetic impurity in the superconducting state
\cite{AVBalatsky:1995}. See also Fig.~\ref{FIG:phasediagram}. The
states generated by the impurity are still given by the poles of the
$T$ matrix:
\begin{equation}
\label{EQ:U} G_0 (\Omega) = \frac{1}{U_0}\;.
\end{equation}
This is an implicit equation for $\Omega$ as a function of $U_0$.
This solution can be complex, indicating the resonant nature of the
virtual state. To solve this equation, we need to know the form of
nonperturbed Green's function on the impurity site, $g_{0}(E)$. To
do so, we split $g_0$ into its imaginary and real part $g_0 = g_0' +
i g_0 ^{\prime\prime}$ with $G_0^{\prime\prime}(\omega) = - \pi
N_0(\omega)$, where $N_0(\omega)$ is the density of states.

Measurements on the electronic specific heat by Loram {\it et
al.}~\cite{JWLoram:2000} show that the normal state pseudo gap opens
abruptly in the underdoped region below a hole doping equal to
$p_{crit} \sim 0.19$ holes/CuO$_2$. Inspired by these data, we will
assume that around the pseudogap region, states are
  partly depleted and the density of states is linear, that is
 $N_{0}(\omega) = N_0  |\omega| / \Delta_{PG}$ for $|\omega| \leq
\Delta_{PG}$
 and $N_{0}(\omega) = N_0$ for  $\Delta_{PG}< |\omega| < W/2$ with $W$
the
 bandwidth.
 This density of states is depicted in Fig.~\ref{FIG:density}(a).
 As it is obvious from
the solution of Eq.~(\ref{EQ:U}), the precise position and the width
of the resonance will depend on the specific form of the PG. We will
use this linearly vanishing PG DOS. Results for other forms of
$N(\omega)$ like a fully gapped DOS or a DOS with a quadratic
dependent gap, can be obtained in the same way  and lead essentially
to similar expressions.~\footnote{We argue that the appearance of
the intragap impurity state is a robust feature of any depleted DOS
around the Fermi surface. We also considered the model DOS with
$N_{0}(\omega) = N_0 [ a +  (1-a) \omega^2 / \Delta_{PG}^2 ]$ which
leads essentially to similar results as a function of the impurity
strength with a resonant state at \mbox{$\Omega = -\Delta_{PG} (1 +
i  \pi a N_0 U_0)/ (4 N_0 U_0(1-a -\Delta_{PG}/W))
 $} \mbox{$\approx  -\Delta_{PG} (1 + i  \pi a N_0 U_0)/ (4 N_0 U_0(1-a)) $} when
 $\Delta_{PG}/W$ is small.
But also a fully gapped DOS equal to $N(\omega) =N_0$ for $|\omega|
\in [ \Delta_{PG}, W/2 ]$ and zero otherwise gives rise to a
comparable expression with a resonant state at $\Omega = -
\Delta_{PG}/(2 U_0 N_0)$.}

\begin{figure}[htbp]


\caption{An impurity state in a high $T_c$ superconductor: (a) The
DOS in the pseudogap regime used in this article (see also $\left[
11 \right]$) and (b) the DOS in the superconducting state as was
used in $\left[ 1 \right]$. In both phases there is a resonant
state.} \label{FIG:phasediagram}
\end{figure}

Using this DOS for $g_0^{\prime\prime}$ and invoking the
Kramer-Kronig relation~\cite{GDMahan:2000}
\begin{equation}
 g_0^{\prime} (\omega) = \frac{1}{\pi} \int_{- \infty}^{\infty} d \omega^{\prime}
 g_0^{\prime\prime}(\omega^{\prime}) P \left(\frac{1}{\omega^{\prime} - \omega}\right),
\end{equation}
with $P$ the Cauchy's principle value, one can obtain the real part
$g_0^{\prime}$ as
\begin{eqnarray}
\label{kramer} g_0^{\prime}(\omega)& =& - N_0 \ln \left|
\frac{\frac{W}{2} -\omega} { \frac{W}{2} +\omega}\right| +N_0 \ln
\left| \frac{\Delta_{PG} -\omega} { \Delta_{PG}+\omega}\right|
\nonumber
\\&-& N_0 \frac{\omega}{\Delta_{PG}} \ln \left|
\frac{\Delta_{PG}^2 -\omega^2} { \omega^2 }\right|.
\end{eqnarray}
This function is plotted in Fig.~\ref{FIG:density}(b) together with
 $1/U_0$.
If $2U_0 N_0 > 1$, one can see from this figure that equation
(\ref{EQ:U}) has four solutions. Since the width of a resonance
state is proportional to $|\Omega|$, the only state with sharp width
is the solution with $|\Omega|$ close to zero and we will only
consider this solution. After expansion in $\omega$ of equation
(\ref{kramer}) we arrive at an expression for this solution $\Omega$
of Eq.~(\ref{EQ:U}):
\begin{equation}
g_0(\Omega) = - \frac{2 \Omega N_0}{\Delta_{PG}} \left[ \ln \left|
\frac{\Delta_{PG}}{\Omega} \right|+1 - \frac{i \pi~ \mbox{sgn}
(U_0)}{2} \right] = \frac{1}{U_0},
\end{equation}
This equation can be solved exactly in terms of LambertW
functions,~\footnote{ The exact solution in terms of a LambertW
function, $Lw(-1,x)$,
 is
$\Omega = - \Delta_{PG} \mbox{sgn}(U_0) \exp\{Lw (-1,-
\mbox{sgn}(U_0) \exp(i \pi /2 -1 )/(2 N_0 U_0)) +1 -i \pi /2\}$,
where $Lw(x)$ is such that $Lw(x) \exp[Lw(x)] = x$. } which, to
logarithmic accuracy with $\ln \vert 2 U_0 N_0\vert
> 1$, gives:~\footnote{The simplest model for thermal broadening is to
assign the
 temperature dependent width: Thermal broadening at high temperatures $T >T_c$
 substantially  broadens the impurity resonance peak $\Omega '' (T)
 =\sqrt{(\Omega ''(T=0))^2 + T^2}$.}
\begin{eqnarray}
 \label{EQ:solution}
 \Omega&=& \Omega^{\prime} + i \Omega^{\prime\prime}
 \nonumber \\
 &=& - \frac{\Delta_{PG}}{2 U_0 N_0} \frac{1}{ \ln |2 U_0
N_0|} \nonumber \\
&& \times \left[ 1 - \frac{1}{\ln|2U_0 N_0|}+ \frac{ i \pi
~\mbox{sign} (U_0)}{2 \ln |2 U_0 N_0|} \right],
\end{eqnarray}
where $\Omega^{\prime}$ is the energy position and
$\Omega^{\prime\prime}$ the decay rate.

Using formula  (\ref{EQ:solution}), and taking $N_0 = 1~
\mbox{state}/\mbox{eV}$, $\Delta_{PG}\sim 300 \mbox{K} \sim 30
\mbox{meV}$ and the scattering potential  $U_0 \approx \pm 2
\mbox{eV},$ we estimate $\Omega \sim \pm 2 \mbox{meV} \sim \pm 20
\mbox{K} $ as was found by Loram {\it et al.}~\cite{JWLoram:2000}.
This energy is close to the Zn resonance energy $\omega_0 = - 16 K$,
seen in the superconducting state \cite{SHPan:2000b}. By combining
these results with the band-structure arguments~\cite{IMartin:2002},
we come to conclusion that the Zn impurity in Bi2212 is strongly
attractive, with $U_0 \sim -2\mbox{eV}$. This result, as we will now
see, may be modified due to the particle-hole asymmetry
characteristic of doped cuprates.

\begin{figure}[htbp]






\caption{(a) The density of states $N(\omega) = -
g_0^{\prime\prime}(\omega) / \pi$. Around the pseudogap states are
only partly depleted e.g. $N_0(\omega) = N_0 |\omega|/\Delta_{PG}$,
where $N_0(\omega) = N_0$ for $\Delta_{PG}< |\omega| < W/2$ with $W$
the
 bandwidth. (b) The real part $g_0^{\prime}(\omega)$ of Green's function together
with $1/U_0$ and $U_0$ positive. $\Omega^{\prime}$ is the real part
of the solution of the equation $g_0 (\Omega) = 1/U_0$ close to zero
and therefore with sharp bandwidth.
 (c) The  impurity induced resonance
at $\Omega^{\prime} = - \Delta_{PG}/2 U_0 N_0 \ln (2 U_0 N_0)$.
Because the other three solutions of Eq.~(\ref{EQ:U}) have much
broader bandwidth, they are not depicted here. All the figures are
taken on the impurity site. From Kruis, Martin and
Balatsky~\cite{HVKruis:2001} } \label{FIG:density}
\end{figure}

In the absence of particle-hole symmetry, a similar calculation can
be done.  The simplest way to introduce the asymmetry is by making
the upper and lower cutoffs in the DOS unequal.  This situation
corresponds to a chemical potential $\mu$, located away from the
center of the band.  Keeping the DOS otherwise unchanged, with the
pseudogap centered at the chemical potential, results only in the
following change in the first logarithmic  term of
Eq.~(\ref{kramer}):
\begin{equation}
- N_0 \ln \left| \frac{\frac{W}{2} - \mu -\omega} { \frac{W}{2} +
\mu +\omega}\right|
\end{equation}
Neglecting the frequency $\omega$ relative to chemical potential
$\mu$ and assuming that $\mu$ is small relative to the bandwidth, we
obtain that the results for the asymmetric case can be obtained from
the symmetric ones by the substitution
\begin{equation}
\frac1U_0 \rightarrow \frac1U_0 - \frac{4N_0\mu}{W}.
\end{equation}
The effect of the asymmetry term can be estimated for the
superconducting cuprates.  For 20\% hole doping, $\mu \sim - (1/5)
W/2 = -W/10$.  Hence, the modified value for the Zn impurity
strength in Bi2212 can be obtained from the symmetric result, $1/U^*
= 1/U_0 + {4N_0\mu}/{W}$.  The new value is $U^* \sim -1\ eV$, which
is a strongly attractive potential, as is expected from the band
structure arguments.

The solution of the impurity state deep in the superconducting
regime involves two aspects: the energy position and the width of
the resonance and secondly, the real space shape of the impurity
state. We have discussed the energy of the impurity state above.
Great advantage of the on-site impurity solution for the local
potential $U_0$ is that only on-site propagator $g_0(\omega)$ enters
into calculation. Hence the knowledge of the DOS was sufficient to
calculate the impurity state. On the other hand, to calculate the
real space image of impurity induced resonance, one  would require
more detailed knowledge of the Green's functions in the PG regime.
Quite generally, one would expect for  a $d$-wave-like PG with
nearly nodal points  along the $(\pm \pi/2, \pm \pi/2 )$ directions,
that the impurity resonance in the pseudogap regime would be
four-fold symmetric, similar to superconducting
solutions~\cite{AVBalatsky:1995}. This calculation  would require a
specific model for the PG state, some of which are considered below.

{\em Impurity state in a DDW system.} The model Hamiltonian for a
clean DDW system is written as:
\begin{equation}
H_0=\sum_{ij,\sigma} [-t_{ij} + (-1)^{i} iW_{ij}]
c_{i\sigma}^{\dagger} c_{j\sigma} -\mu \sum_{i,\sigma}
c_{i\sigma}^{\dagger}c_{i\sigma}\;.
\end{equation}
Here $c_{i\sigma}^{\dagger}$ ($c_{i\sigma}$) is the creation and
annihilation of an electron with spin projection $\sigma$ at the
$i$th site. The DDW order parameter $W_{ij}$ has the value
$W_{i,i\pm \hat{x}}=W_{\pm\hat{x}}=\frac{W_{0}}{4}$ and
$W_{i,i\pm\hat{y}}=W_{\pm\hat{y}}=-\frac{W_{0}}{4}$, while is zero
otherwise. Notice that the prefactor $i=\sqrt{-1}$ before $W_{ij}$.
It indicates that the DDW state breaks the time reversal symmetry.
The quantity $\mu$ is the chemical potential. Without loss of
generality, we have assumed that the impurity is located at the
origin. In the momentum space, the Hamiltonian is given by:
\begin{equation}
H_0= \sum_{k,\sigma}\xi_{k}c_{k\sigma}^{\dagger}c_{k\sigma}
+\sum_{k,\sigma} iW_{k}[c_{k\sigma}^{\dagger}c_{k+Q,\sigma} -
c_{k+Q,\sigma}^{\dagger}c_{k\sigma}]\;.
\end{equation}
Here $c_{k\sigma}^{\dagger}$ ($c_{k\sigma}$) are the creation and
(annihilation) operators of an electron at the wave vetor $k$ and
with spin projection $\sigma$. The single particle energy dispersion
for the normal state is given by Eq.~(\ref{EQ:ENERGYDISP}). The DDW
gap is given by:
\begin{equation}
W_{k}=\frac{W_{0}}{2}(\cos k_x - \cos k_y)\;.
\end{equation}
Here the wave vector $k_x$ and $k_y$ are defined in the first
Brillouin zone.

In view of the fact that the DDW state breaks the translational
symmetry with lattice constant but conserves that by $\sqrt{2}a$
along the diagonals of the square lattice, it is convenient to
rewrite the Hamiltonian in the reduced Brillouin zone with its area
one half of the original by introducing a two-component electron
operator $\Psi_{k\sigma}^{\dagger} = (c_{k\sigma}^{\dagger},
c_{k+Q,\sigma}^{\dagger})$ with $\mathbf{Q}=(\pi,\pi)$:
\begin{equation}
H_0=\sum_{k\in rBZ,\sigma} \Psi_{k\sigma}^{\dagger} \left(
\begin{array}{cc}
\xi_{k} & i2W_{k} \\
-2iW_{k} & \xi_{k+Q}
\end{array} \right)
\Psi_{k\sigma} \;, \label{EQ:DDW_Hamil0}
\end{equation}
where $rBZ$ represents the reduced Brillouin zone.

Accordingly, we can introduce the following Green functions in the
clean limit:
\begin{equation}
\hat{\mathcal{G}}^{(0)}(k;\tau)=\left( \begin{array}{cc}
\mathcal{G}^{(0)}_{11} &
\mathcal{G}^{(0)}_{12} \\
\mathcal{G}^{(0)}_{21} & \mathcal{G}^{(0)}_{22} \end{array}
\right)\;.
\end{equation}
with
\begin{subequations}
\begin{eqnarray}
\mathcal{G}_{11}^{(0)}(k;\tau)&=&-\langle T_{\tau}
[c_{k\sigma}(\tau) c_{k\sigma}^{\dagger}(0)]
\rangle \;,\\
\mathcal{G}_{12}^{(0)}(k;\tau)&=&-\langle T_{\tau}
[c_{k\sigma}(\tau) c_{k+Q,\sigma}^{\dagger}(0)]
\rangle \;,\\
\mathcal{G}_{21}^{(0)}(k;\tau)&=&-\langle T_{\tau}
[c_{k+Q,\sigma}(\tau) c_{k\sigma}^{\dagger}(0)]
\rangle \;,\\
\mathcal{G}_{22}^{(0)}(k;\tau)&=&-\langle T_{\tau}
[c_{k+Q,\sigma}(\tau) c_{k+Q,\sigma}^{\dagger}(0)] \rangle \;.
\end{eqnarray}
\end{subequations}
Here the factor $T_{\tau}$ is a $\tau$-ordering operator as usual,
and $c_{k\sigma}(\tau)=e^{H\tau}c_{k\sigma}e^{-H\tau}$ is the
operator in the Heisenberg representation. Given the Hamiltonian
Eq.~(\ref{EQ:DDW_Hamil0}), with aid of the equation of motion for
the field operator $c_{k\sigma}(\tau)$ and
$c_{k\sigma}^{\dagger}(\tau)$, and by performing a Fourier transform
with respect to $\tau$,
\begin{equation}
\hat{\mathcal{G}}(k;\tau) =k_{B}T \sum_{\omega_{n}}
\hat{\mathcal{G}}(k;i\omega_{n})e^{-i\omega_{n}\tau}
\end{equation}
with $\omega_{n}=(2n+1)\pi k_{B}T$, one can arrive at:
\begin{equation}
\hat{\mathcal{G}}^{(0)}(k;i\omega_{n}) = \left(
\begin{array}{cc} i\omega_{n}-\xi_{k} & -2iW_{k} \\
2iW_{k} & i\omega_{n}-\xi_{k+Q}
\end{array} \right)^{-1}\;.
\label{EQ:Greenk}
\end{equation}

As will be obvious immediately, we also need to calculate the Green
function in real space, which through the Fourier transform is given
by
\begin{eqnarray}
\mathcal{G}^{(0)}(i,j;i\omega_{n}) &=& \frac{1}{N} \sum_{k\in rBZ}
e^{i\mathbf{k}\cdot \mathbf{R}_{ij}}
[\mathcal{G}^{(0)}_{11}(k;i\omega_{n}) +
\mathcal{G}^{(0)}_{22}(k;i\omega_{n}) \nonumber \\
&&+ e^{-i\mathbf{Q}\cdot \mathbf{R}_{j}}
\mathcal{G}^{(0)}_{12}(k;i\omega_{n}) + e^{i\mathbf{Q}\cdot
\mathbf{R}_{i}} \mathcal{G}^{(0)}_{21}(k;i\omega_{n}) ]\;,\nonumber \\
\label{EQ:Greenreal}
\end{eqnarray}
where $\mathbf{R}_{i}$ are the lattice vectors and $\mathbf{R}_{ij}=
\mathbf{R}_{i}-\mathbf{R}_{j}$. Specially, from
Eqs.~(\ref{EQ:Greenk}) and (\ref{EQ:Greenreal}), one can obtain:
\begin{equation}
\mathcal{G}^{(0)}= \frac{1}{N} \sum_{k\in rBZ}  \frac{2i\omega_{n}
-\xi_{k+Q} -\xi_{k}}{(i\omega_{n}-\xi_{k})(i\omega_{n}-\xi_{k+Q}) -
4W_{k}^{2}}\;.
\end{equation}

In the presence of a single non-magnetic impurity in the DDW state,
the $T$-matrix analysis just shows again that the resonance state is
determined by the poles of $T(i\omega_{n}\rightarrow \omega +
i0^{+})$, i.e., the following equation
\begin{equation}
\mathcal{G}^{(0)}(0,0;\omega + i0^{+})=\frac{1}{U_0}\;.
\end{equation}
The existence of the resonant states will directly manifest in the
local density of states:
\begin{equation}
N_{i}(\omega)=-\frac{2}{\pi} \ {\rm Im} \, \mathcal{G}(i,i;
\omega+i\delta) \;.
\end{equation}

\begin{figure}[t]
\caption{DDW-DOS for the clean case (solid line) and in the presence
of a non-magnetic impurity with $U_0=1$ eV: (1) DOS on the impurity
site, (2) DOS on the nearest-neighbor site, and (3) DOS on the
next-nearest-neighbor site. The other parameter values are: $t=300
meV$, $W_0=25 meV$, $t^{\prime}=0$, and $\mu=0$. From
Morr~\cite{DKMorr:2002}.} \label{FIG:DOSDDW}
\end{figure}

\begin{figure}[t]
\caption{{\it (a)} Fermi surface in the DDW state with
$t^\prime=-0.3t$, $\mu=-0.91t$ (corresponding to a hole-doping of
$10\%$) and $W_0=25$ meV. The hole pockets are centered around $(\pm
\pi/2,\pm \pi/2)$. {\it (b)} DOS in the DDW state with the same band
parameters as in {\it (a)}, for the clean case (solid line) and in
the presence of a non-magnetic impurity with $U_0=1$ eV: (1) DOS on
the impurity site, (2) DOS on the nearest-neighbor site, and (3) DOS
on the next-nearest-neighbor site. Inset: SC DOS for the same band
parameters as in {\it (a)}. From Morr~\cite{DKMorr:2002} }
\label{FIG:dkmorrFig4}
\end{figure}

The results are displayed in Figs.~\ref{FIG:DOSDDW} and
\ref{FIG:dkmorrFig4}. As shown in these figures, the electronic
excitation spectrum around the impurity in the DDW state is very
sensitive to the parameter values, which control the band structure.
For $t^{\prime}=0$ and at the half filling ($\mu =0$), the electron
density of states in the clean limit is vanishingly small around the
Fermi energy, which leads to resonance states near the Fermi energy.
With $t^{\prime}=0$ but the system doped away from the half filling,
the resonant peak in the LDOS is shifted away from the Fermi energy.
This is because the energy at which the band DOS vanishes no longer
coincides with the Fermi energy. For a set of more realistic
parameter values, the density of states in the clean limit shows
negligible reduction at low energies, which makes the local density
of states near the impurity not exhibit any signature of a resonance
state. These results were independently obtained by Zhu {\em et
al.}~\cite{JXZhu:2001d}, Wang~\cite{QHWang:2002}, and
Morr~\cite{DKMorr:2002}. The quasiparticle states in the DDW state
with a finite concentration of non-magnetic impurities was
investigated by Ghosal and Kee~\cite{AGhosal:2004}.

{\it Phase-fluctuation scenario}: We now devote to a discussion on
the impurity state in a phase-fluctuating pairing field. For more
details, see~\onlinecite{QHWang:2002}. The effective mean field
Hamiltonian in a two-dimensional square lattice for a $d$-wave
superconductor can be written as:
\begin{equation}
H  =  \sum_{ij} \Psi_{i}^{\dagger} \left( \begin{array}{cc}
-t_{ij} -\mu \delta_{ij} & -\Delta_{ij} \\
-\Delta_{ij}^{*} & -(-t_{ji}-\mu\delta_{ij})
\end{array}
\right) \Psi_{j} \;,
\end{equation}
where $\Psi_{i}^{\dagger} =(c_{i\uparrow}^{\dagger}\;,
c_{i\downarrow})$ is the two-component spinor operator in the Nambu
space, the other notations are the standard except the pairing field
$\Delta_{ij}$ is fluctuating. In the phase fluctuation scenario, the
amplitude of the superconducting order parameter is fixed while its
phase fluctuating. For $d$-wave superconductivity, one can write the
pairing field as:
\begin{equation}
\Delta_{ij}=\frac{\Delta_0 \eta_{ij}}{4} e^{i\varphi_{ij}}
=\tilde{\Delta}_{ij}e^{i\varphi_{ij}}\;,
\end{equation}
with $\eta_{ij}=1 \;(-1)$ for $x\; (y)$ direction bonds and the
phase $\varphi_{ij}=(\varphi_{i}+\varphi_{j})/2$. The spatial
variation of the phase will give rise to a superfluid flow
associated with the Cooper pairs. By performing a gauge
transformation,
\begin{equation}
\tilde{\Psi}_{i} = e^{-i\varphi_{i}\sigma_{3}/2} \Psi_{i}
\end{equation}
with $\sigma_{3}$ the third component of the Pauli matrix, one can
remove the phase factor on the pairing field such that
\begin{equation}
\tilde{H}  =  \sum_{ij} \tilde{\Psi}_{i}^{\dagger} \left(
\begin{array}{cc}
-\tilde{t}_{ij} -\mu \delta_{ij} & -\tilde{\Delta}_{ij} \\
-\tilde{\Delta}_{ij}^{*} & -(-\tilde{t}_{ji}-\mu\delta_{ij})
\end{array}
\right) \tilde{\Psi}_{j} \;,
\end{equation}
where $\tilde{t}_{ij}=t_{ij}e^{-i(\varphi_i -\varphi_{j})/2}$. In
type-II superconductors, the length scale of the phase variation
(the London penetration depth) is much larger than the Fermi
wavelength. Hence one can specify the phase for the Cooper pair
$\varphi_{i}=2\mathbf{q}_{s}\cdot \mathbf{R}_{i}$, where
$\mathbf{q}_{s}$ is the average momentum per electron in the
superfluid state. This ansatz leads to the Green function in the
clean limit:
\begin{equation}
\hat{\mathcal{G}}^{(0)}(k;q_{s};i\omega_{n}) = \left(
\begin{array}{cc}
i\omega_{n}-\xi_{k+q_s} & -\Delta_{k} \\
-\Delta_{k}^{*}  & i\omega_{n}+\xi_{k-q_s}
\end{array} \right)^{-1}\;,
\end{equation}
where $\Delta_{k}=\frac{\Delta_{0}}{2} (\cos k_x -\cos k_y)$ and the
energy dispersion is still given by Eq.~(\ref{EQ:ENERGYDISP}).

In the presence of a non-magnetic single-site impurity at site
$i=(0,0)$ in the 2D system, the
 Green function for the impurity system can be obtained, within the
$T$-matrix approximation,
\begin{eqnarray}
\hat{\mathcal{G}}(i,j;q_s;i\omega_{n}) &=&
\hat{\mathcal{G}}^{(0)}(i,j;q_s;i\omega_{n}) +
\hat{\mathcal{G}}^{(0)}(i,0;q_s;i\omega_{n}) \nonumber
\\ && \times \hat{T}(q_s;i\omega_{n}) \hat{\mathcal{G}}^{(0)}(0,j;q_s;i\omega_{n})\;,
\end{eqnarray}
\noindent with the $T$-matrix given by
\begin{equation}
\hat{T}^{-1}(i\omega _{n};{\bf q}_{s}) =\tau_{3}/U_0
-\hat{\mathcal{G}}^{(0)}(0,0;q_s;i\omega_{n})\;.
\end{equation}
With a fixed $\mathbf{q}_s$, the local density of states (LDOS) at
site $i$ is given by:
\begin{equation}
N(i;q_{s};\omega)=-(2/\pi)\mbox{Im}\mathcal{G}_{11}(i,i;q_{s};
\omega+i0^+)\;.
\end{equation}

For the fluctuating phase, one needs to take average over
$\mathbf{q}_{s}$. If the fluctuation is thermal, the statistical
distribution is Gaussian, as governed by the Kosterlitz-Thouless
(KT)
theory~\cite{JMKosterlitz:1973,JMKosterlitz:1974,DESheehy:2001},
that is $ \rho(q_{s}) =e^{-q_{s}^{2}/2n_{v}}/\sum_{q_{s}}
e^{-q_{s}^{2}/2n_{v}}$, where $n_{v} =\exp[-\sqrt{aT_{c}/(T-T_c)}]$
within the KT theory. In the continuum limit, $\sqrt{\langle
q_{s}^{2} \rangle } =\sqrt{n_v}$. The averaged LDOS is caculated as
$N(i;\omega) = \langle N(i;q_{s};\omega)\rangle$.

\begin{figure}[tbp]
\caption{Local density of states with $\Delta _{0}=0.68t$,
$\protect\mu =-0.3t$ and $U_0=100t$. (a) $N({\bf
r}_{nn},\protect\omega )$ versus $\protect\omega $. Solid lines:
$n_v= 10^{-6},10^{-4},10^{-3}$, and $5\times 10^{-3}$ with
decreasing peaks. The dotted line is the LDOS at $n_v=0$ and $U_0=0$
for comparison. (b) $N({\bf r},0.05t)$ at $n_v=0$. The impurity is
at the center. (c) The same as (b) for $n_v=5\times 10^{-3}$. The
gray scale is the same in (b) and (c). The other parameter
$t^{\prime}=0$. From Wang~\cite{QHWang:2002}.}
\label{FIG:qhwangFig1}
\end{figure}

The results for various values of $n_v$ are shown in
Fig.~\ref{FIG:qhwangFig1}. For small $n_{v}$, the resonance peak is
very sharp and similar to that in the superconducting state
($n_v=0$). When $n_{v}$ is increased, the peak is broadened with
decreasing peak height. Finally, the spectrum at low energies
becomes featureless.

Finally, in the phase fluctuation scenario of the PG state, the
electron excitation spectrum around the impurity is very sensitive
to how far the temperature is away from the superconducting
transition temperature $T_c$. However, in the normal-state ordering
scenario, the resonance states are not sensitive to the temperature
up to the PG critical temperature. Another piece of difference of
the electronic states around the impurity between two scenarios is
the energy position of the resonance state in the phase fluctuation
state is not sensitive to the doping while that in the state with a
normal ordering shifts with the doping. More generally, if the
superconducting fluctuations are
 present, then an additional satellite peak should appear
 at the opposite bias due to the particle-hole nature of
 the Bogoliubov quasiparticles.  The relative magnitude
  of the particle and the hole parts of the impurity
  spectrum can be used to determine the extent to which
  the PG is governed by the superconducting fluctuations.
  In the case of fully non-superconducting PG (e.g., the DDW state),
  there should
  be no observable counterpart state.
Combined with other experimental
proposals~\cite{BJanko:1999,IMartin:2000b}, the impurity state can
help to better understand the mysterious PG state.


\section{Quantum phase transition in  $S$-wave superconductor with magnetic impurity}
\label{sec:QPT}

\subsection{Introduction}

In this Chapter  we revisit the problem of a localized classical
magnetic moment in  a superconductor. A remarkable aspect of this
interaction we will focus on here is the {\em first-order zero
temperature transition} which takes place in an $s$-wave
superconductor as a function of the ''magnetic moment,'' $J_0S$,
where $S$ is the local impurity spin and $J_0$ is the exchange
coupling. In this transition, the spin quantum number $s$ of the
electronic ground state of the superconductor $|\Psi_0\rangle$
changes from zero for a subcritical moment $J_0<J_{crit}$ to $1/2$
for $J_0>J_{crit}$.
 The total
spin becomes $S \pm 1/2$ depending on the sign of the exchange
interaction $J_0$. The first to point out the phase transition was
Sakurai\cite{ASakurai:1970} it corresponds to a level crossing
between two ground states as a function of the exchange coupling
$J_0$. In a singlet superconductor level crossing occurs  between
states where the impurity spin is either unscreened or partially
screened.  The states have  different spin quantum numbers and level
crossings are generally allowed.   These quantum phase transitions
are of the first order and hence do not have divergent time or
length scale associated with them.

\begin{figure}[htbp]
\caption{The local effect of a magnetic moment on the low-energy
spectral density in an $s$-wave superconductor.}
\label{FIG:FigMtalk1}
\end{figure}

We address the above problem at zero temperature by using the
mean-field approximation  within the $T$-matrix formulation and the
self-consistent approach, which takes into account a local
gap-function relaxation. Local Coulomb interaction $U$ which breaks
particle-hole symmetry and leads to an asymmetric spectral density
for the impurity-induced quasiparticle states.
Figure~\ref{FIG:FigMtalk1} illustrates the local effect of a
magnetic moment on the low-energy spectral density in an $s$-wave
superconductor. Since we limit our considerations to a classical
spin, $S\gg 1$, the impurity moment cannot be screened completely by
the quasiparticles. We show that the gross features of the
impurity-induced quasiparticle states in $s$- and $d$-wave
superconductors can be qualitatively understood within the
non-self-consistent $T$-matrix formalism.

The transition is not unique to the classical spin. Similar effect
is found in a Kondo model of a quantum spin, see
Sec.~\ref{sec:Kondo}.

\subsection{Quantum phase transition as a level crossing}

The physical picture of the quantum transition follows from the
behavior of the impurity-induced bound state. This transition is a
consequence of the instability of the spin unpolarized ground state,
because, for a large enough $J_0$, the energy of the
impurity-induced quasiparticle excitation would fall below the
chemical potential.

As we have discussed it in the context of Yu-Shiba-Rusinov solution
for a classical spin, see Sec.~\ref{sec:Shiba}, impurity state has
energy that is always below the gap threshold :
    \beqa \Omega_0/\Delta_0 = \frac{1 -(\pi J_0SN_0)^2}{1+(\pi J_0SN_0)^2}
    \label{EQ:QPT3} \eeqa
and  has particle ($u_{-1}$) and hole ($v_{-1}$) amplitudes at
positive and negative energies. One can see the level crossing and
change of the ground state already from this result. Ignoring the
self-consistency, and using Eq.~(\ref{EQ:QPT3}),  we find that the
singularity occurs at \beqa J_0 = J_{crit} = 1/\pi N_0 S
\label{EQ:QPT3a} \eeqa

For a weak coupling $J_0<J_{crit}$ , the ground state of the
superconductor is a paired state of time-reversed single-particle
states in the presence of the impurity scattering, with ground state
wave function being the BCS like :
    \beqa
|\Psi_0\rangle_{J_0<J_{crit}} \sim \Pi_n[u_n + v_n \psi^\dag_n
\psi^\dag_{-n}]|0\rangle = |\Psi_0\rangle \label{EQ:QPT1} \eeqa
Here, since the translational symmetry is broken by impurity, and we
consider the eigen-states of the scattering problem in the presence
of impurity, states are labeled by a discrete index $n =
1,...\infty$ corresponding to a discrete scattering states that are
the basis of the Bogoliubov Hamiltonian with impurity. The $n=1$
state would correspond to an impurity bound state, localized on
impurity site.  Index $-n$ correspond to a time reversal state, i.e.
the localized state with opposite spin.  The first excited state at
$J_0<J_{crit}$ would correspond to a single quasiparticle state
where one excitation is present. The lowest excited state
corresponds to a intragap Yu-Shiba-Rusinov state at energy
$\Omega_0$, see Fig.~\ref{FIG:QPT-pic}.

\begin{figure}[th]
\caption{Two
variational states  are shown schematically. $|\Psi_0\rangle $ is a
standard BCS wave function that contains only paired particles and
has unscreened impurity spin $S$. $|\Psi_1 \rangle$ is a variational
wave function that describes the formation of the bound state
between particle  with  the spin opposite to the local spin ( for
antiferromagnetic coupling); this state is inherently a non BCS
state and electonic spin quantum number differs by one unpaired spin
compared to $|\Psi_0 \rangle $. } \label{FIG:QPT-pic}
\end{figure}

The excited state can be written as: \beqa
|\Psi_{-1}\rangle_{J_0<J_{crit}} \sim \gamma^{\dag}_{-1}|\Psi_0\rangle =  |\Phi_{-1}\rangle \nonumber\\
|\Phi_{-1}\rangle = \psi^{\dag}_{-1}\prod_{n>1}[u_n + v_n
\psi^{\dag}_{n} \psi^{\dag}_{-n}]|0\rangle \label{EQ:QPT2} \eeqa
with standard quasiparticle definitions of $\gamma_1 = u_1 \psi_1 -
v_{1}\psi^{\dag}_{-1}$ $\gamma^{\dag}_1 = u_1 \psi^{\dag}_1 -
v_{1}\psi_{-1}$,  $\gamma^{\dag}_{-1} = u_1\psi^{\dag}_{-1} +
v_1\psi_1$,  etc, with $u^2_n +v^2_n = 1$. For future use we
introduce \beqa
 \widetilde{|\Psi_0}\rangle = \prod_{n>1}[u_n + v_n\psi^{\dag}_n\psi^{\dag}_{-n}]|0\rangle
\label{EQ:QPT8} \eeqa Then $|\Phi_{-1}\rangle = \psi^{\dag}_{-1}
\widetilde{|\Psi_0}\rangle$. The state
$\gamma^{\dag}_1|\Psi_0\rangle$
 does not appear inside the superconducting gap and hence is not relevant for this discussion.
 Note that $|\Psi_{0}\rangle $ is a true vacuum for all  quasiparticles: e.g. $\gamma_{\pm 1}\prod_{n>0}[u_n + v_n
\psi^{\dag}_{n} \psi^{\dag}_{-n}]\rangle = 0$. \footnote {Here the
spin of the state $n=1$ is determined by
 the sign of exchange coupling $J_0$. We will assume it to be antiferromagnetic.
 So the electronic spin of the state $n=-1$ in
Eq.~(\ref{EQ:QPT2}) is opposite to the local spin $S$  assumed to be
up without loss of generality. Case of ferromagnetic coupling is
similar. Indeed classical spin solution Eq.~(\ref{EQ:QPT3}) is
symmetric between $J_0 \rightarrow -J_0$ as it contains only even
powers of exchange.} This state is a true spin singlet $\langle
\Psi_0| {\bf S}_{electron}|\Psi_0\rangle = 0$. To avoid confusion
with impurity spin $S$ we explicitly indicate that ${\bf
S}_{electron}$ is the net spin of  conduction  electrons. Hence if
$|\Psi_0\rangle_{J_0<J_{crit}} =|\Psi_0\rangle$ is a ground state,
 the total  spin  of electrons is zero, and only the spin of impurity counts.
The first excited state at energy $\Omega_0$ has a spin 1/2
quasiparticle in it: $\langle \Phi_{-1}|
S^z_{electron}|\Phi_{-1}\rangle = -1/2$.

Upon increasing coupling constant $J_0$ one reaches the regime where
these two states cross,  Fig.~\ref{FIG:ELevelsQPT}. At that point
the excited and ground states changes the roles.

\begin{figure}[th]
\caption{Energies
of two variational states  are shown. $|\Psi_0\rangle $ is a
standard BCS state with energy $E_0$.  $ |\Psi_1 \rangle$ is a
variational state  that describes the formation of the bound state
between particle with the spin opposite to the local spin with
energy $E_1$. Level crossing between states with different symmetry
occurs at some critical value of the coupling $J_{crit}$. This is an
example of a first order quantum phase transition with  no divergent
length or time scale associated with it.} \label{FIG:ELevelsQPT}
\end{figure}

 \beqa
|\Psi_0\rangle_{J_0>J_{crit}} = |\Psi_{-1}\rangle = |\Phi_{-1}\rangle \nonumber\\
|\Psi_{-1}\rangle_{J_0>J_{crit}} = |\Psi_0\rangle \label{EQ:QPT4}
\eeqa

Another way to see this quantum phase transition   is to examine
energy levels as a function of $J_0/J_{crit}$. For  variational
wavefunctions $|\Psi_{0,-1}\rangle$ we define the respective
energies as expectation vales of the Hamiltonian: \beqa
E_{0,-1}(J_0/J_{crit}) = \langle \Psi_{0,-1}|H|\Psi_{0,-1}\rangle
 \label{EQ:QPT5}
\eeqa Energy of the first excitation, the impurity bound state is
then simply \beqa
\Omega_0(J_0/J_{crit}) = E_{-1} - E_0, J_0<J_{crit}\nonumber\\
\Omega_0(J_0/J_{crit}) = E_0 - E_{-1}, J_0>J_{crit} \label{EQ:QPT6}
\eeqa

There are several  implications of this result. Firstly, the ground
state of  superconductor with a magnetic impurity in the  strong
coupling limit is a non-BCS state. The ground state is a pair
condensate with a single unpaired quasiparticle.  Similar result was
observed for a Kondo screening in superconductor \cite{OSakai:1993}.
One can easily understand the result by going to a strong coupling
limit $ J_0N_0 \gg 1$. In this case long before any supercondcuting
correlation are established the single electron state bound to the
impurity site will form. This is the  strong coupling limit of a
Kondo screening problem. In our case bound electron will partially
screen  the large impurity spin. In a case of a spin $S = 1/2$ the
screening will be complete and the net spin of the state will be
singlet \cite{OSakai:1993}. Therefore state  would evolve to a
superconducting state with one unpaired electron that is bound to
impurity. Ground state in a strong coupling limit is a non-BCS state
with only pairs present. Ultimately this state is formed by the
energy balance between superconducting and magnetic energies. Single
electron state bound to a local spin provides a much stronger energy
gain $\sim J_0$ compared to the gain due to pairing $\Delta_0$.

Crossing point and related quantities are shown in the
 Fig.~\ref{FIG:SalkolaQPTfig}. This crossing point corresponds to  a true quantum phase
transition.

\begin{figure}[th]
 \caption{
 a)  The bound-state energy $\Omega_0$, ~b) the spectral weight of
 the pole
$Z^{\pm}$ for positive and negative energies   in units of $N_0 J_0,
N_0 = N_F$, ~c) the spin polarization $\langle {\bf s}(\br =
0)\rangle$, and ~d) the gap function $\Delta(\br = 0)/\Delta_0$ at
the impurity site $\br = 0$ as a function of $J_0$ in the s-wave
superconductor. Lines denote the T matrix results for the uniform
order parameter and symbols the self-consistent mean-field results
on a square lattice at half filling. The quantities of the
impurity-induced intragap quasiparticle state belonging to the
branch $J_0 < J_{crit}$ are denoted by solid lines and solid
symbols, whereas those ones belonging to the branch $J_0 > J_{crit}$
are marked by dashed lines and open symbols. Taken from
\cite{MISalkola:1997}. }
 \label{FIG:SalkolaQPTfig}
\end{figure}

Secondly, the  crossing point occurs exactly at critical point in
Eq.~(\ref{EQ:QPT3}) only in a non-self-consistent treatment
 where single particle levels provide the only contribution to total
 energy. In practice the true phase transition occurs
   slightly earlier. The gap
suppression and quasiparticle interaction  also contribute to free
energy and in  self-consistent mean-field approximation, we find
that order-parameter relaxation shifts $J_{crit}$ downwards and  the
energy of the impurity-induced bound state does not reach zero when
a first-order transition between the two ground states occurs.  In
practice the result are qualitatively similar and are typically
within 10 percent of the numerical results obtained in a
selfconsistent treatment~\cite{MISalkola:1997}. In contrast, a
d-wave superconductor has no quantum transition for any value of the
magnetic moment when its quasiparticle spectrum in the normal state
has particle-hole symmetry. The absence of the transition follows
from the behavior of the impurity-induced quasiparticle states which
are pinned at the chemical potential for an arbitrarily large
magnetic moment, see Sec.~\ref{sec:Dwave}. However, if particle-hole
symmetry is broken or if the pairing state acquires a small $s$-wave
component, the transition is again possible for a large enough
moment.  The impurity moment induces two virtual-bound states which
have four-fold symmetry and extend along the nodal directions of the
energy gap.

\subsection{Particle and hole component of impurity bound state}

In this section we show that the excited states inside the gap in
superconducting state appear in pairs at positive and negative
energies. This is a direct consequence of the fact that natural
excitations are Bogoliubov excitations.  Particle and
 hole coefficients of the excited state
$|\Psi_{-1}\rangle_{J_0<J_{crit}}$ are given by the $u$ and $v$
components of the quasiparticle operators $\gamma_{n}$, see
Sec.~\ref{sec:BCS}. To be specific we  confine subsequent discussion
to s-wave case, however the results are applicable to any
superconducting state.

Consider two independent processes: a) electron at energy $\Omega_0$
and spin down, $n=-1$ and b) hole with  spin up, $n=1$ injected in
superconductor
 with the same energy $\Omega_0$. Hole creation means electron
 with spin up is extracted from superconductor.
This can be achieved by reversing the bias of the STM tip for
example and it would correspond to the negative energy axis.
 Variational wave functions that would describe these processes are
\beqa
\psi^{\dag}_{-1}|\Psi_0\rangle_{J_0<J{crit}} = -u_1|\Phi_{-1}\rangle \nonumber\\
\psi_1|\Psi_0 \rangle = v_1|\Phi_{-1} \rangle \label{EQ:QPT7} \eeqa
Here, to be specific we consider the case of $J_0<J_{crit}$. This
illustrates the point that in BCS like ground state the particle
excitation with energy $\Omega_0$ and hole excitation with  negative
energy $-\Omega_0$,  aside from irrelevant normalization factors, is
the same excited state, namely $|\Phi_{-1}\rangle$. Therefore the
poles in the density of states are always coming in pairs at
positive and negative energies. The true
 quasiparticles
in superconducting state are Bogoliubov excitations $\gamma_n$ that
have a finite component of particle and hole with amplitudes $u_n$
and $v_n$. The strength of the electron  absorption and emission
process is controlled by these coherence factors. This is generally
true for BCS superconductor even without impurities. For the case at
hand, impurity states are well distinguished from the continuum. The
two poles at $\pm \Omega_0$ are part of the same physical
excitation. We can write impurity state
  local spectral function $A_1(\br, \omega) = - \mbox{Im} G_{11}(\br, \omega)/\pi$ as
\beqa A_1(\omega) = Z^+ \delta(\omega - \Omega_0) + Z^-
\delta(\omega + \Omega_0) \label{EQ:QPT9} \eeqa and the relative
strength is $Z^+ \sim u^2_{-1}$ and $Z^- \sim v^2_{-1}$, so that the
net strength of the poles  $Z^+ + Z^-> = 1$ as
 it should for a physical excitation. For more details and references reader
  is referred to~\cite{MISalkola:1997}.

Analysis for $J_0>J_{crit}$ is more involved. The ground state wave
function is now $|\Phi_{-1}\rangle$. Injection of the electron
 with  spin opposite  to  the spin of the bound state and
extraction of electron with spin down  will produce respectively
\begin{equation}
 \psi^{\dag}_1| \Phi_{-1}\rangle =
\psi^{\dag}_{1}\psi^{\dag}_{-1}|\widetilde{\Psi_0}\rangle\;,\;\;
\psi_{-1}|\Phi_{-1}\rangle = |\widetilde{\Psi_{0}}\rangle
\label{EQ:QPT28}
\end{equation}
with complementary annihilated states
$\psi^{\dag}_1|\Phi_{-1}\rangle = 0$ and
$\psi^{\dag}_{-1}|\Phi_{-1}\rangle = 0$. Although the two states
written in Eq.~(\ref{EQ:QPT28}) are different, the only difference
is that the number of Cooper pairs in these two states differ by
one. For a macroscopically large system with number of Cooper pairs
$N \gg 1$ this produces negligible difference
 in the energies and matrix elements. Therefore, again, the injection of
 electron with spin up (in our convention) and extraction of electron of spin
down physically will produce the same state. This state will have a
particle and hole projection just as we discussed in case of
$J_0<J_{crit}$.

Similar quantum phase transition occurs in a d-wave superconductor
even for a nonmagnetic impurity. In the case of particle-hole
symmetric band the unitary scattering produces a zero energy state,
see Sec.~\ref{sec:Dwave}, Eq.~(\ref{eq:impdwave1}). However for the
particle hole asymmetric band impurity state will reach zero energy
and eventually will change the sign as a function of impurity
strength. This transition occurs at $U_0 >U_{crit} \sim \mu$, $\mu$
being the chemical potential that   leads to a particle-hole
nonsymmetric band. It is known that single quasiparticle bound state
will form at $U_0
>U_{crit}$ and the ground state wavefunction will have unpaired
single quasiparticle, apart from the BCS pairs,
see~\cite{MISalkola:1996,MISalkola:1997}.

\subsection{ Intrinsic $\pi$ phase shift for $J_0 > J_{crit}$ coupling}

Here we would like to  point a little noticed by important fact
about the impurity induced modifications of the order parameter at
the impurity site, namely  a $\pi$ phase shift  of the order
parameter near impurity.

>From the above Fig.~\ref{FIG:SalkolaQPTfig}, it follows   that at
$J_0>J_{crit}$ the self-consistent solution indicates that  the
phase of the order parameter on the impurity site is shifted by
$\pi$ with respect to the phase in the bulk, see
Fig.~\ref{FIG:IntrPiJunc}.

\begin{figure}[th]
\caption{Cartoon
of the intrinsic $\pi$ junction near magnetic impurity in $s$-wave
superconductor.} \label{FIG:IntrPiJunc}
\end{figure}

Spatial extent of the $\pi$ shifted region in numerical calculation
was  found to be few atomic sites. Such
  a sharp change in the phase of the order parameter would cost a
superconducting condensate energy and would not be preferential
under normal circumstances. In the case at hand however, condensate
energy is secondary to the magnetic exchange at strong coupling
limit near impurity site.  Physics is driven by magnetic
interactions in a strong coupling limit. Even thought eh phase
difference is $\pi$ the phase shift does not lead to any time
 reversal violating observable effects.
One can see that for the phase difference $\pi$ there are no
superconducting currents near the impurity: $I = I_c \sin \phi = 0$.
These  results  were obtained  in the self-consistent treatment
within a negative U model that allows for the on-site
pairing\cite{MISalkola:1997}.

We are not aware of a simple explanation of this effect. It appears
to be general and not related to a particular model. It is connected
to the $\pi$ junctions discussed in the context of tunneling
barrier. The $\pi$ phase shift is preferential in the junction in
the presence of magnetic impurity or ferromagnetic layer. This
subject
 is covered extensively, see e.g.  recent review and other papers
\cite{LNBulaevskii:1977,AIBuzdin:2004,AIBuzdin:1982,BISpivak:1991,LIGlazman:1989}.

\section{Kondo impurity}
\label{sec:Kondo}

In all of previous discussions, we have concentrated on static
impurities. The next two Sections are devoted to the examples when
impurity atoms have their own internal degrees of freedom, the
dynamics of which is coupled to the scattering of conduction
electrons. This dynamical behavior often leads to qualitatively new
results. Quantum dynamics is particularly important for study of
magnetic impurities as the earliest and the best known example of
non-trivial consequences of this dynamics is the Kondo
effect~\cite{JKondo:1964}.

Dilute magnetic impurities doped into an otherwise nonmagnetic
metallic host have dramatic effects on the low temperature
resistivity, susceptibility, and specific heat. These anomalies are
associated with screening of the impurity spin by conduction
electrons. For $S=\frac{1}{2}$ a global singlet is formed by
coupling an electron state to the impurity site; dynamics of spin
flips is crucial for the formation of the singlet. For a single
magnetic impurity in the metallic host, this is manifested as a
crossover from the Curie susceptibility at high temperatures,
$\chi=C/T$ with $C=4\mu_{B}^{2}S(S+1)/3k_{B}$, where $S$ the
magnitude of the spin and $\mu$ the Bohr magneton, to singlet
behavior below a characteristic Kondo temperature $T_K\simeq
W\exp(-1/2J N_0)$, where $W$ is the electron half band width and $J$
is the exchange constant. Two important notes are  that a) Kondo
screening occurs only for antiferromagnetic exchange constant $J>0$;
and b) the process is non-perturbative, as is clear from the
non-analytic dependence of the Kondo temperature on the exchange
constant. Understanding of the single impurity Kondo proble in a
metal required concerted use of the renormalization
group~\cite{PWAnderson:1970a,PWAnderson:1970b}, numerical
renormalization group (NRG)~\cite{KGWilson:1975}, exact solutions
via the Bethe ansatz~\cite{NAndrei:1980,PBWiegmann:1980}, and
large-$N$
expansions~\cite{NRead:1983a,NRead:1983b,NRead:1985,PColeman:1984,PColeman:1985}.
Many important results are summarized in recent reviews
\cite{ACHewson:1993,DLCox:1998}.

Kondo effect depends on the existence of the host electronic
excitations at the Fermi energy. In metals, the density of states is
constant, which simplifies the analysis. If the density of states
varies in the immediate vicinity of the Fermi surface, the effect is
realized rather differently (or not at all). Kondo effect in an
insulator, with a hard band gap, was investigated in
\onlinecite{JOgura:1993}. In superconductors, however, the gap in
the single particle spectrum only arises below the transition
temperature for the Cooper instability driven by the finite DOS.
Consequently, the two effects compete.

Studies of magnetic impurities in superconductors began, and largely
continued, with the investigations of the properties of classical
spin, $S\rightarrow\infty$, for which no reduction in magnitude due
to Kondo screening is
possible~\cite{LYu:1965,HShiba:1968,AIRusinov:1968}. The question of
which conclusion of their analysis are robust for small spin values
and in gapless superconductors remains of intense current interest.

\subsection{Kondo effect in fully gapped superconductors}

In Shiba-Rusinov analysis the sign of the exchange interaction
between the conduction electrons and localized impurity spins is
irrelevant. As discussed above in real metals antiferromagnetic
exchange leads to a complete screening of the impurity below Kondo
temperature $T_K$, while ferromagnetic exchange does not produce
resonance states. Consequently, treatment of quantum impurity spins
has to bring out the differences between two signs of $J$.

For $J>0$ opening of the superconducting gap competes with Kondo
screening as both instabilities are driven by a finite DOS at zero
energy. Qualitatively, it is clear that if $T_K\gg T_c$, the
impurity is completely screened by the time of the onset of
superconductivity. In contrast, for $T_K\ll T_c$ Kondo screening is
suppressed by a decreased density of states upon opening of the
superconducting gap.

Kondo screening can be viewed  as growth and divergence of the {\em
effective} exchange coupling as we look at the properties of the
system at lower and lower energies. This is the essence of the the
NRG method, and the underlying principle of successful analytical
work. Therefore the effective exchange constant, and, with it, the
phase shift of scattering on the impurity depends on the energy of
the incoming electron. Consequently, the effect of scattering varies
with temperature.

We immediately conjecture that, while a localized Shiba-like state
exists in the gap of a superconductor, its energy in the Kondo limit
changes upon lowering the temperature. Therefore the ``final'',
$T=0$, position of the impurity resonance is a complex function of
$T_c$ and $T_K$. Moreover, as results of the previous section imply,
the nature of the ground state, i.e. its spin and degeneracy, depend
on the relation between $T_c$ and $T_K$. As a result, there has been
increased interest in determining the properties of the ground
state, and the localized excited (Shiba-Rusinov) state of a BCS
superconductor with a quantum impurity spin.

\subsubsection{Ferromagnetic exchange}

Early analytical attempts were carried out
\cite{JZittartz:1970,EMullerHartmann:1973} in the framework of
Nagaoka decoupling scheme
\cite{YNagaoka:1965,YNagaoka:1967,DRHamann:1967}. For $J<0$ the
bound state splits off the band edge and was found to move towards
an asymptotic value
\begin{equation}
  \epsilon\equiv\frac{E_0}{\Delta}=\biggl[1+g^2\pi^2
  S(S+1)\biggr]^{-1/2},
\end{equation}
where $g=\lambda N_0$, and $\lambda$ is the superconducting coupling
in the BCS weak coupling theory. Since $g\ll 1$ the bound state
remains close to the gap edge for all values of $J<0$. This
qualitative result was later confirmed by NRG
calculations~\cite{KSatori:1992,OSakai:1993}, which showed that the
binding energy is well approximated by $\epsilon\approx1-\pi^2
J^2_{eff}/8$, where the renormalized exchange constant
\begin{equation}
J_{eff}=\frac{2|J|/W}{1+(2|J| W)\ln(W/\Delta)}.
\end{equation}
Therefore the ferromagnetic case corresponds to weak coupling and
small phase shift of scattering at low temperatures.

The ground state of this system was argued to be a
doublet~\cite{TSoda:1967,KSatori:1992,OSakai:1993}, since the
ferromagnetic interaction renormalizes to weak coupling and the
impurity spin remains essentially free. Recently it was suggested
that, since superconducting interaction is relevant in this system,
and therefore above a critical coupling, $J_C$ that depends on
$\Delta$ ($J_C$ is larger for smaller $\Delta$), the ground state of
the coupled superconductor-impurity system is a triplet ($m_z=0,\pm
1$) \cite{TYoshioka:1998}. This suggestion is worth exploring
further.

\subsubsection{Antiferromagnetic coupling}

If the coupling is antiferromagnetic, in a normal metal Kondo
screening corresponds  to $J\rightarrow\infty$ and to phase shift of
scattering, $\delta\rightarrow \pi/2$. The Hartree-Fock analysis
\cite{HShiba:1973} is insufficient to fully describe this effect.

The limit $T_K\ll \Delta$ was considered following the work of Shiba
\cite{TSoda:1967,JZittartz:1970,EMullerHartmann:1973}, and the
position of the localized excited state was found with various
degrees of accuracy. Notice that in this regime the localized state
lies close to the gap edge as it does for ferromagnetic coupling. In
the opposite limit, $T_K\gg\Delta$ approximate solution for the
position and the residue of the bound state was obtained in
Refs.~\cite{JZittartz:1970,EMullerHartmann:1973}, however, the
results were inexact due to the nature of their approximation.
Later, within the local Fermi liquid appraoch, the energy of the
bound state in this limit was found to be \cite{TMatsuura:1977}
\begin{equation}
  \epsilon=\frac{1-\alpha^2}{1+\alpha^2},
\end{equation}
where
\begin{equation}
  \alpha\approx \frac{\pi\Delta}{4T_K}\ln\frac{4eT_K}{\pi\Delta}.
\end{equation}
This result clearly shows that the phase shift of scattering depends
on the ratio $T_c/T_K$.

The properties of the bound state, including its position and
spectral weight, for arbitrary values of $T_K/T_c$ were obtained
with the help of NRG~\cite{KSatori:1992,OSakai:1993}. They found
level crossing similar to the quantum phase transition of the
previous section at $T_K/\Delta\sim 0.3$. For $T_K/\Delta>0.3$ the
impurity moment is largely quenched by the time the depletion of
states caused by superconductivity affects screening. In that case
the ground state is a Kondo-screened singlet, while the excited
intra-gap state is a doublet with the spectral weight $\alpha\approx
2$ for $T_K\Delta\gg 1$, corresponding to a single-particle state.
Here $\alpha$ is defined from
    \begin{equation}
    -\frac{1}{\pi}{\rm Im} G (\omega+i\delta)/\pi=\frac{\alpha}{2}
    \biggl[\delta(\omega-E_0)+\delta(\omega+E_0)\biggr].
    \end{equation}
On the other hand, for $T_K\Delta<0.3$ the Kondo effect is
suppressed by by the opening of the superconducting gap, the ground
state is a doublet corresponding to a free spin state, while the
bound excited state is a Kondo singlet. The spectral weight,
$\alpha\approx 0.5$ for $T_K\ll\Delta$, and changes discontinuously
at the phase transition point.

Level crossing means that the bound state is at zero energy for
$T_K/\Delta\approx 0.3$, while it is close to the gap edge for both
$T_K\gg\Delta$ and $T_K\ll\Delta$. Numerical results show that the
energy of the bound state is not symmetric with respect to the
crossing point: $E_0/\Delta<0.5$ for $0.03\lesssim
T_K/\Delta\lesssim 1$ \cite{KSatori:1992}.

\subsubsection{Anisotropic exchange and orbital effects}

Several more complicated aspects of Kondo screening in
superconductors attracted attention in recent years, and we review
them briefly, referring the reader to the original papers for
further information. Anisotropic exchange interaction, $J_z\neq
J_\pm$, allows the investigation of the crossover between the Ising
regime, $J_\pm=0$, when the spin-flip is disallowed and there is no
Kondo screening, and the isotropic exchange considered so far. The
main features of the phase diagram are discussed by
\onlinecite{TYoshioka:1998}, and new phases occur on the
ferromagnetic side. In particular, these authors find an extended
regime of Ising-dominated ground state even for $J_\pm\neq 0$. In
addition, they find small regions of the phase diagram around
isotropic ferromagnetic and Ising antiferromagnetic lines, where
there exist two localized intra-gap states. They also obtain a
perturbative analytic expression for the shift of the bound state
energy due to anisotropy of the interaction.

Using the numerical RG approach to analyse Anderson's model allows
to interpolate between asymmetric magnetic scattering,  Kondo
problem, and non-magnetic scattering, including resonance $U=0$
limit~\cite{TYoshioka:2000}. In particular, the crossover from
magnetically induced bound state to the resonance non-magnetic
scattering regime\cite{KMachida:1972} was studied.

Finally, so far we only discussed purely $s$-wave superconductors.
Fully gapped systems include also materials with a complex order
parameter combining two (or more) out of phase unconventional gaps,
such as $d_{x^2-y^2}+i d_{xy}$, or $p_x+i p_y$. In both of these
cases Cooper pairs have orbital degrees of freedom that also couple
to the impurity spins, leading to multichannel Kondo effect. In
addition, for $p$-wave pairing, the total spin of Cooper pairs is
$s=1$, so that non-trivial changes in screening occur depending on
whether the impurity spin $S=1/2$ or $S=1$. The NRG analysis of the
Kondo problem in this system was carried out very recently
\cite{MMatsumoto:2002,MKoga:2002a}. They found that the two order
parameters are indistinguishable when only $l=0$ impurity scattering
partial wave is taken into account, i.e. only the depletion of the
density of states due to the gap, rather the spin structure of the
Cooper pair dictated the Kondo screening. In that case the moment of
the ground state is determined by the orbital structure of the
Cooper pair. However, inclusion of higher harmonics with $l\neq 0$
for scattering (extended impurity potential), leads to some novel
dependencies of the screening and ground states on the exchange
couplings.

\subsection{Kondo effect in gapless superconductors}

Gapless systems like $d$-wave superconductors with the low-energy
quasiparticle density of states following the power law,
$N(E)\propto \vert E \vert^{r}$ with the exponent $r>0$, constitute
a marginal situation. The Kondo effect in these systems where the
host single particle density of states follows a power law,  has
been studied
intensively~\cite{DWithoff:1990,LSBorkowski:1992,YItoh:1993,LSBorkowski:1994,
KChen:1995,RBulla:1997,RBulla:2000,CRCassanello:1996,CRCassanello:1997,
KIngersent:1996,CGonzalez:1998,KIngersent:1998,DELogan:2000,
APolkovnikov:2001,JXZhu:2001a,JXZhu:2001b,GMZhang:2001,
APolkovnikov:2002,MVojta:2001a,MVojta:2001b,QHan:2002,QHan:2004}.~\footnote{There
have also been discussion about  local moment formation in
high-temperature
cuprates~\cite{NNagaosa:1997,GKhaluillin:1997,RKilian:1999,MESimon:1999}.
Discussion on Kondo problem in $s$-wave superconductors, in
unconventional superconductors with time-reversal-symmetry-broken
pairing state and in insulators can be found in the
literatures~\cite{KSatori:1992,OSakai:1993,JOgura:1993,TYoshioka:1998,TYoshioka:2000,
MMatsumoto:2001,MKoga:2002}.} Fradkin and
co-workers~\cite{DWithoff:1990,CRCassanello:1996,CRCassanello:1997}
first employed a combination of the poor man's scaling argument and
the large-$N$ approach to the case of spin-$\frac{1}{2}$ (impurity
degeneracy $N=2$) for $0<r\le 1$, an showed that a Kondo effect
takes place only when the electron-impurity exchange $J$ exceeds a
critical value. Otherwise the impurity decouples from the band.
However, the study based on a nonperturbative renormalization group
approach~\cite{KChen:1995,KIngersent:1996} to a spin-$\frac{1}{2}$
identified particle-asymmetry as a key factor in determining the
low-temperature physics. At small asymmetry, the critical coupling
$J_c$ above which the impurity moment is screened becomes so large
for all $r>\frac{1}{2}$ that the Kondo effect is suppressed. Away
from the particle-hole symmetry, any quenching of the impurity
moment is accompanied by a low-temperature decrease in the impurity
resistivity, rather than the increase found in metals. The
discrepancy between the two categories of work may stem from the
mean-field nature of the large-$N$ method, or from the symmetry
breaking that is implicit, for all $N>2$, in the restriction that
the impurity level be singly occupied. In real systems, the
power-law variation of $N(\epsilon)$ is restricted to an energy
range $\vert \epsilon \vert \leq \Delta_0$, with $N(\epsilon)\approx
N(\Delta)$ for $\Delta_0 < \vert \epsilon \vert \leq W$.  The NRG
approach gave results entirely consistent with those known for
gapped systems (The gap $2\Delta_0$ is for $r=\infty$ limit). At the
particle-hole symmetric case, an impurity in an insulator retains
its moment, no matter how large $J$ is; away from this symmetry, the
spin is screened provided that $J>J_c \approx
2W/\ln(W/\Delta_0)$~\cite{KTakegahara:1992}.

Notice that considering the Kondo effect in a system with the power
law dependence of the DOS is not the same as analysing the
competition between superconducting and Kondo correlations. Within
the Kondo exchange model, the Hamiltonian of the single magnetic
impurity in a medium other than a superconductor, can be written as:
\begin{eqnarray}
H&=&\sum_{\sigma} \int_{-\infty}^{\infty} d\epsilon N(\epsilon)
\epsilon c^{\dagger}_{\epsilon\sigma} c_{\epsilon\sigma} +
\frac{1}{N_{L}}\sum_{k,k^{\prime}} [(U_0+
\frac{J}{2})c_{k\uparrow}^{\dagger} c_{k^{\prime}\uparrow}\nonumber \\
&& + (U_0 -\frac{J}{2})c_{k\downarrow}^{\dagger}
c_{k^{\prime}\downarrow}] +
\frac{J}{2}\sum_{k,k^{\prime}}[c_{k\uparrow}^{\dagger}
c_{k^{\prime}\downarrow}S_{-} + c_{k\downarrow}^{\dagger}
c_{k^{\prime}\uparrow}S_{+}]\;,\nonumber \\ \label{EQ:KondoPG}
\end{eqnarray}
where $N(\epsilon)$ is the electron density of states, $N_{L}$ is
the lattice size, and all operators are for electrons. In contrast,
for a superconductor, we need to rewrite the interaction via the
Bogoliubov quasiparticles, and enforce the self-consistency
condition on the gap. The resulting hamiltonian is rather lengthy,
and follows straightforwardly from symmetric Anderson or Kondo
hamiltonian, so that we do not give it here. We note that the
formation and screening of the local moments in $d$-wave
superconductors has been investigated using the variational wave
function approach \cite{MESimon:1999}.

The interest in the Kondo impurities in the high-temperature
cuprates is motivated by the recent STM and NMR experiments around
single impurities. Unlike most dopants, Zn, Ni are believed to
substitute for Cu on the cooper-oxide plane and causes effective
changes to the local electronic structure without much change of
hole concentration. Simple valence counting suggests that if the Zn
and Ni impurities maintain a nominal Cu$^{2+}$ charge, the Zn$^{2+}$
would have a $(3d)^{10}$, $S=0$ configuration and acts as a
nonmagnetic impurity while the Ni$^{2+}$ would have a $(3d)^{8}$,
$S=1$ configuration and acts a magnetic impurity. Although it is
natural that the spin-1 impurities carry an on-site magnetic
moment~\cite{PMendels:1999} expected to give rise to the Kondo
physics, the behavior associated with the nonmagnetic impurities is
completely unexpected. Nuclear magnetic resonance (NMR) experiments
performed with nonmagnetic spin-0 (Zn,Li,Al) in doped
cuprates~\cite{HAlloul:1991,KIshida:1993,
AVMahajan:1994,KIshida:1996, PMendels:1999,AVMahajan:2000} showed
clearly that each impurity, itself carrying no magnetic moment,
induces a local $S=\frac{1}{2}$ moment sitting on the
nearest-neighbor Cu orbitals. It was also demonstrated that the
magnetic properties associated with the substitution of these
impurities strongly depends on the hole doping: In the underdoped
regime, the moments retain their Curie law below the superconducting
transition temperature $T_c$ while near the optimal doping, the
Kondo screening might persist even to $T\rightarrow 0$ though
strongly reduced.
>From the NMR, it is known that the induced moment is distributed
around the impurity. We would like to emphasize that this moment is
merely a particular bound state of conduction electrons near the
impurity and the precise form of the interaction between the induced
moment with other conduction electrons is {\em a priori} unknown. At
this stage, we are unable to make a definitive conclusion about its
importance for our understanding of high-temperature cuprates.
However, in a broader sense, the magnetic impurity embedded in
superconductors is a very-well defined theoretical issue. Generally,
the system Hamiltonian with a magnetic impurity consists of a
$d$-wave BCS state $\mathcal{H}_{BCS}$, a potential scattering term
$\mathcal{H}_{pot}$, and a magnetic term $\mathcal{H}_{mag}$. The
magnetic term can described by the Anderson impurity model or Kondo
spin exchnange model, and the impurity spin can be either coupled to
a single site or be spatially distributed. For the Anderson model
with the single-site coupling, the magnetic term is given by:
\begin{equation}
\mathcal{H}_{mag}=\sum_{k\sigma} [V_{kd}
c_{k\sigma}^{\dagger}d_{\sigma} + H.c.] + \epsilon_{d} \sum
d_{\sigma}^{\dagger} d_{\sigma} + U_d n_{d\uparrow}
n_{d\downarrow}\;. \label{EQ:HamilAndersonS}
\end{equation}
In the strong $U_d$ limit, the Anderson model can be mapped onto a
Kondo s-d exchange model through the Schrieffer-Wolff
transformation:
\begin{equation}
\mathcal{H}_{mag}=J\mathbf{s}_0\cdot  \mathbf{S}\;,
\end{equation}
where $\mathbf{s}_0=\frac{1}{2} \sum_{\sigma\sigma^{\prime}}
c_{0\sigma}^{\dagger} \bm{\sigma}_{\sigma\sigma^{\prime}}
c_{0\sigma^{\prime}}$ is the spin operator for the conduction
electron at the impurity site. The corresponding models for the
multi-site coupling case are given by:
\begin{equation}
\mathcal{H}_{mag}=\sum_{I\sigma}
[V_{Id}c_{I\sigma}^{\dagger}d_{\sigma} + H.c.] +
\epsilon_{d}\sum_{\sigma} d_{\sigma}^{\dagger}d_{\sigma} + U_d
n_{d\uparrow} n_{d\downarrow} \;, \label{EQ:HamilAndersonM}
\end{equation}
and
\begin{equation}
\mathcal{H}_{mag}= \sum_{I} J_{I} \mathbf{s}_{I}\cdot \mathbf{S}\;,
\label{EQ:HamilKondoM}
\end{equation}
respectively. We point out that in the large $U_d$ limit, the
Schrieffer-Wolff transformation will map the model as described by
Eq.~(\ref{EQ:HamilAndersonM}) onto to a Hamiltonian nonequivalent to
the Kondo model described by Eq.~(\ref{EQ:HamilKondoM}). Therefore,
Eq.~(\ref{EQ:HamilKondoM}) comes from different origin.

The Anderson impurity model for a single-site coupling in $d$-wave
superconductors, Eq.~(\ref{EQ:HamilAndersonS}), was studied by
Zhang, Hu, and Yu~\cite{GMZhang:2001}. A sharp localized resonance
above the Fermi energy, showing a marginal Fermi liquid behavior,
was predicted for the impurity states. The same logarithmic
dependence of self-energy and a linear frequency dependence of the
relaxation rate were also obtained, indicating a new universality
class for the strong coupling fixed point. Almost at the same period
of time, the multi-site coupling Anderson impurity model,
Eq.~(\ref{EQ:HamilAndersonM}) was considered by Zhu and
Ting~\cite{JXZhu:2001b,JXZhu:2001c} while the multi-site coupling
Kondo impurity was studied by Polkovnikov, Sachdev, and
Vojta~\cite{APolkovnikov:2001,APolkovnikov:2002}. All these works
show the existence of Kondo resonance. However, the low energy
structure of spectral weight of the conduction electrons depends
delicately on the local environment surrounding the dynamic
impurity. The on-site potential scattering was  taken to be either
zero~\cite{GMZhang:2001} or very
weak~\cite{APolkovnikov:2001,APolkovnikov:2002} so that the
resonance peak is located very close to the Fermi energy. Zhu and
Ting~\cite{JXZhu:2001b} took into account the quasiparticle
scattering from a geometrical hole, where electrons are allowed to
hop onto the four neighbors of the impurity site, and obtained a
double-peak structure around the Fermi energy. They
further~\cite{JXZhu:2001c} considered  the potential scattering term
to be in the unitary limit ($U\rightarrow \infty$), and found that
the Kondo resonance effect is weaved into that from the strong
potential scattering to determine the low energy quasiparticle
states. The delicate influence of the potential scattering on the
Kondo physics as well as the local electronic structure in $d$-wave
superconductors has been re-emphasized by Vojta and
Bulla~\cite{MVojta:2001b}.

To be concrete, we present a discussion based on the multi-site
coupling Kondo impurity model, as given by
Eq.~(\ref{EQ:HamilKondoM}). As demonstrated in previous sections,
the problem of a single-site potential scattering can be exactly
solved. In the Nambu space, the Green's function is given by
\begin{equation}
\mathcal{G}(i,j;i\omega_{n}) = \mathcal{G}^{0}(i,j;i\omega_{n}) +
\mathcal{G}^{0}(i,0;i\omega_{n})T(i\omega_n)\mathcal{G}^{0}(0,j;i\omega_{n})
\end{equation}
where the $T$-matrix due to the potential scatterer is
\begin{equation}
T^{-1}(i\omega_{n})=\tau_{3}/U-\mathcal{G}^{0}(0,0;i\omega_{n})\;,
\end{equation}
and $\mathcal{G}^{0}$ is the Green's function for the system in the
absence of impurities and has been given in early discussions. In
the presence of both a potential scattering and a Kondo impurity,
the system Green's function is found to be:
\begin{eqnarray}
\tilde{\mathcal{G}}(i,j;i\omega_{n})&=&\mathcal{G}(i,j;i\omega_{n})
+ \sum_{l,l^{\prime}} \varphi_{l}\varphi_{l^{\prime}}
\mathcal{G}(i,l;i\omega_{n}) \mathcal{T}_{K}(i\omega_{n})\nonumber
\\
&&\times \mathcal{G}(l^{\prime},j;i\omega_{n})\;.
\end{eqnarray}
Here $l$ and $l^{\prime}$ label the sites neighboring to the
impurity site at (0,0), and $\mathcal{T}_{K}$ is the $T$-matrix for
the Kondo impurity. The variables $\varphi_{l}$ have different
meaning depending on the approach to the $\mathcal{T}_{K}$. In the
large-$N$ approximation (equivalent to the slave-boson mean-field
approximation), where
\begin{equation}
\mathcal{T}^{-1}_{K}=i\omega_{n}-\lambda \tau_{3} -
\sum_{l,l^{\prime}}\varphi_{l}\varphi_{l^{\prime}} \tau_{3}
\mathcal{G}(l,l^{\prime};i\omega_{n})\tau_{3} \;,
\end{equation}
where $\varphi_{l}$ are the complex Hubbard-Straonovich fields, and
are determined, together with the Lagrange multiplier $\lambda$, by
the saddle point solution. However, in the numerical renormalization
group technique, when the strongest $d$-wave-like channel is
considered, the variables are taken to be $\varphi_{l}=+(-)1$
depending on the bond orientation. Note that this $d$-wave-like
pattern is merely a band structure effect and has nothing to do with
the $d$-wave symmetry of the superconducting order parameter of the
host. The LDOS in the presence of both types of impurities is
obtained as:
\begin{equation}
\rho_{i}(\omega) = -\frac{1}{\pi} \mbox{Im}\{ \mbox{Tr}\biggl{[}
\tilde{\mathcal{G}}(i,i;\omega+i 0^{+})\frac{1+\tau_{3}}{2}
\biggr{]}\}\;.
\end{equation}

\begin{figure}[t]
\caption{
Calculated tunneling density of states for the four-site Kondo
impurity model at 15\% hole doping with a realistic band structure
($t=0.15$ eV, $t^{\prime} = -t/4$, $t^{\prime\prime} = t/12$),
$\Delta_0 = 0.04$ eV, and $\mu = -0.14$ eV. The Kondo coupling is
$J=0.09$ eV, the potential scattering $U=0$. Top: Local DOS vs.
energy for the impurity site (red) and the nearest (blue) and second
(green) neighbor sites. Bottom: Spatial dependence of the local DOS
at $\omega = -2$ meV. Left: Local DOS in the CuO$_2$ plane. Right:
Local DOS after applying the filter effect proposed by Martin,
Balatsky, and Zaanen~\cite{IMartin:2002}. From Vojta and
Bulla~\cite{MVojta:2001b} } \label{FIG:VojtaStm1}
\end{figure}

\begin{figure}[t]
\caption{
Same as Fig.~\protect\ref{FIG:VojtaStm1}, but with potential
scattering $U=t=0.15$ eV. Here, $J = 0.065$ eV. The lower panel
shows the local DOS at $\omega = +2$ meV. From Vojta and
Bulla~\cite{MVojta:2001b}. } \label{FIG:VojtaStm2}
\end{figure}

\begin{figure}[t]
\caption{
Same as Fig.~\protect\ref{FIG:VojtaStm1}, but with potential
scattering $U=4t=0.6$ eV. Here, $J = 0.04$ eV. The lower panel shows
the local DOS at $\omega = +3$ meV. From Vojta and
Bulla~\cite{MVojta:2001b}. } \label{FIG:VojtaStm3}
\end{figure}

Figures~\ref{FIG:VojtaStm1} through~\ref{FIG:VojtaStm3} show the
LDOS for a four-site Kondo impurity different various strength of
the potential scattering, calculated using the NRG
technique~\cite{MVojta:2001b}, which removes some artifacts of the
slave-boson method. It is shown clearly that the Kondo effect is
very sensitive to the strength of the potential scattering. In the
absence of the potential scattering, sharp resonance peak show up
directly on the impurity site, and on its next-nearest neighbors
with reduced intensity, which is consistent with the experimental
observation~\cite{SHPan:2000b}. For a moderate value of the
potential scattering, as shown in Fig.~\ref{FIG:VojtaStm2}, the
global particle-hole asymmetry changes its sign and the Kondo peak
appears at the opposite side of the Fermi level compared to
Fig.~\ref{FIG:VojtaStm1}. For a strong potential scattering, the
resonance peak directly from the impurity scattering becomes
dominant, and the Kondo effect is weaved into the overall structure
of the LDOS. In this case, the intensity of the peak is strongly
suppressed by the on-site potential scattering and a double-peak
structure with enhanced intensity is seen in the LDOS at the sites
nearest-neighboring to the impurity site. This result was also
obtained by Zhu and Ting~\cite{JXZhu:2001b} based on the Anderson
impurity model. It is expected that, in the unitary limit of the
potential scattering, the LDOS has a zero intensity at the impurity
site and a sharp single peak at its nearest neighbors. Consequently,
the spatial shape of the resulting pattern is more compatible with
the experiment after the filter
effect~\cite{IMartin:2002,JXZhu:2000b}, as seen in
Sec.~\ref{sec:STM}, is taken into account. It is also shown in this
simple model that the large LDOS from the resonance state induced by
the strong potential scatterer reduces dramatically the critical
Kondo coupling, indicating the determination of the Kondo effect by
the local rather than global environment in which the magnetic
impurity is embedded.

\section{Dynamical impurities}
\label{sec:DynamicalImp}

\subsection{Inelastic scattering from a single spin in $d$-wave
superconductors}

We will address the inelastic tunneling features due to the
scattering off a local spin impurity. Assume that we have localized
magnetic atom with spin $S$ on a surface of a d-wave superconductor.
We will treat the problem following
\cite{DKMorr:2003,AVBalatsky:2003}. Electrons in a superconductor
interact with the localized spin via point-like exchange interaction
at one site $J {\bf S} \cdot \bsig$:
\begin{eqnarray}\label{eq:Ham}
H = \sum_{\bk} \xi({\bk}) c^{\dag}_{\bk \sigma} c_{\bk \sigma} +
\sum_{\bk} [\Delta(\bk) c^{\dag}_{\bk \uparrow}c^{\dag}_{-\bk
\downarrow} + h.c.] \nonumber
\\+ \sum_{\bk,\bk', \sigma, \sigma'} J {\bf S}\cdot c^{\dag}_{\bk \sigma}
\bsig_{\sigma \sigma'} c_{\bk' \sigma'} + g \mu_B {\bf S}\cdot {\bf
B}\;,
\end{eqnarray}
where $c_{\bk \sigma}$ is annihilation operator for the conduction
electron of spin $\sigma$, $\xi(\bk)$ is the energy of the
electrons, $\Delta(\bk) = \frac{\Delta}{2} (\cos k_x  - \cos k_y)$
is the d-wave superconducting gap of magnitude $\Delta \simeq 30
meV$ in typical high-T$_c$ materials.  The local spin ${\bf S}$ is a
$|S|=1/2$. We focus here on the effect of the Zeeman splitting of
the otherwise degenerate local spin state in the external magnetic
field $B$ with splitting energy $\omega_0 \equiv \omega_L = g \mu_B
B$. Below we use a mean field description of superconducting state
at low temperatures $T\ll T_c$. Assuming field $B \ll H_{c2}$ we
will ignore the orbital and Zeeman effect of the field on the
conduction electrons~\footnote{To minimize the orbital effect of
magnetic field one can apply it parallel to the surface of
superconductor. The magnetic field is penetrating the surface sheath
on the scale of penetration depth so that its effect on the
conduction electrons for d-wave SC is small.}.

We are  interested in a local effect of  inelastic scattering of
electrons. Thus  only local properties will determine the conduction
electron self-energy. Results we obtain will also hold for a normal
state with linearly vanishing DOS, such as a pseudogap state of
high-Tc superconductors.  In the  case of a normal state one would
model normal pseudogap state  with a single particle Hamiltoninan $
H_0 = \sum_{\bk}  \epsilon({\bk}) c^{\dag}_{\bk \sigma}c_{\bk
\sigma}$ with $N(\omega) \sim \omega$.

Because of the  vanishing DOS in a d-wave superconducting state
Kondo singlet formation occurs only for a coupling constant
exceeding some critical
value~\cite{DWithoff:1990,CRCassanello:1996,CGonzalez:1998}. For a
particle-hole symmetric spectrum Kondo singlet is not formed for
arbitrarily large values of $J$. Another situation where Kondo
effect is irrelevant is the case of ferromagnetic coupling $J$. This
allows us, quite generally, to consider a  single spin in a d-wave
superconductor that is not screened and we ignore the Kondo effect.

In the presence of magnetic field $\mathbf{B}||\hat{\mathbf{z}}$
spin degeneracy is lifted and  components of the spin
$\mathbf{S}||\hat{\mathbf{z}}$ and $\mathbf{S}\perp \mathbf{B}$ will
have different propagators. It is obvious that only transverse
components of the spin will contain information about level
splitting  at $\omega_0 = \omega_L$. We have therefore focused on
$S^+, S^-$ components only. The propagator in imaginary time $\tau$
is $\chi(\tau) = \langle T_{\tau} S^+(\tau)S^-(0)\rangle$ with
Fourier transform and continuing to real frequency $\chi_{0}(\omega
) = \frac{\langle S^{z}\rangle}{\omega_0^{2}-(\omega
+i\delta)^{2}}$. For free spin we have $\langle S_z\rangle = \tanh
(\omega_0/2T)/2$. For more general case of  magnetic anisotropy this
does not have to be the case. To be general we will keep $\langle
S_z\rangle$.

We begin with evaluation of the DOS correction due to coupling to
 localized spin.
 Self-energy correction is:
\begin{equation}\label{EQ:Selfenergy1}
  \Sigma(\omega_l) = J^2 T \sum_{\bk,\Omega_n}G(\bk, \omega_l -
  \Omega_n) \chi^{+-}(\Omega_n)\;,
\end{equation}
where $G^0(\bk, \omega_l)  = [i\omega_l - \xi(\bk)][ (i\omega_l)^2 -
\xi^2(\bk) - \Delta^2(\bk)]^{-1} $ is the particle Green's function
in  d-wave superconductor, $G^{-1} = G^{(0)-1} - \Sigma$,
$F^0(\bk,\omega_l) = [\Delta(\bk)][ (i\omega_l)^2 - \xi^2(\bk) -
\Delta^2(\bk)]^{-1}$; $\Omega_l = 2 \pi l T$ is the bosonic
Matsubara frequency and $\omega_l = (2l+1) \pi T; l=0,1,2...$ is the
fermionic frequency. Using spectral representation and analytical
continuation onto real axis $i \omega_n \rightarrow \omega + i\delta
$ we find for imaginary part of self energy $\Sigma(\omega)$ :
\begin{equation}\label{EQ:Selfenergy2}
  \mbox{Im} \Sigma(\omega) = -J^2  \langle S_z\rangle   \mbox{Im}
  G(\omega-\omega_0)[n_F(\omega - \omega_0)- n_B(\omega_0) - 1]\;,
\end{equation}
where $n_F(\omega) = 1/[1+\exp(\beta \omega)], n_B(\omega)=
1/[\exp(\beta \omega)-1]$ are Fermi and Bose distribution functions.
This local self-energy leads to the modifications of the DOS. In
this solution we treat the self-energy effects in G to all orders,
i.e. $G$ in Eq.~(\ref{EQ:Selfenergy2}) is  full Green's function
$G^{-1} = G_0^{-1} - \Sigma(\omega)$ and solution for $\Sigma$ is
found self-consistently for  a local vibrational mode. The
modifications of the superconducting order parameter and bosonic
propagator were ignored in this calculation. Results are presented
in Fig.~\ref{FIG:DOScomplete}. To proceed with analytic treatment,
unless stated otherwise, we limit ourselves below to second order
scattering in $\Sigma$. Difference between self-consistent solution
and second order calculation are only quantitative and small for
small coupling. Corrections to the Green's function $ G({\bf r},{\bf
r'},\omega) = G^0({\bf r},{\bf r'},\omega) + G^0({\bf r},0,\omega)
\Sigma(\omega) G^0(0,{\bf r'}, \omega) + F^0({\bf r},0,\omega)
\Sigma(\omega) F^{*0}(0,{\bf r},\omega)$. For simplicity we define
$K(T, \omega,\omega_0) = -[n_F(\omega - \omega_0)- n_B(\omega_0) - 1
\simeq \Theta(\omega - \omega_0)]$ which becomes a step function at
low $T\ll \omega_0$, the limit we will focus on hereafter.
Correction to the local density of states as a function of position
comes from the correction to the bare Green's function $G^0: \delta
N({\bf r}, \omega) = 1/\pi \mbox{Im}[G^0({\bf r},0,\omega)
\Sigma(\omega) G^0(0,{\bf r}, \omega) \pm F^0({\bf r}
,0,\omega)\Sigma(\omega) F^{*0}(0,{\bf r},\omega) ]$, where keeping
it general, the plus sign corresponds to the coupling to the local
vibrational mode and minus -- to the spin scattering respectively.
The strongest effect will be at the impurity site. For on-site
density of states we have:
\begin{eqnarray}\label{EQ:DOS1}
&&\frac{\delta N({\bf r} = 0, \omega)}{N_0} = \frac{\pi^2}{2}
(JSN_0)^2
\frac{\omega - \omega_0}{\Delta}K(T,\omega,\omega_0)\nonumber\\
&&\times \left(\frac{2\omega}{\Delta}
\ln\left(\frac{\Delta}{\omega}\right)\right)^2, \ \ \omega \ll
\Delta\;,
\end{eqnarray}
\begin{eqnarray}\label{EQ:DOS2}
&&\frac{\delta N({\bf r} = 0, \omega)}{N_0} = 2\pi^2
(JSN_0)^2K(T,\omega,\omega_0) \ln^2\left(\frac{|\omega -
\Delta|}{4\Delta}\right)\nonumber\\
&&\times \ln\left(\frac{4\Delta}{|\omega + \omega_0 -
\Delta|}\right) + (\omega_0 \rightarrow -\omega_0),\ \omega \simeq
|\Delta|\;,
\end{eqnarray}
where we used  for on-site Green's function $G^0(0,0,\omega) = N_0[
\frac{2\omega}{\Delta}\ln(\frac{4\Delta}{\omega}) + i\pi
\frac{|\omega|}{\Delta}]$, for  $ \omega \ll \Delta$ and  we
retained only dominant real part of $G^0$. In opposite limit $\omega
\simeq \Delta$  we    retained only dominant imaginary part of
$G^0(0,0,\omega) = i\pi N(\omega) = -2iN_0 \ln(\frac{|\omega -
\Delta|}{4\Delta})$. At ${\bf r} = 0$ we have $F^0(0,0,\omega) = 0$.
Complete DOS $N(\omega)$ and derivative $\frac{dN(\omega)}{d\omega}$
are shown on Fig.~\ref{FIG:DOScomplete}. For arbitrary position
$N({\bf r},\omega)$ we would have to add a Friedel oscillation
factor $ \Lambda({\bf r}) = [|G^{0}({\bf r},\omega)|^{2} \pm
|F^0({\bf r}, \omega)|^2] \sim \frac{\sin(k_F r)}{(k_Fr_{||})^2 +
(r_{\perp}/\xi)^2}$ that describes the real space dependence of the
Green's function on distance for small $\omega \ll \Delta$. Here
$\mathbf{r}_{\perp}|| \mathbf{k}_{F \perp}$ is the component of $\bf
r = (r_{\perp}, r_{||})$ that is along the Fermi surface near the
nodal point of the gap and $\mathbf{r}_{||}|| \mathbf{k}_{F ||}$ is
the component perpendicular  to the Fermi surface at nodal point.
Existence of the nodes in $d$-wave case results in the power law
decay of $\Lambda({\bf r})$ in all directions and it has a four fold
modulation due to gap anisotropy (See detailed discussions  in
\cite{MISalkola:1997}), also Sec.~\ref{sec:Dwave}). The final result
is our Eqs.~(\ref{EQ:DOS1}-\ref{EQ:DOS2}).

\begin{figure}
\caption{The solution for the DOS (black line) and its energy
derivative (red line) are presented for a local boson mode
scattering in a d-wave superconductor. The normal self-energy was
treated self-consistently as a full solution of the
Eq.~(\ref{EQ:Selfenergy2}), ignoring vertex corrections and gap
modifications. Apart from the feature at $\omega = \omega_0$ we get
also strong satellite peaks at $ \Delta + \omega_0$ that are a
consequence of a coherence peak in DOS of a $d$-wave superconductor.
These satellites are a specific property of a superconducting state
and will not be present in a pseudogap state. These features are
best seen in $\frac{dN}{d\omega}$. Energy scale is given in units of
$\Delta$, the dimensionless coupling constant is taken to be 1. For
comparison we plot the results for local mode frequency
$\omega_0/\Delta = 0.2, 0.4, 0.6 $ in the first three panels. The
lower panel gives the results for the asymptotic analytic solution,
that assumes $\omega_0 \ll \Delta$ using Eq.~(\ref{eq:Ham}), for
$\omega_0 = 0.4$. The overall features are similar for both cases,
however the analytic solution shows a somewhat larger feature. Taken
from \cite{AVBalatsky:2003}} \label{FIG:DOScomplete}
\end{figure}

It follows immediately that
\begin{eqnarray}\label{EQ:d2Id2V}
&\delta \frac{dI}{dV}/\frac{dI}{dV} \sim \delta N({\bf r} = 0,
V)/N_0 \sim (JSN_0)^2 \frac{V - \omega_0}{\Delta} \Theta(V -
\omega_0)\;,\nonumber\\
&\delta \frac{d^2I}{dV^2} \sim (JSN_0)^2\Theta(V - \omega_0)\;.
\end{eqnarray}
Here we have used the fact that the derivative of
$(\omega-\omega_0)\Theta(\omega - \omega_0)$ with respect to
$\omega$ yields $\Theta(\omega-\omega_0)$. Thus in a d-wave
superconductor and in a metal with vanishing DOS $N(\omega) = N_0
\frac{\omega}{\Delta}$ one should expect a {\em step discontinuity}
in $d^2I/dV^2$ at the energy of a local mode with the strength
$J^2N^2_0$ (see Fig.~\ref{FIG:DOScomplete}). This result is
qualitatively different from the case of conventional metal. For
metal with energy independent DOS we have from Eq.~(\ref{EQ:DOS1})
for $T\ll\omega_0$
\begin{equation}\label{EQ:DOSMETAL}
\frac{dI}{dV} \sim \delta N({\bf r}=0,V)  \sim J^2 N^3_0 \Theta(V -
\omega_0)\;,
 \end{equation}
 and the second derivative will reveal a delta function $d^2I/dV^2
 \sim J^2 N^3_0 \delta(\omega
- \omega_0) $ The effect in d-wave superconductor is clearly smaller
than correction to DOS in a normal metal with the same coupling
strength.

For completeness we also give the result for inelastic scattering in
a metal with the more general DOS $N(\omega) = 1/\pi \mbox{Im}
G^0(0,0,\omega) =  (\omega/\Delta)^{\gamma}N_0$ with power $\gamma
> 0$ that is determined by the microscopic properties of the
material. Then, from Eqs.~(\ref{EQ:Selfenergy2}-\ref{EQ:DOS1}) we
have for $\omega \ll \Delta$:
\begin{eqnarray}\label{EQ:DOSFRAC}
&\delta \frac{dI}{dV}/\frac{dI}{dV} \sim \delta N({\bf r} = 0,
V)/N_0 \sim  (V - \omega_0)^{\gamma} \Theta(V -
\omega_0)\;,\nonumber\\
&\delta \frac{d^2I}{dV^2} \sim (V - \omega_0)^{\gamma -1}\Theta(V -
\omega_0)\;.
\end{eqnarray}
Depending on the value, we get divergent singularity at $\omega_0$
for $\gamma < 1$, or a power law rise for $\gamma \geq 1$. In case
of $\gamma = 1$ we recover the result  for d-wave superconductor and
for a pseudogap normal state.

\begin{figure}[htbp]
\caption{Appearance of the satellite peaks for an impurity resonance
$\omega_{imp}$ at $\omega_{imp}\pm \omega_0$ is shown schematically.
The satellites will have different spectral weight. Imagine we
inject into system an electron at energy $\omega_{imp}+ \omega_0 $.
To create a peak at $\omega_{imp}$ one need to excite local mode and
the energy of the electron will be equal to the difference between
local state and local mode energies. Similarly, to obtain the peak
at $\omega_{imp}$ from injected electron at energy $\omega_{imp} -
\omega_0$ one needs to add local mode energy to an electron. For
this process to occur the local mode has to be excited to begin with
and hence this process will have very low weight at low T.  These
two processes will also have different matrix elements. Overall
relative weight of the side peaks is proportional to $J^2N^2_0$
which we assumed to be small. In case of magnetic scattering, when
$\omega_0 = g \mu_B B$ the splitting will be tunable by the field.
Taken from~\cite{AVBalatsky:2003}} \label{FIG:satellite}
\end{figure}

Quite generally one can express the results in terms of the spectrum
of superconductor.  We can write $ \mbox{Im} \Sigma(\omega)$ using
spectral representation for $G(\mathbf{r}, \omega)$. In
superconducting case, using Bogoliubov $u_{\alpha}(\mathbf{r}),
v_{\alpha}(\mathbf{r})$ for eigenstate $\alpha$, we have
\begin{equation} G(\mathbf{r}, \omega) =
\sum_{\alpha}\biggl{[}\frac{|u_{\alpha}(\mathbf{r})|^2}{\omega -
E_{\alpha} + i\delta} + \frac{|v_{\alpha}(\mathbf{r})|^2}{\omega +
E_{\alpha} + i\delta}\biggr{]}.  \end{equation} Taking imaginary
part of $G(\mathbf{r}, \omega)$
 we arrive  for $T \ll \omega_0$ at:
\begin{eqnarray}
\label{EQ:Selfenergy3} \mbox{Im} \Sigma(\omega) = \frac{\pi J^2}{2
\omega_0} \langle S_z\rangle
[|u_{\alpha}(\mathbf{r}=0)|^2\delta(\omega-\omega_0-E_{\alpha})
\nonumber \\
+|v_{\alpha}(\mathbf{r}=0)|^2\delta(\omega-\omega_0+E_{\alpha})]\;
 \  , \omega >0 .
\end{eqnarray}
At negative $\omega < 0$ one has to replace $\omega_0 \rightarrow
-\omega_0$ in Eq.(\ref{EQ:Selfenergy3}). For example, consider a
magnetic impurity resonance in d-wave superconductor at energy $
\omega_{imp}$, such as a Ni induced
resonance~\cite{MISalkola:1997,EWHudson:2001}. Then only the term
with resonance level $E_{\alpha} = E_{imp}$ will dominate the sum
over eigenstates $\alpha$ in the vicinity of impurity site.
Inelastic scattering off this impurity induced resonance will
produce additional satellite {\em split away from the impurity level
by
 $\omega_0$}, see Fig.~\ref{FIG:satellite}.
   Sharp coherence peaks will also produce split satellites.  Again, for a
  local phonon mode one gets a similar
splitting of impurity level with $\omega_0$ now being the phonon
energy.

These results  suggest the possibility of {\em single  spin
detection} as one monitors the feature in $d^2 I/dV^2$ as a function
of position and external magnetic field. If we take experimentally
seen DOS $N_0 \simeq  1/eV$ with $JN_0 \simeq 0.14, \Delta = 30
meV$~\cite{EWHudson:2001} and assuming the field of $\sim 10 T$ we
have $\omega_0 = 1 meV$ (corresponding to the Zeeman splitting of
$\sim 1 meV$ in a magnetic field, we have from
Eqs.~(\ref{EQ:DOS1}-\ref{EQ:d2Id2V})
\begin{equation}\label{EQ:estimate}
 \delta N({\bf r} = 0, \omega)/N_0 \simeq  10^{-2}
\frac{\omega -\omega_0}{\Delta} \Theta(\omega-\omega_0)\;.
\end{equation}
Result is expressed in terms of the relative change of DOS of a
metal $N_0$. For observation of this effect one would have to sample
DOS in the vicinity of $eV = \omega_0 \propto B$. Assuming $\omega -
\omega_0 = \omega_0$ we have from Eq.~(\ref{EQ:estimate}) $\delta
\frac{dI}{dV}/\frac{dI}{dV} \sim 10^{-2}$. Expressed as a relative
change of {\em DOS of a superconductor} $N(\omega) = N_0
\omega/\Delta$ effect is:
\begin{equation}
\delta \frac{dI}{dV}/\frac{dI}{dV} \sim \delta N({\bf r} = 0,
\omega)/N(\omega_0) \sim 10^{-2} \frac{\omega -\omega_0}{\omega_0}
\Theta(\omega-\omega_0)\;.
\end{equation}
It is of the same order of magnitude as the observed vibrational
modes of localized molecules in inelastic electron tunneling
spectroscopy STM, IETS-STM~\cite{BCStipe:1998,JRHahn:2001}. The
satellites at $ \Delta + \omega_0$ produce the effect on the scale
of unity and clearly seen even for small coupling.  The important
difference is that for localized spin the kink in DOS is {\em
tunable} with magnetic field and this should make its detection
easier.

\subsection{Localized vibrational modes in $d$-wave
superconductors}

When a localized vibrational mode is coupled to electrons in a
superconductor, the Hamiltonian is given by:
\begin{eqnarray}
H&=&\sum_{k\sigma} \xi_{k} c_{k\sigma}^{\dagger} c_{k\sigma}
+\sum_{k} [\Delta_{k} c_{k\uparrow}^{\dagger}
c_{-k\downarrow}^{\dagger} + h.c.]\nonumber \\
&&  + g\sum_{\sigma} (b^{\dagger} + b) c_{0\sigma}^{\dagger}
c_{0\sigma} \;,
\end{eqnarray}
 Up to a second order in coupling constant $g$ calculations are very
 similar if not identical to the ones in the previous section. For more
 detailed analysis see \cite{DKMorr:2003}.

The proposed extension of the inelastic tunneling spectroscopy on
the strongly correlated electrons states, such as a d-wave
superconductor and pseudogap normal state would open up
possibilities to study the dynamics of local spin and vibrational
excitations. The DOS in these systems has a nontrivial energy
dependence of general form $N(\omega) \sim \omega^{\gamma}, \gamma
> 0$. This technique could allow for a Zeeman level spectroscopy
of a single magnetic center, thus, in principle,  allowing a single
spin detection. There is a feature in $dI/dV \sim (\omega -
\omega_0)^{\gamma - 1} \Theta(\omega - \omega_0)$ near the threshold
energy $\omega_0$ that indicates the inelastic scattering. One also
finds strong satellite features near the gap edge due to coherence
peak for a superconducting case. The singularity is a power law and
qualitatively different from the results for a simple metallic DOS
\cite{BCStipe:1998,JRHahn:2001}. For the relevant values of
parameters for high-$T_c$ the feature is on the order of several
percents and makes the feature observable in these materials.
Similar predictions are also applicable to the local vibrational
modes, where $\omega_0$ becomes a vibrational mode frequency.

\section{Interplay between collective modes and impurities in  $d$-wave superconductors}
\label{sec:Interplay}

Conventional wisdom dictates that local probes have poor momentum
resolution, and therefore cannot identify collective modes that
exist at specific wave vectors. However, recent progress in relating
the STM observations with ARPES \cite{KMcElroy:2003} by using a
Fourier transform of the image taken over a large area opens up the
possibility to study connections of STM measurements with other
spectroscopies. Since the ability of the STS to identify the Fermi
wave vector is most naturally explained by the sensitivity of the
technique to Friedel oscillations in the electron density due to
impurities, we now look into the DOS features arising from the
interplay between dynamic scattering of
 of the collective mode and static disorder.

Neutron scattering is one of the spectroscopic measurements which
revealed a resonance excitation, the so called 42 meV peak that is
commonly present in a number of materials. It has been proposed that
STM may be used to detect neutron mode~\cite{JXZhu:2004}. The main
tool for such a measurement is the Inelastic Electronic Tunneling
Spectroscopy (IETS), see Sec.~\ref{sec:DynamicalImp}. Specifically
the proposal is to use Fourier transform (FT) tunneling maps and
search for features that represent an IETS signature.

We limit our consideration to the example of a well defined mode at
wave vector $(\pi, \pi)$ with energy $\omega_0$. The ultimate goal
is to detect the bosonic spectral function, be it magnetic spin mode
or some lattice modes. Recent efforts indicate possible connection
between the kink in ARPES data on quasiparticle dispersion and
phonon modes and, possibly,
superconductivity~\cite{ALanzara:2001,ADamascelli:2003,GHGweon:2004}.
It was suggested that the full Eliashberg function in frequency {\em
and momentum} space may be extracted from ARPES data
\cite{IVekhter:2003a}, and the challenge is to design a similar
procedure for its determination from the FT IETS STM. Efforts to
relate the data from ARPES, STM, and transport measurements in
cuprates have recently intensified
\cite{LZhu:2004a,DJScalapino:2004}.

We need the impurity scattering to produce features that can be
analyzed using Fourier transform. We consider  weak Born scattering
from distributed scalar potential $U(\br_i) = U_i$, $i$ is the
lattice site index. We find indeed that disorder potential can
strongly modify the features as seen in local DOS. One of the
interesting findings is that characteristic wave vectors of the
impurity potential $U_{\mathbf{q}} = \sum_{i}
U_{i}\exp(i\mathbf{q}\cdot \mathbf{r}_{i})$ do play a crucial role
in defining characteristic wave vectors of the DOS modulation.

To explain the method we point out that FT STM data do contain
momentum information. However all the analysis up to date on FT STM
data was done within the framework of elastic scattering that
considers the natural quasiparticle excitations at the Fermi surface
\cite{JEHoffman:2002b}. No inelastic scattering processes off the
collective mode were included in the analysis. To consider the
scattering of quasiarticles off the collective mode   one has to
explicitly keep track of the self-energy effects that result from
scattering. In this case simple noninteracting quasiparticle picture
is not adequate. Inelastic scattering of quasiparticles requires us
to consider excitations off shell, for example to consider
excitations at energies that are typically $\Delta + \Omega_0 \sim
70 meV$. At these energies the combination of the Fermi surface
effects, typical wavevectors of the collective mode and typical
wavevectors of the random potential all enter in determining the
momentum of the inelastic tunneling features as seen in FT STM.

We limit ourselves to the second order scattering between carriers
and bosonic excitations and at this level there is no conceptual
difference in the method as applied to spin or phonon bosonic mode.

We start with a model Hamiltonian describing two-dimensional
electrons coupled to a collective spin mode and in the presence of
inhomogeneity:
\begin{equation}
\mathcal{H}=\mathcal{H}_{BCS}+ \mathcal{H}_{sp} +
\mathcal{H}_{imp}\;.
\end{equation}
Here the BCS-type Hamiltonian is given by $
\mathcal{H}_{BCS}=\sum_{\mathbf{k},\sigma}
(\varepsilon_{\mathbf{k}}-\mu) c_{\mathbf{k}\sigma}^{\dagger}
c_{\mathbf{k}\sigma} +\sum_{\mathbf{k}}(\Delta_{\mathbf{k}}
c_{\mathbf{k}\uparrow}^{\dagger}c_{-\mathbf{k}\downarrow}^{\dagger}
+\Delta_{\mathbf{k}}^{*}
c_{-\mathbf{k}\downarrow}c_{\mathbf{k}\uparrow}) $, where
$c_{\mathbf{k}\sigma}^{\dagger}$ ($c_{\mathbf{k}\sigma}$) creates
(annihilates) a conduction electron of spin $\sigma$ and wavevector
$\mathbf{k}$, $\varepsilon_{\mathbf{k}}$  is the normal state energy
dispersion for the conduction electrons, $\mu$ the chemical
potential, and $\Delta_{\mathbf{k}}=\frac{\Delta_{0}}{2}(\cos k_x
-\cos k_y)$ the $d$-wave superconducting energy gap. The coupling
between the electrons and the resonance mode is modeled by an
interaction term $\mathcal{H}_{sp}=g\sum_{i} \mathbf{S}_{i}\cdot
\mathbf{s}_{i}$, where the quantities $g$, $\mathbf{s}_{i}$, and
$\mathbf{S}_{i}$ are the coupling strength, the electron spin
operator at site $i$, and the operator for the collective spin
degrees of freedom, respectively. The dynamics of the collective
mode will be specified below. The quasiparticle scattering from
impurities in the Hamiltonian is given by,
$
H_{imp}=\sum_{i\sigma} U_{i} c_{i\sigma}^{\dagger} c_{i\sigma} \;,
$
where $U_{i}$ is the strength of the impurity scattering potential.
For simplicity, only the case of nonmagnetic scattering is
considered here. By introducing a two-component Nambu spinor
operator, $\Psi_{i}=(c_{i\uparrow}, c_{i\downarrow}^{\dagger})^{T}$,
one can define the matrix Green's function for the full Hamiltonian
system, $\hat{G}(i,j;\tau,\tau^{\prime})=-\langle T_{\tau}
[\Psi_{i}(\tau) \otimes \Psi_{j}^{\dagger}(\tau^{\prime})]\rangle$.
An algebra leads to the full electron Green's function with impurity
scattering:
\begin{eqnarray}
\mathcal{G}_{11}(i,j;i\omega_{n}) &=& \tilde{\mathcal{G}}_{11}^{(0)}(i, j;i\omega_{n}) \nonumber \\
&&+ \sum_{j^{\prime}}
U_{j^{\prime}}[\tilde{\mathcal{G}}_{11}^{(0)}(i,j^{\prime};i\omega_{n})
\tilde{\mathcal{G}}_{11}(j^{\prime},j;i\omega_{n}) \nonumber \\
&&-\tilde{\mathcal{G}}_{12}^{(0)}(i,j^{\prime};i\omega_{n})
\tilde{\mathcal{G}}_{21}(j^{\prime},j;i\omega_{n})]\;.
\end{eqnarray}
Here $\tilde{\mathcal{G}}^{(0)}$ is the dressed Green's function,
with its Fourier component given by:
\begin{equation}
[\tilde{\mathcal{G}}^{(0)}]^{-1}(\mathbf{k};i\omega_{n})=\left(
\begin{array}{cc}
i\omega_{n}-\xi_{\mathbf{k}}-\Sigma_{11} &
-\Delta_{\mathbf{k}}-\Sigma_{12} \\
-\Delta_{\mathbf{k}}-\Sigma_{21} &
i\omega_{n}+\xi_{\mathbf{k}}-\Sigma_{22}
\end{array}
\right)\;,
\end{equation}
where $\xi_{\mathbf{k}}=\varepsilon_{\mathbf{k}}-\mu$,
 $\omega_{n}=(2n+1)\pi T$ is the fermionic
Matsubara frequency for fermions. When the inelastic scattering of
quasiparticles from the collective mode occurs, the self-energy is
obtained to the second order in the coupling constant as:
\begin{equation}
\hat{\Sigma}(\mathbf{k};i\omega_{n})=\frac{3g^{2}T}{4}\sum_{\mathbf{q}}
\sum_{\Omega_{l}} \chi(\mathbf{q};i\Omega_{l})
{G}_{0}(\mathbf{k}-\mathbf{q};i\omega_{n}-i\Omega_{l})\;,
\end{equation}
where $\chi(\mathbf{q};i\Omega_{l})$ is the dynamical spin
susceptibility $\chi_{ij}(\tau)=\langle T_{\tau}
(S^{x}_{i}(\tau)S^{x}_{j}(0))\rangle$ and
 $\Omega_{l}=2l\pi T$ the bosonic
Matsubara frequency, and $G_0$ is the bare superconducting Green's
function.
We have also assumed that the $d$-wave pair potential is real. For a
single-site impurity, the equation of motion for the full Green's
function can be exactly solved. For the case of multiple impurities
and especially the inhomogeneous situation, some approximation needs
to be taken. We consider here the case of the weak impurity
scattering limit so that the Born approximation can be used. With
this ansatz, we arrive at:
\begin{equation}
\mathcal{G}_{11}(i,j;i\omega_{n}) =
\tilde{\mathcal{G}}_{11}^{(0)}(i, j;i\omega_{n}) + \delta
\mathcal{G}_{11}(i,j;i\omega_{n})
\end{equation}
with
\begin{eqnarray}
&\delta \mathcal{G}_{11}(i,j;i\omega_{n}) = \sum_{j^{\prime}}
U_{j^{\prime}}
[\tilde{\mathcal{G}}_{11}^{(0)}(i,j^{\prime};i\omega_{n})
\tilde{\mathcal{G}}_{11}^{(0)}(j^{\prime},j;i\omega_{n})
&\nonumber \\
&-\tilde{\mathcal{G}}_{12}^{(0)}(i, j';i\omega)
\tilde{\mathcal{G}}_{21}^{(0)}(j^{\prime},j;i\omega_{n})]\;.&
\end{eqnarray}
The LDOS at the $i$-th site, summed over two spin components, is
\begin{equation}
\rho(\mathbf{r}_{i},E)=-\frac{2}{\pi} \mbox{Im}
\mathcal{G}_{11}(i,i;E+i\gamma)\;, \label{ldos}
\end{equation}
where $\gamma=0^+$.
  We are especially interested
in the correction to the LDOS from the impurity scattering:
\begin{equation}
\delta\rho(\mathbf{r}_{i},E)=-\frac{2}{\pi} \mbox{Im} \delta
\mathcal{G}_{11}(i,i;E+i\gamma)\;. \label{ldos_correction}
\end{equation}
Its Fourier transform is:
\begin{eqnarray}
\delta \rho(\mathbf{q},E) &=& \sum_{i} \delta \rho(i,E)
e^{-i\mathbf{q}\cdot \mathbf{r}_{i}}  \nonumber \\
&=& -\frac{U_{\mathbf{q}}}{N\pi i} \sum_{\mathbf{k}}
[\tilde{\mathcal{G}}_{11}^{(0)}(\mathbf{k}+\mathbf{q};E+i\gamma)
\tilde{\mathcal{G}}_{11}^{(0)}(\mathbf{k};E+i\gamma) \nonumber \\
&&- \tilde{\mathcal{G}}_{11}^{(0)*}(\mathbf{k}-\mathbf{q};E+i\gamma)
\tilde{\mathcal{G}}_{11}^{(0)*}(\mathbf{k};E+i\gamma) \nonumber \\
&& -\tilde{\mathcal{G}}_{12}^{(0)}(\mathbf{k}+\mathbf{q};E+i\gamma)
\tilde{\mathcal{G}}_{21}^{(0)}(\mathbf{k};E+i\gamma) \nonumber \\
&&+ \tilde{\mathcal{G}}_{12}^{(0)*}(\mathbf{k}-\mathbf{q};E+i\gamma)
\tilde{\mathcal{G}}_{21}^{(0)*}(\mathbf{k};E+i\gamma)]\;,
\end{eqnarray}
where the Fourier transform of the scattering potential is
\begin{equation}
U_{\mathbf{q}}=\sum_{i} U_{i} e^{-i \mathbf{q}\cdot
\mathbf{r}_{i}}\;.
\end{equation}
The corresponding
 Fourier (wave-vector) spectrum is defined as
\begin{equation}
P(\mathbf{q},E)=\vert \delta \rho(\mathbf{q},E) \vert\;.
\end{equation}

Up to now discussion and formulation are quite general and can be
used to study the effects of any dynamic mode once the
susceptibility $\chi$ is known. Consider now specific case of
magnetic mode with  susceptibility taking  a phenomenological form
(based on the inelastic neutron scattering observations), see
also~\cite{MEschrig:2000}:
\begin{equation}
\chi(\mathbf{q};i\Omega_{l})=-\frac{\delta_{\mathbf{q},\mathbf{Q}}}{2}
\left[\frac{1}{i\Omega_{l}-\Omega_{0}}-\frac{1}
{i\Omega_{l}+\Omega_{0}}\right]\;,
\end{equation}
where we denote the wavevector $\mathbf{Q}=(\pi,\pi)$ and the mode
energy by $\Omega_{0}$. This form captures the essential feature of
resonant peak observed by neutron scattering experiments in the
superconducting state of cuprates~\cite{JXZhu:2004}. For the
normal-state energy dispersion, we use $\varepsilon_{\mathbf{k}}=-2t
(\cos k_x + \cos k_y) -4t^{\prime} \cos k_{x} \cos k_y$, where $t$
and $t^{\prime}$ are the nearest and next-nearest neighbor hopping
integral. Unless specified explicitly, the energy is measured in
units of $t$. We choose $t^{\prime}=-0.2$ to model the band
structure of the hole-doped cuprates.  Since the maximum energy gap
for most of the cuprates at the optimal doping is about $30
\;\mbox{meV}$ while the resonance mode energy is in
 the range between $35$ and $47$ meV,
we take $\Delta_0=0.1$ and $\Omega_0=0.15$ (i.e., $1.5 \Delta_{0}$).
The chemical potential ($\mu\approx -1.15$) is tuned to give an
optimal doping value 0.16. To mimic the intrinsic life time
broadening, in our numerical calculation we take $\gamma$ of
Eq.~(\ref{ldos}) to be $0.08\Delta_{0}$. A system size of
$N=N_x\times N_y=256 \times 256$ is taken in the numerical
calculation.
\begin{figure}[th]
\caption{The
density of states for a clean system with $g=0$ and $0.2$.}
\label{FIG:BAND_DOS} \label{FIG:SPECTRUM}
\end{figure}

Figure~\ref{FIG:BAND_DOS} plots the density of states in a clean
system for various values of coupling constant $g$. Without the
electron-mode coupling ($g=0$), the density of states is peaked only
at the maximum gap edges $\pm \Delta_0$. When there exists the
electron-mode coupling, for example $g=0.2$ as shown in the figure,
several new effects emerge. As a result of the additional anomalous
self-energy introduced through the coupling, the maximum gap edge is
renormalized to $\Delta_{ren}$, which is larger than $\Delta_0$.
More importantly, the singularity in the quasiparticle self-energy
causes additional poles in the Green's function, and new peaks show
up at the energy $\pm E_r =\pm (\Delta_r + \Omega_0)$. A strong
implication of this result is that regardless of the renormalization
of the energy gap, the position of the new peaks relative to the
superconducting coherent peak is shifted by $\Omega_0$. In addition,
with the appearance of the peaks away from the gap edge, the
intensity of the superconducting coherent peaks is reduced such that
the sum rule is obeyed. The intensity of the peaks at negative
energies is stronger than that at positive energies since the van
Hove singularity is below the Fermi energy. These results, for the
clean case, are consistent with earlier studies of the
ARPES~\cite{DSDessau:1991,ZXShen:1997,JCCampuzano:1999,MRNorman:1998b,
ArAbanov:2002,HYKee:2002,MEschrig:2000} and
DOS~\cite{ArAbanov:2000}. The shift of states due to inelastic
scattering is also expected for scattering off of local
mode~\cite{AVBalatsky:2003}.

We now turn to the Fourier spectrum in the presence of disorder. An
accurate description of this problem requires an extremely high
energy and spatial resolution. Therefore, a very large system size
should be considered. For the quasiparticle scattering off a single
impurity, as considered in the work by Zhu {\em et
al.}~\cite{JXZhu:2004}, one can first calculate the LDOS within a
small window around the impurity site in a very big system size
(e.g., $N=1000\times 1000$), then perform a Fourier transform over
the window size and even with masking of sites. This procedure and
flexibility does not exist in the case of disorder and inhomogeneity
with multiple scattering centers. Summation over the wave-vector in
the Brillouin zone constrains us to consider a moderate system size.

\begin{figure}[th]
\caption{Fourier spectrum at $E=-E_r$ for the constant $g=0.2$ and
the structure factor $U_{\mathbf{q}}=U_0$. The parameter
$U_{0}=0.3$.} \label{FIG:FOURIER-1}
\end{figure}

\begin{figure}[th]
\caption{Fourier spectrum at $E=-E_r$ for the constant $g=0.2$ and
the structure factor $u_{\mathbf{q}}=U_0 q_{0}^{2} /\{q_{0}^{2} +
4[\cos^{2}(q_{x}/2) + \cos^{2}(q_y/2)]\}$. The parameter $U_{0}=0.3$
and $q_0=0.5$.}
 \label{FIG:FOURIER-2}
\end{figure}

In Fig.~\ref{FIG:FOURIER-1}, we show the Fourier spectrum at the
energy $-E_r$ for the coupling constant $g=0.2$ with a structurless
scattering potential $U_{\mathbf{q}}=U_0$. This structureless
$u_{\mathbf{q}}$ corresponds to a single-site impurity in real
space.  In the absence of the electron-mode coupling ($g=0$), the
Fourier spectrum has a strongest intensity at $\mathbf{q}=(0,0)$ and
its equivalent points, and a moderately strong weight along the
edges of the square around $\mathbf{q}=(\pi,\pi)$. When there exists
the electron-mode coupling, as shown in Fig.~\ref{FIG:FOURIER-1},
the spectrum has the strongest intensity point at four corners of
the square, and a moderately strong intensity at the four ridges of
the square. This implies that as one takes line cuts along the
diagonal the first feature at $\pm E_r$ always will be at the
wavevector $(\pi - \delta, \pi - \delta)$ before the $(\pi,\pi)$
point.  Regardless of whether the electrons are coupled to the
collective mode, the spectrum has a minimum in intensity at
$\mathbf{q}=(\pi,\pi)$, which is different from the results when the
pre-dominant Friedel oscillation is filtered out of the Fourier
transform. When the scattering potential has a clear stucture, which
might be relevant to the inhomogeneity in high-$T_c$ cuprates, the
pattern of the Fourier spectrum changes dramatically. As an ansatz,
we propose the following stucture for the scattering potential:
\begin{equation}
U_{\mathbf{q}}=\frac{U_0 q_{0}^{2}}{q_{0}^{2} + 4 [\cos^{2}(q_x/2) +
\cos^{2}(q_y/2)]}\;,
\end{equation}
where the parameter $q_0$ describes the extent of the peak at
$\mathbf{q}=(\pi,\pi)$. This structure factor has a highly
nontrivial consequence on the Fourier spectrum of the local density
of states. This is because overall modulation $\delta \rho({\bf q},
\omega) \propto U_{\mathbf{q}}$ and for $U_{\mathbf{q}}$ peaked at
$(\pi,\pi)$ FT DOS $\delta \rho({\bf q})$ will also be peaked at
this wavevector.  As shown in Fig.~\ref{FIG:FOURIER-2}, the Fourier
spectrum now has a strongest intensity at the four ridges, which are
located along the diagonals of the first Brillouin zone, and a
secondly strongest intensity at $\mathbf{q}=(\pi,\pi)$. Except in
the four small lobes around the corners, the spectrum has moderate
intensity inside the square.

Here we have considered the situation where all impurities have
identical potential scattering strength. If one dopes a strong
scattering impurity, for example Zn substituted for Cu, into a
high-$T_c$ cuprate which has inhomogeneity coming from weak
potential scatterers, the pattern of the Fourier spectrum of the
local density of states is mostly determined by the scattering off
the strong impurity.

To  summarize,  the main results of this section are : 1. the energy
of the inelastic feature is at $E_r = \Delta_0 + \Omega_0 \sim 70
meV$ for optimal doping. Given that the gap is position dependent in
observed spectra, this energy will be position dependent. 2. the
typical wave vectors along diagonals where the inelastic features
are seen are determined by a number of factors: the momentum of the
disorder potential $U({\bf q})$, the doping and positions of the
``diamonds'' seen in Figs.~\ref{FIG:FOURIER-1} and
\ref{FIG:FOURIER-2}. 3. the first feature that is seen in FT IETS
STM in our calculation is occurring at wavevectors that are inside
$(\pi,\pi)$ vectors.

\section{Scanning tunneling microscopy results}

\label{sec:STM}

\subsection{STM results around a single impurity}

\begin{figure}[tbp]
\caption{Left
panel: The
 $dI/dV$ spectra measured near (A) Mn, (B)
Gd, and (C) Ag atoms and  far away from the impurity where  local
density of states can be fit by the BCS theory. Right panel:
Constant-current topographs and simultaneously acquired $dI/dV$
images show the spatial extent of the bound state near Mn and Gd
adatoms.
 (A) Constant-current (32 $\AA$ by 32 $\AA$) topograph of a Mn adatom.
 (B) Image of $dI/dV$ near the Mn adatom, acquired simultaneously with the topograph in
 (A) by using an ac detection.
 The areas where $dI/dV$ is reduced (dark) show the extent of the bound state.
  This reversed contrast comes about because a dc bias voltage was chosen
  well above the energy of the bound state, where the bound state affects
  $dI/dV$
  only indirectly by contributing to the total current $I$.
     (C) Constant-current (32 $\AA$ by 32 $\AA$) topograph of a Gd adatom.
  (D) Image of dI/dV near the Gd adatom, acquired simultaneously with the topograph in (C).
  From Yazdani~\cite{AYazdani:1997}.} \label{FIG:Yazdani}
\end{figure}

The STM has established itself as a remarkably powerful and
versatile tool for studying the electronic properties of solids. Its
remarkable energy and spatial resolution makes it particularly well
suited for materials characterized by small energy and short length
scales. It measures the tunneling current varying with the voltage
bias and the tip positions. In the tunneling Hamiltonian formalism,
the differential tunneling conductance---the derivative of the
current with respect to the voltage bias, is given by
\begin{equation}
\frac{dI}{dV} \propto -\int d\omega \sum_{\mathbf{k},\sigma} \vert
T_{\mathbf{k}}\vert^{2} A_{\sigma}(\mathbf{k},\omega)
f_{FD}^{\prime}(\omega-eV)\;,
\end{equation}
where $f_{FD}$ is the Fermi distribution function, and
$A_{\sigma}(\mathbf{k},\omega)$ is the electron spectral function of
the sample. The tunneling matrix element, $\vert
T_{\mathbf{k}}\vert^{2}= \sum_{\mathbf{q}} \vert
M_{\mathbf{kq}}\vert^{2} A_{tip}(\mathbf{q},\omega)$, where
$M_{\mathbf{kq}}$ is the matrix element representing the overlap of
the electronic states on the tip and sample. Using a tip with
featureless DOS around the Fermi energy, we can assume $\vert
T_{\mathbf{k}}\vert^{2}$ is energy independent. If we further assume
a $\mathbf{k}$-independent tunneling matrix element, one can find
that the tunneling conductance is proportional to the local density
of state at the tip position, which we have chosen to be the origin,
$\rho(eV) = -\int d\omega \sum_{\mathbf{k},\sigma}
A(\mathbf{k},\omega) f^{\prime}(\omega-eV)$. At zero temperature, it
is simply given by the imaginary part of the electronic Green's
function we have used heavily  for the discussion throughout the
work, that is $\rho(\mathbf{r},eV) = -\frac{1}{\pi}\sum_{\sigma}
\mbox{Im} G_{\sigma}(\mathbf{r},\mathbf{r};\omega=eV)$. Here we have
labeled the tip position by $\mathbf{r}$.

The experimental attempts to detect and accurately resolve the
sub-gap features in the density of states in superconductors with
impurities have a long history. This feature was observed in the
planar junctions doped with magnetic impurities in earlier years.
However, a direct observation of the sub-gap states induced by a
magnetic impurity did not occur until late 1990's. In 1997, Yazdani
and co-workers~\cite{AYazdani:1997} deposited adatoms, Mn, Gd and
Ag, on the (110)-oriented surface of a superconducting Nb sample,
and examined the local electronic structure around them.
Figure~\ref{FIG:Yazdani} shows their STM tunneling spectrum
measurement. The main fundings are: (1) The local density of states
has no much difference for the tunneling through Ag impurity atoms
and far away from them, since Ag impurity atoms are believed to be
non-magnetic in nature; (2) The LDOS at energies less than the Nb's
gap is enhanced when the tunneling is through Mn and Gd magnetic
atoms. The enhancement happens at the length scale of 10\AA,
indicating the bound nature of the impurity states; (3) The LDOS
spectra are asymmetric about the Fermi energy. Within the framework
of the BdG theory, the authors used a two-parameter magnetic
impurity model, where the electrons are coupled with the impurity
through an magnetic exchange interaction $J$ and experience a
nonmagnetic potential scattering $U$. The obtained results,
consistent with the Yu-Shiba-Rusinov prediction and more recent
theoretical works, fit the experimental data. However, the model
calculation required the value of $J$ of the order of
$4\;\mbox{eV}$, in the strong coupling limit, and failed to capture
the detailed spatial dependence of the spectra around Gd site.

The pioneering STM experimental research on the local electronic
structure around single defects and impurities in high-$T_c$
cuprates was carried out by two groups led by Eigler at IBM Almaden
Research Center~\cite{AYazdani:1999} and Davis at UC
Berkeley~\cite{EWHudson:1999}. Since the high-$T_c$ cuprates have a
$d$-wave pairing symmetry, even nonmagnetic impurity scattering
would affect the superconductivity. Byers, Flatte, and
Scalapino~\cite{JMByers:1993} were the first to suggest the use of
STM to study the spatial variations of the tunneling conductance
near impurities. In particular, it was theoretically
predicted~\cite{AVBalatsky:1995,MISalkola:1996} that quasiparticle
resonance states are induced around a nonmagnetic impurity in a
$d$-wave superconductor, in striking contrast to $s$-wave systems.
The sample used by Yazdani and co-workers is overdoped
Bi$_{2}$Sr$_{2}$CaCu$_{2}$O$_{8}$ with a superconducting transition
temperature of 74K and a transition width of 3K. The sample used by
the Berkeley group is Bi$_{2}$Sr$_{2}$CaCu$_{2}$O$_{8+\delta}$ with
a transition temperature of 87K and a transition width of 5K. The
STM experiments were operated at 5K and 4.2K, respectively. The STM
spectroscopy on these samples, which were nominally undoped with
known impurities, shows clearly the enhancement of the local density
of states near the zero voltage bias in regions where the chemically
induced defects in the sample are located. The experiments provided
a strong evidence for the existence of low-energy quasiparticle
resonance states around single nonmagnetic impurities, as predicted
theoretically. The asymmetric or splitting of the measured resonance
near the zero bias may come from the fact that  the particle-hole
symmetry may be broken by impurities and defects locally or the
underlying realistic band structure of
BSCCO~\cite{MEFlatte:1998,JXZhu:2000a}. However, in these two
experiments, the location in the crystal and the identity of these
scattering centers are unknown. Moreover, since the enhancement of
the local density of states at these scattering centers is not
dramatically large and the coherence of high-Tc superconductors is
so short, it is very difficult to investigate in detail the local
electronic structure around them at an atomic scale.

\begin{figure}[tbp]
\caption{Differential
tunneling spectra taken at the Zn-atom site (open circles) and a
location far away from the impurity (filled circles). Note that even
on the impurity site one has peaks at both positive and negative
bias  albeit of very different magnitude that are reflection of the
particle hole character of the impurity resonance.  To fit the data
one can use a simple potential scattering model with essentially
unitary scattering phase shift $\theta = 0.48 \pi$. Phase shift is
related to a impurity potential $U_0$ via simple formula: $\cot
\theta = \frac{1}{\pi N_F U_0}$. From Pan {\em et
al.}~\cite{SHPan:2000b}.} \label{FIG:Pan_Zn}
\end{figure}

\begin{figure}[tbp]
\caption{
Differential conductance spectra above the Ni atom and at several
nearby locations. Differential conductance spectra obtained at four
positions near the Ni atom showing the maxima at $eV=\pm
\Omega_{1}$. Intensity as a function of position relative to
impurity site reverses upon change of the bias sign. This effect is
explained as a result of particle and hole components of the
impurity state. From Hudson {\em et al.}~\cite{EWHudson:2001}.}
\label{FIG:Hudson_Ni}
\end{figure}

\begin{figure}[tbp]
\caption{
High-spatial-resolution image of the differential tunneling
conductance at a negative  tip voltage bias $eV=-1.2mV$ at a
$60\times 60 \AA^{2}$ square. Also shown d-wave gap nodes
orientation and lattice sites to indicate that impurity state is
registered to lattice. From Pan {\em et al.}~\cite{SHPan:2000b}.}
\label{FIG:Pan_Zn_Image}
\end{figure}

New STM study on the impurity effects in BSCCO was reported by the
Berkeley group~\cite{SHPan:2000b}. The samples were
Bi$_{2}$Sr$_{2}$Ca(Cu$_{1-x}$Zn$_{x}$)$_{2}$O$_{8+\delta}$ single
crystals with intentionally doped with $x=0.6\%$ Zn. The crystals
have the transition temperature of 84K and a width of 4K. To search
for low-energy quasiparticle states associated with the Zn atoms,
the authors first mapped the differential tunneling conductance at
zero sample bias in a larger window of the surface and found a
number of randomly distributed bright sites corresponding to the
areas of high LDOS. Then they measured the tunneling spectroscopy
exactly at the center of a bright scattering site. As shown in
Fig.~\ref{FIG:Pan_Zn}, the spectrum showed a very strong DOS peak at
the energy $\Omega =-1.5 \pm 0.5 \;\mbox{meV}$. The peak intensity
can be up to six times greater than the normal-state conductance. At
the same time, the intensity at the superconducting coherence peak
is strongly suppressed, indicating the almost complete local
destruction of superconductivity. These phenomena is consistent with
the theoretically predicted characteristics of quasiparticle
scattering off a nonmagnetic unitary impurity in a  $d$-wave
superconductor. The strong intensity of the near-zero-bias peak
allows the authors to give a close inspection of the electronic
structure around the Zn impurity. As shown in
Fig.~\ref{FIG:Pan_Zn_Image}, the STM differential conductance
imaging at $\Omega=-1.5 meV$ exhibits two novel features: Firstly,
it has strongest intensity directly at the impurity site and local
maxima at the sites belonging to the sublattice containing the
impurity site, while local minima at the sites belonging to the
other sublattice. Secondly, the intensity decays much faster along
the gap nodal direction than along the bond direction.  These new
features are totally unexpected. The theory based on a potential
scattering model would predict a vanishingly small intensity at the
impurity in the unitary limit. The first feature motivated theorists
to study the electronic structure around a Kondo impurity in a
$d$-wave
superconductor~\cite{APolkovnikov:2001,GMZhang:2001,JXZhu:2001b},
and consider the importance of the BiO layer which is the exposed
surface~\cite{JXZhu:2000b,IMartin:2002,JXZhu:2001c}.


\begin{figure}[tbp]
\caption{Tunneling DOS for tunneling on Ni impurity site. Note that
there are always states at opposite bias as well. The peak intensity
is largest on either positive or negative bias depending on the
position. To fit the data one need to use both $U_0$ and $J$. From
Hudson {\em et al.}~\cite{EWHudson:2001}.}
\label{FIG:Hudson_Ni_Image}
\end{figure}

The STM study on the local electronic structure around a magnetic Ni
atom in BSCCO was also reported by the same Berkeley
group~\cite{EWHudson:2001}. It was found that there two
spin-resolved resonance states induced by the Ni atom, in contrast
to the case of Zn atom in previous experiment where only a
spin-degenerate resonance state is induced. The energy of four
resonance peaks in the tunneling spectrum are, $\pm \Omega_{1}$ and
$\pm \Omega_{2}$, with $\Omega_{1}=9.2 \pm 1.1 \;\mbox{mev}$ and
$\Omega_{2}=18.6\pm 0.7\;\mbox{meV}$. The experimental result is
reasonable agreement with a theoretical model with both nonmagnetic
and magnetic scattering~\cite{MISalkola:1997,HTsuchiura:2000}. By
substituting the values of $\Omega_{1,2}$ and the maximum
superconducting energy gap $\Delta_{0}=28\;\mbox{meV}$ into the
theoretical formula~\cite{MISalkola:1997}:
\begin{equation}
\Omega_{1,2} = -\frac{\Delta_{0}}{2N_{F}(U_0\pm J)\ln \vert
8N_{F}(U\pm W)\vert}\; \label{EQSTMMagnImp}
\end{equation}
with $N_{F}$ the normal-state density of states at the Fermi energy,
$U$ and $W$ the strength of nonmagnetic and magnetic scattering, one
can find $N_{F}U=-0.67$ and $N_{F}W=0.14$. This result indicates
that, despite Ni atoms possess a magnetic moment, the scattering off
them is dominated by potential interactions. In addition, the
experiment also showed that the intensity at the gap edge in the
tunneling conductance directly at the Ni impurity site is almost
unaffected, in comparison with that far away from the impurity,
supporting the scenario that the high-Tc superconductivity is
magnetically mediated~\cite{DPines:1997}.

\subsection{Spatial distribution of particle and hole components}

Spatial distribution of tunneling intensity clearly exhibits
alternation between positive and negative bias, see for example,
Fig.~\ref{FIG:Hudson_Ni}. It appears as a rotation of an impurity
induced cross upon changing the sign of the bias. Since the effect
is so explicit in the images we will address it here.

Apparent rotation of the impurity intensity  can be understood as a
result of interplay between particle and hole components of the
Bogoliubov quasiparticle. This effect is a general property of
superconductivity and is seen in both $s$-wave and $d$-wave
superconductors \cite{SHPan:2000a,EWHudson:2001,AYazdani:1997}, see
also Sec.~\ref{sec:QPT}. The Bogoliubov quasiparticles, that are
native excitations in superconductor, have both particle and hole
component. The sites where there is a large particle components will
have large intensity on positive bias site and hence will be show up
as bright sites on positive bias. Sites with large hole component
will be bright on negative bias, Fig.~\ref{FIG:YazdaniScience1}.

\begin{figure}[tbp]
\caption{The particle and hole components of the impurity wave
function for a magnetic impurity in a s-wave superconductor is
shown. A) Impurity wave function $\Psi_{B}(r)$ and B)
$r^2\Psi_{B}(r)$ are shown. The maxima of particle and hole
components occur at different positions. This results in the
different image of the impurity state, seen on positive and negative
bias. This effect is a general property of a superconductor
regardless of the symmetry of the pairing state. From
~\cite{AYazdani:1997}.} \label{FIG:YazdaniScience1}
\end{figure}

Let us define the respective amplitudes of particle and hole
amplitudes  of the Bogoliubov quasiparticle, $u_n(i)$ and $v_n(i)$
for site i and for particular eigenstate n. They obey the
normalization condition $\sum_n |u_n(i)|^2 + |v_n(i)|^2 = 1$ for any
fixed site i. Consider now a site where, say, $u_n(i)$ is large and
close to 1 for particular eigenvalue. It follows therefore that for
the same site the $v_n(i)$ would have to be small, since the
normalization condition is almost exausted by $|u_n(i)|^2$ term
alone. Similarly, for the sites where $v_n(i)$ has large magnitude,
$u_n(i)$ would have to be small. Large $u_n(i)$ component would mean
that quasiparticle has a large electron component on this site.
Hence the electron will have large probability to tunnel into
superconductor on this site and the tunneling intensity for
electrons {\em positive sample bias} will be large. Conversely, for
those sites the hole amplitude is small $|v_n(i)| « |u_n(i)|$ and
the hole intensity {\em negative sample bias} will be small.
Similarly, for sites with large hole amplitudes $|v_n(i)| »
|u_n(i)|$ the electron amplitude will be suppressed and this site
will be bright on the hole bias. Therefore if there is a particular
pattern for the large particle amplitude (sampled on positive bias)
on certain sites i, the complimentary pattern of bright sites for
hole tunneling (on negative bias) will develop as a consequence of
the inherent particle-hole mixture in superconductor. This is the
physics behind what appears as the cross rotation upon bias switch,
seen in experiments \cite{SHPan:2000a,EWHudson:2001}, see
Fig.~\ref{FIG:Hudson_Ni}.

\subsection{Fourier-transformed STM Measurement}
The ingredient of the Fourier-transform STM technique is to collect
a large set of tunneling conductance data (at a fixed voltage bias)
in the real space, and then to perform a Fourier transform. This
technique was first applied by~\onlinecite{JEHoffman:2002a} to study
the quasiparticle states generated by a quantized magnetic vortex in
the mixed state of slightly overdoped high-$T_c$ superconductor,
Bi$_2$Sr$_2$CaCu$_2$O$_{8+\delta}$. A Cu-O bond-oriented
``checkboard'' pattern with $4a_0$ periodicity.  The $4a_0$
modulation periodicity is one half of that ($8a_0$) of the
field-induced SDW modulation observed in neutron
scattering~\cite{BLake:2001,BLake:2002,BKhaykovich:2001} on other
cuprate materials. This field-induced ``checkerboard'' pattern has
been interpreted as the induction of two-dimensional spin density
wave around the vortex core where the superconductivity is
suppressed~\cite{JXZhu:2002,MTakigawa:2003,BMAndersen:2003a}, the
nucleation of the antiferromagnetic order brought about by local
quantum fluctuations of a vortex~\cite{MFranz:2002}, and the frozen
of $d$-wave hole pairs into a crystal by the magnetic
field~\cite{HDChen:2002}. Most of the theories rely on the proximity
of the system to a quantum critical point so that it is very
sensitive to external perturbations. The same kind of checkerboard
pattern has also been predicted around a single strong impurity with
induced local moment in the optimally doped
cuprates~\cite{JXZhu:2002,SDLiang:2002,HYChen:2003,YChen:2004}.

\begin{figure}[tbp]
\caption{ A
representative set of seven scattering vectors $\mathbf{q}_{i}(E)$
of the `octet' model. Reproduced with permission
from~\cite{KMcElroy:2003}.} \label{FIG:octet}
\end{figure}

The challenge comes from the experimental observation of a similar
checkerboard pattern even at zero field in the same doping
regime~\cite{JEHoffman:2002b,KMcElroy:2003,CHowald:2003}. In the
experiments by~\onlinecite{JEHoffman:2002b,KMcElroy:2003}, the
Fourier analysis of the images of the energy-dependent modulations
yields the dispersion of wavevectors. Instead, Howard and co-workers
observed the existence of static striped density of electronic
states, i.e., the four-period peaks in the Fourier transform of the
data are present at all energies, including very low energies. One
can understand these two effects separably. Those peaks showing
energy dispersion comes from the a quasiparticle scattering from
impurities~\cite{JMByers:1993,QHWang:2003,DZhang:2003,DZhang:2004}.
A heuristic model based upon the electronic band structure is as
follows~\cite{KMcElroy:2003}: In BSCCO, four nodes exist in the
superconducting gap $\Delta_{k}$. Below the gap maximum $\Delta_0$,
the contours in $k$-space along which quasiparticle exist at a given
energy are banana-shaped, as shown in Fig.~\ref{FIG:octet}. The
quasiparticle density of states at energy $E=\omega$,
$\rho(E=\omega)$ is proportional to $\int_{E_k=\omega} \vert
\nabla_{k}E_{k}\vert^{-1} dk$, where the integral is performed over
the contour $E_{k}=\omega$. Each `banana' exhibits its largest rate
of increase with energy, $\vert \nabla_{k} E_{k}\vert^{-1}$, near
its two ends. Therefore, the primary contributions to $\rho(E)$ are
from the octet of momentum-space regions centered around
$\mathbf{k}_{j}(E),\;j=1,2,\dots,8$, at the end of the `banana's.
(Red circles in Fig.~\ref{FIG:octet}.) In the presence of
impurities, quasiparticles will be elastically scattered. A
quasiparticle located in momentum-space near one element of the
octet is highly likely to be scattered to the vicinity of another
element of the octet, because of the large density of final states
there. For each $\mathbf{k}_{j}$ in a representative octet, there
are seven characteristic scattering wavevectors. This octet-model
then predicts a total of 56 scattering wavevectors. Of these, 32
constitute a complete set of inequivalent wavevectors and therefore
16 distinct $\pm \mathbf{q}$ pairs can be detected by
Fourier-transformed scanning tunneling spectroscopy. The
experimental data~\cite{KMcElroy:2003} are in good agreement with
this model. From the material point of view, although no external
impurities were introduced in a controlled manner into the sample of
these experiments, the source of quasiparticle scattering may be
closely related to the experimentally observed nanoscale
inhomogeneity~\cite{SHPan:2001,CHowald:2001,KMLang:2002}.

In contrast, the observed non-dispersive LDOS
modulations~\cite{CHowald:2003} should be interpreted by invoking a
static (or fluctuating) charge- or spin-ordered
state~\cite{APolkovnikov:2003,SAKivelson:2003,DPodolsky:2003}. The
emergence of a competing ordering is due to the quantum criticality
with or without the aid of the inhomogeneity in the sample. An even
stronger evidence of the competing ordering is provided by recent
observation that the electronic states at low energies within the
pseudogap state in Bi$_2$Sr$_2$CaCu$_2$O$_{8+\delta}$ exhibit
spatial modulations with an energy-independent incommensurate
periodicity~\cite{MVershinin:2004}. Theoretically, a complete
microscopic model with all these elements has not yet been
developed.

\subsection{Filter}

We point out  that for  unitary scattering impurity in any model it
is difficult if not impossible to produce large intensity on the
impurity site. Unitary scattering produces a node in the wave
function. Yet, experimentally, the impurity site is bright
\cite{SHPan:2000b}. One explanation is that the image seen by STM is
not the real intensity of the impurity state that is buried below
the exposed layer in STM experiments. One needs to have a model on
how intensity is transmitted to the top layer. The idea of a filter
then comes in naturally. Martin {\em et al.}~\cite{IMartin:2002}
proposed an idea of filter that intensity as seen at the top layer
by STM is a convolution of initial intensity due to impurity
scattering and filter function that comes from the effective hopping
matrix element between CuO planes, $t_k \propto |\cos k_x -\cos
k_y|^2 $.

The reasoning goes as follows. In order to tunnel between layers it
is advantageous  to involve tunneling between $s$-wave orbitals that
extend out of the Cu-O plane. These orbitals are off the chemical
potential and virtual hopping on these orbitals would bring large
energy  denominators in any perturbation scheme. Still  it pays to
engage $s$-wave orbitals of Cu because one gains on the exponential
overlap factors between s-orbitals in adjacent planes. The
electronic orbitals near chemical potential are essentially
$d_{x^2-y^2}$ orbitals of Cu (hybridized with p-orbitals but we
ignore this hybridization). The $d_{x^2 - y^2}$ orbitals on the
impurity site are orthogonal to the s-orbital of the impurity site.
The next available s orbitals are on the nearest Cu sites. Therefore
electron hops virtually on to $p_x $ or $p_y$ orbitals of nearest O
and then onto s-orbital. The amplitude for the  hops $\mbox{Cu}\;
d_{x^2-y^2} \rightarrow \mbox{O}\; p_{x,y} \rightarrow \mbox{Cu}\;
s$ would be different for hops along horizontal and vertical
directions, as one can verify from Fig.~(\ref{FIG:dps}).

\begin{figure}[tbp]
\caption{The real space
image of different orbitals on Cu, nearest O and nearest Cu sites
are shown. Quantum mechanical interference produces the filter
effect that changes the  distribution of the impurity state
intensity~\cite{IMartin:2002}.} \label{FIG:dps}
\end{figure}

For example the hopping to the Cu site on the right one would get
for and amplitude $A_{i, i+x(y)}$ :
 \beqa
 A_{i, i+x}
\propto \frac{\langle d_i|p_x \rangle \langle p_x|s_{i+x}
\rangle}{[E_p - E_d][E_s- E_p]} \sim  \frac{(-1) \exp(ik_x a)}{[E_p
- E_d][E_s- E_p]}
\nonumber\\
A_{i, i+y} \propto \frac{\langle d_i|p_y \rangle \langle p_y|s_{i+y}
\rangle}{[E_p - E_d][E_s- E_p]} \sim  \frac{(+1) \exp(ik_x a)}{[E_p
- E_d][E_s- E_p]}\;, \label{EQ:A1} \nonumber \\ \eeqa
 in the second equation we considered a plane waves that
 describe the states without impurity scattering. One immediately can
 see that the signature for the horizontal and
vertical amplitudes is opposite in sign regardless of the phase
assignment of $p$-orbitals for pure case and for general amplitudes
of the states produced by impurity scattering. For a quantum
mechanical process to hop from one site to nearest neighbor
$s$-orbitals one would have to add the amplitudes: \beqa A_{tot} &=&
A_{i, i+x} +
A_{i, i-x} + A_{i, i+y} + A_{i, i-y} \nonumber \\
&& \sim \cos(k_xa)  - \cos(k_ya)\;. \label{EQ:A2}\eeqa Again, second
equation refers to the pure plane wave analysis to make contact with
the bands structure calculations for the tunneling matrix
element~\cite{OKAndersen:1995}. Upon hopping on the $s$-orbitals
electron hops to the next layer and retraces its path exactly in
reversed sequence as described above. Therefore the amplitude for
the hopping will be proportional to the {\em square} of the
$A_{tot}$. The net hopping matrix element has the from consistent
with $|d_{x^2-y^2}|$ modulations: \beqa |A_{tot}|^2 \propto |A_{i,
i+x} + A_{i, i-x} + A_{i, i+y} + A_{i, i-y}|^2 \label{EQ:A3} \eeqa
This particular filter is directly connected to the interplane
hopping matrix element obtained within the  band structure
calculation \cite{OKAndersen:1995}. However the idea that one has to
involve the $s$-orbitals of the Cu-O plane is relevant also for an
exposed Cu-O layer as one would need to tunnel from the $s$-orbitals
of the tip onto relevant $s$-orbitals of the Cu-O
plane~\cite{SMisra:2002}.

Alternative filter due to blocking of certain hopping matrix
elements has been considered by Zhu  et al.~\cite{JXZhu:2000b}. For
an analysis of the local effects of impurity one need to consider a
local tunneling matrix elements that has to connect impurity
orbitals to $s$-orbitals on neighboring Cu atoms that have a
greatest overlap between Cu-O layers. The net effect of the filter
is to produce large spectral intensity on  an impurity site and
nearest neighbor sites to be dark. More recently, the measured
quasi-continuous data has been converted to a set of LDOS defined on
a two-dimensional lattice~\cite{QWang:2004}, which is suitable for a
rigorous comparison between the tight-binding model studies and the
STM experimental data.

Another important observation one can make by comparing STM and
local NMR results available in Li doped YBCO superconductor
\cite{JBobroff:2001}. In case of Li impurity NMR revealed that
maximum intensity in NMR signal comes from four nearest neighbor Cu
sites and is quite localized near impurity. This observation would
be consistent with the notion that strongly scattering impurity
produces  large density of states on nearest sites. The crucial
difference between NMR and STM is that NMR observation does not
require {\em electronic tunneling}.  Magnetic field is measured
instead. Hence there is no filter to apply to native electronic
states in Cu-O plane to obtain NMR real space distribution.
Therefore,depending on the type of measurement one might need  or
need not to use the filters. Details depend on the nature of the
measurement.

\section{Average density of states in superconductors with impurities}

\label{sec:AverDOS}

The Green's function formalism is well suited to the analysis of the
combined effect of many uncorrelated impurities in the bulk of a
superconductor. The first treatment of the superconducting
properties using this technique was given by Abrikosov and Gor'kov
\cite{AAAbrikosov:1960} in a pioneering paper. The basic assumptions
underlying such calculations were given in Sec.~\ref{ImpAvBasic}.
After averaging over different impurity distributions following
Eq.~(\ref{Gimp}), the translational symmetry in the system is
restored, and therefore the Green's function takes the general form
\begin{eqnarray}
    \label{DysonImpGen}
  \widehat G^{-1}({\bf k}, \omega)&=&i\omega_n -\xi({\bf k})\tau_3-
  \Delta_0\sigma_2\tau_2-\widehat\Sigma
  \\
    &\equiv&
    i\widetilde\omega -\widetilde\varepsilon({\bf k})\tau_3-
  \widetilde\Delta\sigma_2\tau_2.
\end{eqnarray}
Here the second line explicitly takes into account the matrix
structure of the self-energy, $\widehat\Sigma$. The superconducting
gap in the presence of impurities is determined by the
self-consistency condition, Eq.~(\ref{Eq:DeltaSelfConGen}), which
reads here
\begin{equation}
    \label{GapEq}
  \Delta(\widehat\Omega)=\pi T N_0\sum_{\omega_n}\int d\widehat\Omega^\prime
  V(\widehat\Omega,\widehat\Omega^\prime)
  \frac{\widetilde\Delta(\widehat\Omega^\prime)}{\sqrt{\widetilde\omega_n^2
  +\Delta^2(\widehat\Omega^\prime)}}.
\end{equation}
The transition temperature is the temperature at which a non-trivial
solution of the self-consistency equation first appears. Together
with the recipe for computing the self-energy these equations form a
general basis for treating ensembles of impurities in
superconductors. We note here that we always ignore the contribution
of $\Sigma_3$, which is equivalent to the renormalization of the
chemical potential. This is always allowed in computing the density
of states, although the corrections may need to be taken into
account in evaluating the response functions
\cite{PJHirschfeld:1988}.  The basic assumption for computing the
self-energy is that, in addition to neglecting the interaction
between spins on different impurity sites, see
Sec.~\ref{ImpAvBasic}, we can neglecting the interference effects of
scattering on different impurities (which the order $(p_Fl)^{-1}$,
where $l$ is the mean free path).

\subsection{$s$-wave}

\subsubsection{Born approximation and the AG Theory}

In a seminal paper Abrikosov and Gorkov analysed the effect of the
impurity scattering on superconductivity in the Born approximation.
We briefly review this analysis to compare its outcome with the
results of theories going beyond Born approximation. We follow the
treatment of \onlinecite{KMaki:1969}.

Consider a general impurity potential combining the potential and
the magnetic scattering,
\begin{equation}
  \widehat U_{imp}({\bf k}-{\bf k}^\prime)=U_{pot}({\bf k}-{\bf k}^\prime)\tau_3 +
  J ({\bf k}-{\bf k}^\prime){\bm
  S}\cdot{\bm\alpha},
\end{equation}
where ${\bm\alpha}$ is defined in Eq.~(\ref{MatrixAlpha}). AG
considered the self energy in the Born approximation,
\begin{equation}
  \widehat\Sigma(\omega,{\bf k})=n_{imp}
  \int\frac{d{\bf k}^\prime}{(2\pi)^3}
  \widehat U_{imp}({\bf k}-{\bf k}^\prime)
  \widehat G({\bf k}^\prime, \omega)
    \widehat U_{imp}({\bf k}^\prime-{\bf k}).
\end{equation}
Integrating over ${\bf k}^\prime$ we find
\begin{eqnarray}
  \widetilde\omega&=&\omega_n+\frac{1}{2}
  \biggl(\frac{1}{\tau_p}+\frac{1}{\tau_s}\biggr)
  \frac{\widetilde\omega}{\sqrt{\widetilde\omega_n^2
  +\Delta^2}},
  \\
  \widetilde\Delta&=&\Delta +
  \biggl(\frac{1}{\tau_p}-\frac{1}{\tau_s}\biggr)
  \frac{\widetilde\Delta}{\sqrt{\widetilde\omega_n^2
  +\Delta^2}},
\end{eqnarray}
where the potential ($\tau_p$) and spin-flip ($\tau_s$) scattering
times are given by
\begin{eqnarray}
  \frac{1}{\tau_p}&=&n_{imp} N_0\int d\widehat\Omega
  |U_{pot}({\bf k}-{\bf k}^\prime)|^2,
    \\
    \label{BornAlpha}
    \frac{1}{\tau_s}&=&n_{imp} N_0 S(S+1)\int d\widehat\Omega
  |J({\bf k}-{\bf k}^\prime)|^2.
\end{eqnarray}
Here we averaged over all possible directions of the impurity spin.

In the absence of spin-flip scattering both $\Delta$ and $\omega$
are renormalized identically, and it follows from Eq.~(\ref{GapEq})
that the gap remains unchanged compared to the pure case. This is in
accordance with Anderson's theorem. The spin flip scattering time
(which violates the time-reversal symmetry) enters the equations for
$\widetilde\omega$ and $\widetilde\Delta$ with the opposite sign.
Therefore, introducing $u=\widetilde\omega/\widetilde\Delta$, we
find
\begin{equation}
    \label{Eq:BornU}
  \frac{\omega}{\Delta}=u\biggl(1-\frac{(\Delta\tau_2)^{-1}}{\sqrt{1+u^2}}\biggr).
\end{equation}
It follows that the gap in the single particle spectrum is
$E_{gap}=\Delta(1-(\Delta\tau_s)^{-2/3})^{3/2}$ for
$\Delta\tau_s>1$, and vanishes for $\Delta\tau_s<1$. This gapless
region starts at the value of pairbreaking parameter $\alpha$
\begin{equation}
  \alpha^\prime=\tau_s^{-1}=\Delta_{00}\exp(-\pi/4),
\end{equation}
where $\Delta_{00}$ is the gap in the pure material at $T=0$.

The transition temperature is determined from
\begin{equation}
 \psi\biggl(\frac{1}{2}+\frac{1}{2\pi\tau_s T_c}\biggr)-
    \psi\biggl(\frac{1}{2}\biggr)=
    \ln\frac{T_{c0}}{T_c},
\end{equation}
where $\psi(x)$ is the digamma function and  $T_{c0}$ is the
transition temperature of the pure material. Consequently,
superconductivity is destroyed ($T_c=0$) when
\begin{equation}
  \alpha_c=\tau_s^{-1}=\pi T_{c0}/2\gamma=\Delta_{00}/2>\alpha^\prime,
\end{equation}
where $\gamma\approx 1.78$. As $\alpha^\prime\approx 0.912\alpha_c$
AG predicted that a regime of gapless superconductivity exists for a
range of impurity scattering \cite{AAAbrikosov:1960}. This was first
confirmed in experiments by Woolf and Reif \cite{MAWoolf:1965}.

The evolution of the density of states with increasing disorder was
investigated in detail \cite{VAmbegaokar:1965,SSkalski:1964}, and is
shown in Fig.~\ref{fig:skalski}. For $\alpha<\alpha^\prime$ a hard
gap in the single particle spectrum persists up to the critical
impurity concentration, as shown in Fig.~\ref{fig:skalski-gapl}.
This result is clearly at odds with our discussion in
Sec.~\ref{sec:Shiba}, which shows that even a single magnetic
impurity creates a localized state in the superconducting gap.
\begin{figure}
  \caption{Density of states in the Abrikosov-Gorkov theory of
  magnetic impurities in superconductors. Here
  $\Gamma=\tau_s^{-1}$. Reproduced with permission from
  \cite{SSkalski:1964}.}
  \label{fig:skalski}
\end{figure}

\begin{figure}
  \caption{Plot of the dependence of the order parameter,
  $\Delta$, transition temperature, $T_c$, and the single particle
  spectral gap, $\Omega_G$ here, on the scattering rate
  $\Gamma=\tau_s^{-1}$. Reproduced with permission from
  \cite{SSkalski:1964}.}
  \label{fig:skalski-gapl}
\end{figure}

\subsubsection{Shiba impurity bands}

In the AG theory the impurity concentration and the strength of the
exchange coupling contribute to the suppression of superconductivity
as a single pairbreaking parameter, $\alpha=\tau_s^{-1}=(2n_{\rm
imp}/\pi N_0) \sin^2\delta_0 \propto n_{imp}J^2 S(S+1)$ for
isotropic exchange, see Eq.~(\ref{BornAlpha}). This is a result of
the Born approximation; in general, the phase shift $\delta_0$  and
the concentration of impurities $n_{\rm imp}$ are separate variables
that control different aspects of impurity scattering. For example,
in the limit of dilute concentration of strong magnetic impurities,
the AG approach yields a small scattering rate, and a
single-particle spectral gap virtually identical to that in a pure
limit. On the other hand, we have learned that in this regime each
impurity is accompanied by a bound state with the energy below the
gap, and therefore we expect a finite number of these sub-states to
exist in a superconductor. This section addresses this dichotomy.

Analysis of the strong scattering regime requires going beyond the
Born approximation, and here we use the self-consistent $T$-matrix
approach~\cite{SSchmitt-Rink:1986,PJHirschfeld:1986}, where the
self-energy $\widehat\Sigma({\bf p},\omega)=n_{imp}\widehat T_{{\bf
p,p}}$, and
\begin{equation}
  {\widehat T}_{{\bf p, p}^\prime}={\widehat U}_{{\bf p, p}^\prime} +
  \int d{\bf p}_1 {\widehat U}_{{\bf p},{\bf p}_1}
  {\widehat G}({\bf p}_1, \omega)
  {\widehat T}_{{\bf p}_1, {\bf p}^\prime}.
\end{equation}

Following the treatment described in  Sec.~\ref{sec:Shiba}, we
analyse the pairbreaking in different angular momentum channels. The
effective pairbreaking parameter in the $l$-th channel is
$\alpha_l=n_{imp}(1-\epsilon_l^2)/(2\pi N_0)$, where $\epsilon_l$ is
the position of the corresponding bound state, see
Eq.~(\ref{Eq:EnergyOfShibaStates}). In analogy with the AG
treatment, we find that the ratio
$u_n=\widetilde\omega_n/\widetilde\Delta(\omega_n)$ satisfies the
equation \cite{AIRusinov:1969,ANChaba:1972}
\begin{equation}
  \frac{\omega_n}{\Delta}=u_n
  \biggl[1-\sum_{l=0}^\infty (2l+1)\frac{\alpha_l}{\Delta}
  \frac{\sqrt{1+u_n^2}}{\epsilon_l^2+u_n^2}\biggr],
\end{equation}
where the gap is determined self-consistently from
\begin{equation}
  \Delta=2\pi T N_0 {\rm g}\sum_n(1+u_n^2)^{-1/2}.
\end{equation}
This equation should be contrasted with Eq.~(\ref{Eq:BornU}). The
pairbreaking parameter, $\alpha_l$ now depends {\it separately} on
the position of the single-impurity resonance state, $\epsilon_l$
and the impurity concentration, in contrast to the AG theory.

The growth of the impurity band has been investigated for the
spherically symmetric case of purely magnetic scattering
\cite{HShiba:1968,AIRusinov:1969,ANChaba:1972}. The critical
concentration of impurities at which the transition temperature
vanishes is obtained by setting $T_c= 0$ in the gap equation,
\begin{equation}
  \ln\frac{T_{c0}}{T_c}=\psi(1/2 +\alpha/2\pi T_c)-\psi(1/2),
\end{equation}
where now \cite{DMGinzberg:1979}
\begin{equation}
  \alpha=\sum_l (2l+1)\alpha_l.
\end{equation}
Since the gap equation is identical to that considered by AG, the
critical pairbreaking, $\alpha_{cr}=\Delta_0/2$. However, now the
critical concentration of impurities depends on the phase shift of
scattering by individual impurities, and on the position of the
single impurity resonance, see Fig.~\ref{fig:Shiba-Shiba},
\begin{equation}
  n_{cr}=\pi N_0\Delta_0\biggl[\sum_l
  (2l+1)(1-\epsilon_l^2)\biggr]^{-1}.
\end{equation}

The width of the gapless regime now also depends on the details of
scattering. For $l=0$ channel only the gap vanishes when the
pairbreaking exceeds the value \cite{HShiba:1968,AIRusinov:1969}
\begin{equation}
  \frac{\alpha^\prime}{\alpha_{cr}}=
  2\epsilon_0^2\exp[-\pi\epsilon_0^2/2(1+\epsilon_0)].
\end{equation}
In the Born approximation (weak scattering) the bound state moves to
the gap edge, $\epsilon_0=1$, and we regain the result of Abrikosov
and Gorkov. For stronger scattering, $\epsilon_0<1$, the realm of
gapless superconductivity is enhanced compared to the AG theory. As
higher order harmonics are included, the threshold at which the
density of states at the Fermi energy becomes non-zero shifts even
lower \cite{DMGinzberg:1979}.

For $l=0$ in the limit $\alpha_0\ll\Delta$ the width of the impurity
band around $E_0$ is estimated to be
$W=(8\alpha_0\Delta)^{1/2}(1-\epsilon_0)^{1/4}$, and therefore
varies as $n_{imp}^{1/2}$ \cite{HShiba:1968}. Therefore if the
resonance state at $E_0$ is sufficiently close to the gap edge, the
concentration, $c_0$, at which the top of the impurity band merges
with the continuum above $\Delta$ is smaller than the critical
concentration, $c^\prime$, at which the bottom of the impurity band
reaches the Fermi surface and the superconductor becomes gapless
\cite{HShiba:1968}, see Fig.~\ref{fig:Shiba-AG}. In fact, the AG
result is an extreme example of this behavior when the states due to
individual impurities are infinitely close to the gap edge, and
therefore the effect of increasing impurity concentration is an
apparent decrease of the gap until the onset of the gapless
behavior.

\begin{widetext}
  \begin{figure}
    \caption{Evolution of the spectral gaps and density of states
    for strong magnetic impurities ($\epsilon_0\ll \Delta_0$).
    Left panel shows the available states (shaded) as a function
    of the impurity concentration. Right panel shows the
    qualitative features of the density of states for different
    values of the impurity concentration, labeled by vertical
    lines $A,B,C,D$ on the left. Critical concentration of
    impurities corresponds to line $B$, when the impurity band
    first touches $\omega=0$. At the same time, spectral gap
    between the top of the impurity band and the bottom of the
    continuum states persists to higher impurity concentration
    (line $D$).}\label{fig:Shiba-Shiba}
  \end{figure}

  \begin{figure}
    \caption{Evolution of the spectral gaps and density of states
    for weak magnetic impurities ($\epsilon_0\lesssim \Delta_0$).
    Left panel shows the available states (shaded) as a function
    of the impurity concentration. Right panel shows the
    qualitative features of the density of states for different
    values of the impurity concentration, labeled by vertical
    lines $A,B,C$ on the left. The impurity band and the continuum
    above the gap merge at a low impurity concentration,see line $B$,
    and further evolution of the density of states is very close
    to the predictions of the AG theory. Critical concentration
    (line $C$) marks the onset of gapless superconductivity.}
    \label{fig:Shiba-AG}
  \end{figure}

\end{widetext}

\subsubsection{Quantum spins and density of states}

In the fully quantum treatment of the impurity spin,
Sec.~\ref{sec:Kondo}, we discussed the competition between gapping
the density of states due to superconductivity, and the onset of the
Kondo screening of the impurity moment. The main conclusion was
that, in contrast to classical impurity spin, the position of the
bound state is not simply given by the value of the bare exchange
coupling but depends sensitively on the ratio $T_K/T_c$. Once the
position of the bound state is established, in the limit of
independent impurities one can consider the growth of the impurity
band in analogy with the previous section. As discussed previously,
for ferromagnetic coupling of the impurity to the conduction
electrons, the bound state is always close to the gap edge, the
scattering is weak, and we can expect that the Abrikosov-Gor'kov
theory gives correct results.

When Kondo screening is effective, for antiferromagnetic coupling,
the behavior of the density of state and the transition temperature
was studied a series of papers by M\"uller-Hartmann and co-workers
\cite{EMullerHartmann:1971,JZittartz:1972,EMullerHartmann:1973,BSchuh:1978}.
The main new result was the prediction of the re-entrant behavior
for small $T_K/T_c\lesssim 1$. In that case the phase shift of the
scattering increases upon lowering temperature, but remains moderate
at $T_c$ enabling the transition to the superconducting state. Upon
further decrease in temperature, scattering becomes stronger and
suppresses superconductivity in a range of phase diagram of
Fig.~\ref{fig:KondoTc}. Finally, at lowest temperatures below $T_K$,
the system re-enters local Fermi liquid regime with weak scattering
and superconductivity may re-appear. While further work
\cite{TMatsuura:1977a,MJarrell:1990} cast doubt on the existence of
the third transition, region of two solutions for $T_c(n_{imp})$ was
confirmed by theoretical studies. In particular, a combination of
quantum Monte Carlo technique with Eliashberg equations gave the
dependence of the re-entrance transition on the electron-phonon
coupling constant, while accounting non-perturbatively for the Kondo
effect \cite{MJarrell:1990}, see Fig.~\ref{fig:KondoTc}. Moreover,
the initial decrease of $T_c$ with increasing impurity concentration
is fast \cite{EMullerHartmann:1971,MJarrell:1990}, and depends on
the coupling strangth \cite{MJarrell:1990}. The behavior of the
density of states in this limit was investigated in detail
\cite{NEBickers:1987,MJarrell:1990a}. The overall shape of the
transition temperature as a function of impurity concentration with
re-entrant transition was observed in (LaCe)Al$_2$ alloy series
\cite{MBMaple:1973}.

\begin{figure}
  \caption{Reduced transition temperature normalized to pure
  system as a function of the impurity concentration for different
  eletcron-phonon coupling, $\lambda_0$. The impurity
  concentration $\bar{c}=n_{imp}/(2\pi)^2 N_0 T_{c0}$. From
  \cite{MJarrell:1990}.}
  \label{fig:KondoTc}
\end{figure}

\subsection{$d$-wave}

For completeness we briefly consider the growth of the impurity band
with finite concentration of impurities.  As was mentioned above,
scalar (non-magnetic) impurities are pair-breakers for  any
nonconventional superconductor, and substantially change the
low-energy spectrum of superconducting quasiparticles. This problem
has been addressed in great detail in the framework of the
self-consistent $T$-matrix approximation (for example, see
\cite{LPGorkov:1985,PJHirschfeld:1986,CJPethick:1986,SSchmitt-Rink:1986,PJHirschfeld:1988,
PALee:1993,PJHirschfeld:1993,AVBalatsky:1994}), which leads to the
finite density of states at the Fermi level. Here we briefly review
the main steps and give results for the quasiparticle scattering
rate and low-energy density of states for completeness.

For finite impurity concentration, the self-consistent Green's
function, averaged over impurity positions, was given in
Eq.~(\ref{G-SCTM}) as
 \beqa
 \widehat{G}^{-1} ({\bk},\omega) =
\widehat{G}_0^{-1}({\bk},\omega)- \widehat{\Sigma}(\omega). \eeqa
with $\widehat{\Sigma}(\omega) = n_{\rm imp}\widehat{T}(\omega)$. In
the case of particle-hole symmetry \cite{PJHirschfeld:1988}, and
unconventional gap (defined by us as having a zero average over the
Fermi surface, see Sec.~\ref{sec:INTRODUCTION}) the only
non-vanishing component of the $T$-matrix is proportional to
$\tau_0$,
\begin{equation}
    T_0(\omega)=\frac{g_0(\omega)}{c^2-g_0(\omega)}.
\end{equation}
The $T$-matrix has to be determined self-consistently with
$g_0(\omega)\rangle = \frac{1}{2\pi N_0} \sum_{\bk} {\rm Tr}
\widehat{G}({\bk}, \omega) \widehat{\tau}_0$.

Solution of this equation leads to a finite density of states at the
Fermi level. This result was first obtained for Born scattering
\cite{LPGorkov:1985,KUeda:1985}, leading to an exponentially small
$N(0)/N_0\approx 4\tau^2\Delta_0^2\exp(-2\Delta_0\tau)$, where
$\tau$ is the normal state scattering rate. The results are much
more dramatic for unitarity scattering ($c=0$)
\cite{PJHirschfeld:1986,SSchmitt-Rink:1986}, when straightforward
algebra yields \beqa \gamma \simeq \sqrt{n_{\rm imp}(\Delta_0 /\pi
N_0)}, \label{eq:impdwave9} \eeqa where $\gamma = - {\rm Im}
\,\Sigma(\omega\rightarrow 0)$ is the scattering rate for low-energy
quasiparticles. For $\omega \lesssim \gamma$, the density of states
is determined by impurities and is finite: $N_{\rm imp}(0)/N_0 =
2\gamma/\pi\Delta_0$. The characteristic width of the
impurity-dominated region is $\omega^* \simeq \gamma \propto
\sqrt{n_{\rm imp}}$.

The origin of the finite density of states   is the impurity band,
grown from the impurity-induced  states (consider $c = 0$). Scaling
of the impurity bandwidth $\gamma \propto \sqrt{n_{\rm imp}}$ has
been obtained earlier for the case of paramagnetic impurities in an
$s$-wave superconductor \cite{HShiba:1968}. The fact that $\gamma
\propto \sqrt{n_{\rm imp}}$ is obeyed in the case of a $d$-wave
superconductor with scalar impurities as well is consistent with the
claim that the low-energy states in a disordered $d$-wave
superconductor are indeed formed from the bound states at finite
concentration. Many questions about the exact nature of the
interference between impurity sites in unconventional
superconductors remain as of now unanswered. We briefly reviewed
some of the relevant work in the introduction, but do not discuss it
in depth here.

Notice that the results above are for isotropic impurity scattering.
Anisotropic impurities may preferentially scatter electrons between
regions with the same, or close values of the gap, so that the
scattering is inefficient in suppressing $T_c$. For general impurity
phase shifts this has been considered by \onlinecite{CHChoi:1999},
while for the model with dominant small angle scattering in cuprates
\cite{EAbrahams:2000} the effect was considered by
Kee~\cite{HYKee:2001}.

\section{Optimal fluctuation}
\label{sec:OF}

\subsection{Introduction}

So far we concentrated on discussing the effect of a single impurity
on its immediate surrounding and on the combined effect of an
ensemble of scattering centers on the spatially averaged properties
of a superconductor. In the case of a single pairbreaking impurity
the characteristic length is simply the superconducting coherence
length, $\xi_0$. In the Abrikosov-Gorkov approach the gap is assumed
to be uniformly suppressed. If the coherence length is short, this
assumption breaks down as the energy cost of local suppression of
the order parameter becomes smaller than the cost of uniform
reduction of the gap. In this case again the length scale of this
suppression is of the order of $\xi_0$. These results are obtained
by carrying out a standard impurity averaging procedure at the mean
field level, i.e. averaging over all the possible configurations of
impurity atoms \cite{AAAbrikosov:1963}.

It is clear, however, that some physics is missing in such an
approach. Among all the realizations of the impurity distribution in
a sample of size $L_0$ there exist regions where the {\it local}
impurity concentration, on some characteristic scale $L\ll L_0$,
differs significantly from the average concentration, $n$. If the
local impurity concentration is sufficiently high, for $L>\xi_0$
superconductivity may be locally destroyed of sufficiently
suppressed to generate a bound quasiparticle state at an energy
$E\ll \DD$.

Of course, such regions are rare. There is a high entropy cost in
creating an impurity droplet with the concentration significantly
different from the average and hence the probability of encountering
these regions is small. However, the states localized in these
droplets make a non-perturbative contribution to the density of
states averaged over the entire sample, $N(E)$, and qualitatively
modify its behavior compared to the mean field (Abrikosov-Gorkov and
Shiba) treatment. Quite dramatically, they make any $s$-wave
superconductor with a small concentration of magnetic impurities
($\Delta\tau_s\gg 1$) {\it gapless} \cite{AVBalatsky:1997}. It is
due to such a dramatic modification that the interest in these
``tail'' states stretching below the mean field gap edge has peaked
in recent years.

The problem of tail states did not originate in the study of
superconductivity. The contribution of regions of anomalous impurity
concentration to the net density of states below the gap edge was
first considered in doped semiconductors by Lifshitz
\cite{IMLifshitz:1964,IMLifshitz:1965,IMLifshitz:1968}. He was the
first to show that such rare impurity configurations create a local
profile in the Coulomb potential that can have bound states, and
therefore gives rise to the non-vanishing density of states below
the bottom of the band, $E_g$. Henceforth the states localized in
the droplets of impurities have become known as `Lifshitz tails',
and have been extensively studied
\cite{BIHalperin:1966,JZittarz:1966,PVanMieghem:1992}.

While, in retrospect, it is natural that inhomogeneities lead to a
low-energy tail in the density of states in superconductors in much
the same way, little attention has been paid to this problem until
the paper by Balatsky and Trugman \cite{AVBalatsky:1997}. Their
study was stimulated by the experimental observations that the
tunneling density of states in $s$-wave superconductors with
magnetic impurities is far greater at low energies than the
Abrikosov-Gorkov theory suggests
\cite{MAWoolf:1965,ASEdelstein:1967,SDBader:1975}. A number of
theoretical studies of the tail states followed, and this topic is
now a subject of active interest.

Below we first briefly review the physical picture of the tail
states in semiconductors, and then describe how it is applied to the
subgap states in superconductors with impurities.

\subsection{Tail states in semiconductors and optimal fluctuation}
\label{sec:tails1}

In a semiconductor there are two distinct situations: a) heavily
doped, and b) lightly doped with impurity atoms. In the former case
a localized tail state with energy $E<E_g$ forms in the
impurity-rich region, and the extent of its wave function greatly
exceeds the average distance between individual shallow sites.
Therefore the exact impurity potential can be replaced by a smooth
function, averaged over regions containing many impurities. The
probability of realization of the potential with the ``right''
energy of the bound state among all the possible impurity
distributions determines its contribution to the DOS. In the latter
case the number of impurity sites needed to form a bound state
depends on how deep below the band edge the energy of such a state
is. For example,  if each impurity binds an electron at energy
$E_1$, while $E_2$ is the energy  of the state bound by two
impurities on neighboring lattice sites, to obtain a localized state
below $E_1$ but above $E_2$, one simply needs to find a region where
the two impurities are at a particular finite distance from each
other. The probability of finding such an impurity pair determines
the density of states \cite{IMLifshitz:1965,IMLifshitz:1968}. As we
go to energies below $E_2$ we need to position three impurities etc.

For energy, $E$, the most probable (albeit still very rare)
configuration of impurities that creates a potential $U$, with a
bound state from the solution of Schr\"odinger's equation
$[H_{band}+ U]\psi={\cal E}[U]\psi$, such that  ${\cal E}[U]=E$, and
therefore contributes the most to $N(E)$ is called the optimal
fluctuation. Given the probability density for the impurity
potential, $P[U]$, and the density of states in this potential,
\begin{equation}
    \label{OptDos}
  N(E)=\int {\cal D}U P[U]\delta(E-{\cal E}[U]),
\end{equation}
the optimal fluctuation is obtained by using the saddle point
approximation and minimizing the resulting functional with respect
to $U$. This approach finds the cheapest (from the entropy
consideration) impurity potential that creates a bound state at $E$.
Therefore it optimizes the non-uniform impurity distribution
(fluctuation from the uniform average) to the given energy, hence
the name ``optimal fluctuation''. The general technical difficulty
of minimization lies in its essential nonlinearity: the optimal
potential depends on the wave function of the particle in this
potential.

Let us consider the example of many uncorrelated shallow impurity
centers forming an extended potential. It is described by  the
Gaussian probability density,
\begin{equation}
  \label{GaussP}
    P[U]\propto \exp\biggl[-\frac{1}{2U_0}\int d^d{\bf r} U^2({\bf
    r})\biggr].
\end{equation}
Saddle point approximation for Eq.(\ref{OptDos}) gives
\begin{equation}
  \ln\frac{N(E)}{N_0}\approx -{\cal S}[U_{opt}],
\end{equation}
where the optimal fluctuation is obtained by minimizing the
functional
\begin{equation}
  \label{Sopt}
  {\cal S}[{U}]=\frac{1}{2U_0^2}\int d^d{\mathbf r}
{U}^2(\mathbf{r}) +
    \lambda \biggl({\cal E }[{ U}] -E\biggr)
\end{equation}
with respect to the potential $U$ and the Lagrange multiplier
$\lambda$. At the simplest level it is sufficient to consider only
the potentials where ${\cal E}[U]=E$ is the lowest energy state in
the potential $U$; fluctuations where $E$ coincides with the higher
eigenstates are exponentially less probable. In a semiconductor the
kinetic energy of the quasiparticles is $p^2/2m^*$, where $m^*$ is
the effective mass. Consequently, in a potential well of depth $U$
(all energies are measured from the band edge) and size $L$ the
energy of the localized state is of the order of $U+1/(mL^2)=E$
($\hbar=1$). In the optimal fluctuation  $E\sim U\sim L^{-2}$, so
that the action for such fluctuation is ${\cal S}[U]\approx
L^dU^2/U_0^2$, or $\ln\left[ N(E)/N_0 \right]\approx
-|E|^{2-d/2}/U_0^2$ \cite{BIHalperin:1966,IMLifshitz:1965}.
Importantly, the characteristic size of the optimal fluctuation,
$L\propto |E|^{-1/2}$ increases as the energy approaches the band
edge, while its depth, $|U|\sim |E|$ decreases: the potential
becomes more shallow and extended.

More formally, note that the energy of the bound state is the
expectation value of the hamiltonian over the wave function of the
bound state, $\psi({\bf r})$, is equal to $E$,
\begin{equation}\label{VarE}
{\cal E}[U]=\langle \widehat H \rangle=\langle\psi | \frac{{\bf
p}^2}{2m^*} + U |\psi \rangle=E
\end{equation}
Minimization of the action in Eq.(\ref{Sopt}) with respect to $U$
dictates that
\begin{equation}
  \label{equationU}
  U(x)=-\lambda U_0^2 \langle\psi| \frac{\delta
\widehat H}{\delta U}|\psi \rangle=
    -\lambda U_0^2 \psi^2(x),
\end{equation}
while minimization with respect to $\lambda$ dictates that, in this
potential, the bound state is at energy $E$, i.e. (setting $m^*=1$
for simplicity)
\begin{equation}\label{EqOnpsi}
\Biggl[ -\frac{1}{2}\nabla^2-\lambda U_0^2 \psi^2({\bf r})\Biggr]
\psi({\bf r}) = E\psi({\bf r}).
\end{equation}
In one dimension this equation is exactly solved to give
\cite{BIHalperin:1966}
\begin{eqnarray}
  \psi(x)=\sqrt{\frac{\kappa}{2}}{\rm sech} \kappa x,
  \\
  \lambda U_0^2=8\kappa,
\end{eqnarray}
with $E=-\kappa^2/2$. Therefore the ``optimal action'' ${\cal
S}(U_{opt})\simeq \kappa^2/U_0^2\sim |E|^{3/2}$ as expected.

In higher dimensions the corresponding equation is not solvable,
however, one can extract the energy dependence of the action by
assuming a spherically symmetric optimal fluctuation and an
exponentially decaying at large distances bound state to find the
Lifshitz tail $N(E)\propto \exp(-|E|^{2-d/2})$
\cite{IMLifshitz:1965,IMLifshitz:1988}. To obtain the
pre-exponential factor one needs to consider all the wave functions
in the potential, and the corresponding analysis has only been
carried out in low dimensions \cite{BIHalperin:1966}.

\subsection{S-wave superconductors}

\subsubsection{Magnetic and non-magnetic disorder}

Since the effect of the tails is most dramatic for fully gapped
superconductors, most studies focused on conventional, $s$-wave
superconductors with pairbreaking magnetic impurities. The general
route followed in all the investigations is similar to the approach
described above: given the probability density of different impurity
configurations, and the hamiltonian of the system with the potential
of each impurity distribution, we find the most probable
configuration of impurities that gives rise to a state at a given
energy within the gap. Technical implementations of this algorithm
vary depending on the specifics of the problem at hand, see below.

There are important differences between the physics of the optimal
fluctuation in such a superconductor and an optimal potential well
for quasiparticles below the band gap discussed in the previous
section. First, since the superconducting quasiparticles consist of
electron pairs close to the Fermi surface, their kinetic energy is
not simply that of a band particle, but is given instead by the
Hamiltonian
\begin{equation}
  \widehat H=\widehat\xi\tau_3+\Delta({\mathbf{r}})\tau_1\sigma_2,
\end{equation}
where $\tau_i$ and $\sigma_i$ are the Pauli matrices in the
particle-hole and the spin space respectively, so that $\tau_i
\sigma_j$ is a 4$\times $4 direct product. Therefore, while the
envelope of the tail state wave function still varies smoothly over
the length scale of inhomogeneities in the impurity distribution,
the rapid oscillations on the atomic scale associated with the Fermi
surface have to be taken into account. As will be seen below, these
considerations substantially modify the behavior of the tail states.

Second, the scattering potential is a matrix in particle-hole and
spin space. In general, an impurity site acts both as a potential
and a magnetic scatterer, so that the total scattering potential is
\begin{eqnarray}
\widehat {\bf U}({\mathbf{r}})=\sum_i \biggl[ U_0 \tau_3\delta({\bf
r}-{\bf r}_i)+ J({\bf r}-{\bf r}_i){\bf S}_i\cdot
{\bm{\alpha}}\bigr],
\end{eqnarray}
using the Nambu notations. The potential part of the scattering,
$U_0$, is not pairbreaking in accordance with Anderson's theorem.
However, since the size of the optimal fluctuation is large compared
to the correlation length, it is necessary to distinguish between
the cases where the motion of quasiparticles within the optimal
fluctuation is diffusive (strong potential scattering,
$\Delta\tau\ll 1$, $\tau\ll\tau_s$, where $\tau$ is the transport
lifetime) and ballistic (weak potential scattering,
$\tau\gg\tau_s$). Moreover, we should also distinguish between
strong and weak {\it magnetic} scattering: if the magnetic
scattering is strong there are resonance (Shiba-Rusinov) states in
the gap, and the tails stretch not from the mean-field gap edge, but
from the localized impurity band. If the magnetic scattering can be
treated in the self-consistent Born approximation, the tail states
emerge below the Abrikosov-Gorkov renormalized single particle
spectral gap, $\Delta_0=\Delta (1-(\Delta\tau_s)^{-2/3})^{3/2}$,
where $\Delta$ is the self-consistent value of the superconducting
order parameter. In the AG limit the probability density for the
magnetic impurity potential is gaussian, as it is averaged over a
large number of impurity sites. In contrast, in the unitarity
scattering limit there are subgap states localized on one or a few
impurities; consequently, we deal with the Poisson density
distribution. These various possibilities provide for a rich variety
of behavior that is still a subject of active interest.

All models of tail states due to magnetic impurities studied so far
ignore interactions between the impurity spins: it was shown in Ref.
\cite{AILarkin:1972} that the RKKY interaction and glassy behavior
of impurity spins modify the AG results very weakly. The models also
treat impurity spins as classical Heisenberg spins, and therefore
cannot account for the Kondo effect. This is justified either when
the Kondo temperature of individual impurity sites is much smaller
than the superconducting transition temperature, $T_K\ll T_c$ (and
depletion of states at the Fermi level prevents screening of the
local moment), or in the opposite limit, $T_K\gg T_c$, when the
moments are quenched already in the normal state
\cite{EMullerHartmann:1971}.

To our knowledge, the first paper discussing the influence of
non-uniform impurity distribution on the transition temperature in
analogy with Lifshitz's work appeared in 1968 \cite{IOKulik:1968}.
These authors found that, in the limit of  average impuritity
concentration $n\ll n_{cr}$ of the Abrikosov-Gorkov theory, there
are localized regions that become superconducting at a temperature
$T_c^\prime > T_c(n)$, where $T_c(n)$ is the corresponding AG
transition temperature. The difference between the two was evaluated
for parabolic one-dimensional variation of the effective impurity
potential. \onlinecite{IOKulik:1968} also noted that their results
will be modified if there is non-magnetic as well as magnetic
scattering, but did not address this question further.

\subsubsection{Diffusive limit, weak magnetic scattering}

If the scattering on individual magnetic impurities is weak, the
optimal fluctuation is created by large droplets of these scattering
centers. Since the impurities are uncorrelated, the probability
density for the impurity potential is Gaussian, which greatly
simplifies the analysis.

Historically, most of the studies have been carried out in the
diffusive limit. One of the first papers investigated the smearing
of the gap edge due to local fluctuations in the effective
interaction between electrons \cite{AILarkin:1972}. If the
correlation length of the inhomogeneities, $r_c\gg\xi$, where $\xi$
is the coherence length of the dirty superconductor,
$\xi\sim(D/\Delta)^{1/2}$, and $D$ is the diffusion constant, the
order parameter simply locally adjusts to the value of the
interaction and the density of state is determined by the local gap
amplitude,
\begin{equation}
  N(E)=\int_0^\infty N(E,\Delta) W(\Delta) d\Delta,
\end{equation}
where $W(\Delta)$ is the probability density of the gap.

In the opposite limit of short-range correlations in the pairing
interaction, the finite density of states below the mean field gap
edge is due to the states spatially localized in correlated droplets
of size $r_0\sim \xi [(\Delta_0-E)/\Delta]^{-1/4}$ (increasing
rapidly as $E\rightarrow\Delta_0$ as in a semiconductor), which
leads to $N(E)\propto \exp(-[(\Delta_0-E)/\Delta]^{5/4}$ in $d=3$.
As in semiconductors, the high entropy cost of a large droplet is
offset by the lowering of the kinetic energy of the bound state.
Indeed, in a clean system with $\Delta\tau_s\gg 1$, and therefore
$\delta_0\approx \Delta$, we find the characteristic kinetic energy,
$D/r_0^2\simeq \sqrt{\Delta_0^2-E^2}$.

Recently it was argued that the above result is flawed since it does
not account properly for the rapid oscillations of the wave function
of the bound state on the scale of the Fermi
wavelength~\cite{JSMeyer:2001}. These authors used a
field-theoretical approach that maps the disordered superconducting
system onto a non-linear $\sigma$-model (for a review, see
\onlinecite{AAltland:2000}) to show that, while the droplet size for
the optimal fluctuation is identical to that obtained by Larkin and
Ovchinnikov, the subgap density of states is
$N(E)\propto\exp\{-[(\Delta_0-E)/\Delta]^{(6-d)/4}\}$, which gives
the exponent $3/4$, rather than $5/4$, for $d=3$.

The paper that brought the investigation of the subgap states in
superconductors into the limelight after a quarter-century-long
hiatus was the study of the density of states due to regions where
the impurity concentration is sufficient to locally destroy
superconductivity \cite{AVBalatsky:1997}. In that case the spectrum
of the fluctuation region is similar to that of a disordered
metallic grain of the same size, $L$, and depends on the mean level
spacing of the grain, $\delta_L$. The average density of states was
obtained in two steps. First, an average over all realizations of
disorder for grains of size $L$ yielded $N_L(E)\sim \delta_L^{-1}$.
Second, the probability of finding a fluctuation region of size $L$
with the critical concentration of impurities, $n_c$, for a given
average impurity concentration, $n$, $P_L(n_c;n)$ was used to define
the average density of states over the entire sample, $N(E)\sim\int
dV P_L(n_c;n) N_L(E)$. This integral was estimated to find
\begin{equation}
  N(E)\sim \delta_{L_0}^{-1}\exp[-L_0^d(n_c\ln(n_c/n)-n_c+n)],
\end{equation}
as $E\rightarrow 0$. Here $L_0=(\xi_0 l)^{1/2}$ is of the order of
the coherence length in a dirty superconductor with $l\ll\xi_0$.

At energies closer to the gap edge, in the spirit of optimal
fluctuation, it is not necessary to destroy superconductivity
completely to generate the tail states. Using the instanton approach
for the nonlinear $\sigma$-model, Lamacraft and Simons demonstrated
how these states arise out of inhomogeneous instanton configurations
for the action \cite{ALamacraft:2000,ALamacraft:2001}. The resulting
optimal action reads
\begin{equation}
{\mathrm S}_0=a_d (\DD\tau_s)^{2/3}
(1-\DD\tau_s)^{-2/3})^{-(2+d)/8}\biggl(\frac{\DD-E}{\Delta}\biggr)^{(6-d)/4}.
\end{equation}
density of states, varies as $N(E)\sim \exp[-4\pi g(\xi/L)^{d-2}
{\mathrm S}_0]\sim \exp\{-[(\Delta_0-E)/\Delta]^{(6-d)/4}\}$. Here
$g$ is the bare conductance and $a_d\sim 1$

That approach appears sufficiently general to analyse nucleation of
domains in a variety of systems. It was used to re-derive within
this framework \cite{ALamacraft:2001} the universal gap fluctuations
in small metallic grains, first obtained using random-matrix theory
\cite{MGVavilov:2001}, namely $N(E)\sim \exp[-(\Delta_0-E)^{3/2}]$,
valid for $\Delta_0-E\ll\Delta_0$. In this regime the spatial extent
of the optimal fluctuation is greater than the size of the grain, so
that effectively we are in zero dimensions, $d=0$, and the exponent
$3/2$ agrees with the general result of Lamacraft and Simons,
$(6-d)/4$. In the same zero-dimensional limit, but at lower
energies, $E\ll \Delta_0$, the random matrix theory gives $N(E)\sim
(|E|/\delta^{3/2}\Delta^{1/2})\exp[-\pi\tau_s(\Delta_0-E)^2/\delta]$,
where $\delta$ is the mean level spacing in the grain
\cite{ISBeloborodov:2000}.

\subsubsection{Diffusive limit, strong scattering}

Recently the field theoretical treatment has been extended to the
case of strong scatterers \cite{FMMarchetti:2002}. In that case the
probability distribution of scattering strength is Poissonian rather
than Gaussian. In the field theoretical language this implies that
the action cannot be expanded to second order in the magnetic
potential, as it was for the weak potential. Marchetti and Simons
circumvented this difficulty by considering the dominant
contribution of droplets densely populated by magnetic impurities,
so that $\xi\ll l_s\ll l$. As we saw above, an impurity band emerges
within superconducting gap in the limit of near-unitary scattering
already at the level of the mean field theory. Consequently, the
tail states extend from the edge of the continuum above $\Delta_0$
as well as from the top and bottom of the impurity band, see
Fig.~\ref{fig:OF-DOS}. According to Marchetti and Simons in all
these cases the density of states varies as $N(E)\propto
\exp[-(|E-E_i|/\Delta)^{(6-d)/4}$, where $E_i$ is the appropriate
band edge. Therefore the exponent of the action is identical to that
found in other systems in the diffusive limit.

\subsubsection{Ballistic limit, weak scattering}

It was noticed early on that in some systems the magnetic scattering
is dominant: upon increasing the concentration of impurities the
increase in residual resistivity ratio correlates with the
suppression of the superconducting transition temperature
\cite{ASEdelstein:1967}. Since both magnetic and nonmagnetic
scattering contribute to the resistivity, but only the magnetic part
suppresses $T_c$, this is an indication of almost purely
spin-dependent scattering. Shytov and co-workers
\cite{AVShytov:2003} considered the subgap states in this limit in
clean ($l\gg \xi_0$ or $\Delta\tau_s\gg 1$) limit, when the spectral
gap obtained in the self-consistent Born approximation nearly
coincides with the order parameter, $\Delta_0\approx\Delta$.

Once again, since the impurities are assumed to be weak, the optimal
fluctuation for states not too far from the gap edge is large and
shallow, and the spin-dependent potential has the Gaussian
probability density. When the size of the optimal fluctuation is
much greater than the superconducting coherence length, $l\gg
L\gg\xi_0$, the motion of the quasiparticles in this potential is
ballistic. As a result, direct mapping on the non-linear
$\sigma$-model is not feasible, and the problem requires quantum
mechanical treatment akin to that in a semiconductor discussed
above.

As in that case, we first consider the one-dimensional problem. An
important assumption (discussed below) is that a ferromagnetic
fluctuation maximizes the effect of the impurity potential. Choosing
the direction of the impurity spins along the $y$ axis, performing
rotation $\sigma_2\rightarrow\sigma_3$, we remove the vector
character of the slowly varying potential ${\bf U}$, and consider
the hamiltonian
\begin{equation}\label{NewHamiltonian}
\widehat H_\pm= \widehat\xi\tau_3\pm\Delta_0\tau_1 \pm
 U (\mathbf{r}).
\end{equation}
The hamiltonian, however, still remains a matrix in the
particle-hole space, and the wave functions of the optimal
fluctuation are the Nambu spinors $\Psi$.

Let us again discuss the physical behavior of the optimal
fluctuation qualitatively. We linearize the kinetic energy near the
Fermi surface,  so that typical kinetic energy in an OF of size $L$
is $\xi\simeq v_F/L$. Then the energy of a quasiparticle in the
optimal fluctuation (measured from the Fermi energy) is $E\simeq U+
\sqrt{\DD^2 + v_F^2/L^2}$. For the energies close to the
superconducting gap, $(\DD-E)/\DD \ll 1$, the OF is large ($L\gg
\xi_0=v_F/\DD$) and shallow ($|U|/\DD\ll 1$), so that $E-\DD\approx
U+ v_F^2/(\DD L^2)$. Introducing the dimensionless energy
$\epsilon=E/\DD$, we obtain, in analogy with the arguments above,
$|U|/\DD\simeq \xi_0^2/L^2\simeq 1-\epsilon$. Notice that the size
of the fluctuation is $L\simeq \xi_0/\sqrt{1-\epsilon}\gg \xi_0$. As
a result, we find ${\cal S}[U]\approx
LU^2/U_0^2=\DD^2\xi_0(1-\epsilon)^{3/2}/U_0^2$. From the definition
of $U_0$ it follows that
\begin{equation}
    \label{OptimAction1D}
    -\ln\frac{N(E)}{N_0}\approx {\cal S}[U_{opt}]\simeq
    (\DD\tau_s) (1-\epsilon)^{3/2}.
\end{equation}
The energy dependence in Eq.~(\ref{OptimAction1D}) is identical to
the result of Lifshits in $d=1$, despite the linear, rather than
quadratic, dependence of the kinetic energy on the size of the
droplet. This follows from the smallness of this energy compared to
the gap: even though $\xi\propto 1/L$, the expansion is in $\xi^2$.

The minimization of the saddle-point action proceeds exactly
following the steps in section \ref{sec:tails1}. For spin ``up''
particles ${\cal E}_+ [U]=\langle\Psi | \widehat H_+ |\Psi \rangle$.
Minimization with respect to $U$ gives
\begin{equation}
\label{equationU1} U(x)=-\lambda U_0^2 \langle\Psi| \frac{\delta
\widehat H_+}{\delta U}|\Psi \rangle.
\end{equation}
In principle this variational derivative includes the effect of the
self-consistent suppression of the gap. However, it can be
explicitly demonstrated \cite{AVShytov:2003} that the effect of
self-consistency is small. Then, in exact analogy to the
semiconductor problem, $U(x)=-\lambda U_0^2 (\Psi^\star(x)\Psi
(x))$, where $(\Psi^\star\Psi)$ denotes the scalar product in
particle-hole space. In turn, Schr\"odinger equation takes the form
\begin{equation}\label{EqOnPsi}
\Biggl[ -i v_F \frac{\partial}{\partial x}\tau_3 +\DD \tau_1
-\lambda U_0^2 (\Psi^\star\Psi)\Biggr] \Psi = E\Psi.
\end{equation}
This equation is solved by introducing the bilinear forms
$\Psi^\star(x)\tau_i\Psi(x)$. These bilinears play the role of the
Halperin-Lax wave function in the Nambu space, and yield
\begin{eqnarray}
  R_0&=&\frac{1-\epsilon^2}{\xi_0\arccos\epsilon} \
\frac{1}{\epsilon + \cosh (2x\sqrt{1-\epsilon^2}/\xi_0)},
\\
    R_1&=&R_0 (\epsilon + \xi_0 R_0\arccos\epsilon),
\\
    R_2&=&\sqrt{R_0^2-R_1^2},
\end{eqnarray}
and $R_3=0$ \cite{AVShytov:2004}. The physical potential of the
optimal fluctuation is \cite{AVShytov:2003}
\begin{equation}\label{OptimalU}
\frac{U(x)}{2\DD}=-\frac{1-\epsilon^2}{\epsilon + \cosh
(2x\sqrt{1-\epsilon^2}/\xi_0)}.
\end{equation}
which corresponds to the value of the action
\begin{equation}
\label{OptimalAction1D} {\cal S}[U]=8\pi (\DD\tau_s) \left[
\sqrt{1-\epsilon^2}-\epsilon\arccos\epsilon\right].
\end{equation}
For $\epsilon\approx 1$ the length scale of the optimal fluctuation
is $\xi_0/\sqrt{1-\epsilon^2}$, its depth is $U\sim \DD
(1-\epsilon^2)$, and the density of states $N(E)\sim
\exp[-(1-\epsilon^2)^{3/2}]$, in complete agreement with qualitative
estimates.

The most important observation of \onlinecite{AVShytov:2003} is that
in higher dimensions the optimal fluctuation is strongly
anisotropic, in contrast to both the conventional semiconductors and
superconductors in the diffusive limit. This is a direct consequence
of the composite nature of superconducting quasiparticles: they are
made out of objects that move with the Fermi velocity. The wave
function of the subgap state is concentrated along the
quasiclassical trajectory, which is a chord in a potential of any
shape. Consequently, there is little energy cost in reducing the
size of the OF in the ``transverse'' direction, while the smaller
volume makes such fluctuations more probable, see
Fig.~\ref{fig:OF_Nmet}. As a result, the optimal fluctuation is
strongly elongated in one ($x$) direction. The wave function of the
bound state can be written as $\Psi(x, {\bf y})=\exp(ik_F x)\Phi(x,
{\bf y})$, where ${\bf y}$ denotes the transverse $d-1$ coordinates,
and $\Phi$ is a slowly varying function. The kinetic energy of the
quasiparticle is
\begin{equation}
    \widehat\xi\Psi \approx -e^{ik_F x}\left( i v_F
\frac{\partial}{\partial x} + \frac{\nabla_y^2}{2m}\right)\Phi
    \sim \left( \frac{v_F}{L_x}+\frac{1}{mL_y^2}\right) \Psi.
\end{equation}
The transverse size of the fluctuation can therefore be reduced
until the second term becomes comparable to the first, i.e.
$L_y\simeq (\lambda_F L_x)^{1/2}$, where $\lambda_F\simeq k_F^{-1}$
is the Fermi wavelength. Consequently, $|U|/\DD\sim 1-\epsilon$ and
$L_x\sim \xi_0/\sqrt{1-\epsilon}$, and
\begin{equation}
    \label{anisAction}
    {\cal S}[U_{opt}]\simeq L_x L_y^{d-1}\frac{U^2}{U_0^2}\simeq
    (\DD\tau_s)\left(\frac{E_F}{\DD}\right)^{\frac{d-1}{2}}
    \left(1-\epsilon\right)^{\frac{7-d}{4}},
\end{equation}
where $E_F$ is the Fermi energy. Consequently, the density of
states, $N(E)\sim\exp[-(1-\epsilon)^{(7-d)/4}]$.

\begin{figure}
  \caption{The spatial structure of the optimal fluctuation in the
  ballistic and the diffusive limits.}
  \label{fig:OF_Nmet}
\end{figure}

The action for this anisotropic fluctuation is smaller than that for
an isotropic droplet with the same energy of the bound state, by a
factor of $(E_F/\DD)^{(d-1)/2} (1-\epsilon)^{-(d-1)/4}$, so that the
corresponding DOS is exponentially higher.

Since the optimal fluctuation is a result of a saddle point
approximation for the functional integral, Eq.~(\ref{OptDos}), it is
only valid when ${\cal S}[U_{opt}]\gg 1$, or
\begin{equation}
    \label{Valid1}
    1-\epsilon \gg (\DD\tau_s)^{\frac{4}{d-7}}
    \left(\frac{\DD}{E_F}\right)^\frac{2(d-1)}{7-d}.
\end{equation}
For $d=1$ this condition becomes $1-\epsilon\gg (\DD\tau_s)^{-2/3}$,
while for $d=3$ it does not depend on the gap amplitude,
$1-\epsilon\gg (k_F l)^{-1}$.

It is possible to compare the densities of states given by different
approaches at the crossover scale between the diffusive and the
ballistic regimes \cite{IVekhter:2003}. A transition to the
diffusive regime occurs when the size of OF $L\geq v_F\tau_s$, or
$1-\epsilon\leq (\DD\tau_s)^{-2}$. The result of
Ref.\cite{ALamacraft:2000} for $\DD\tau_s\gg 1$ is ${\mathrm
S_D}=(\DD\tau_s)^{5/3}(E_F/\DD)^{d-1}(1-\epsilon)^{(6-d)/4}$.
Consequently, at the crossover point the action from
Eq.~(\ref{anisAction}) is smaller, ${\mathrm S_D/S_0}\simeq
(E_F/\DD)^{(d-1)/2}(\DD\tau_s)^{7/6}\gg 1$, and the OF found by
Shytov et al. corresponds to a greater DOS. Therefore the structure
of the OF near the crossover between the ballistic and diffusive
regimes still resembles closely that given above. As the size of the
OF increases even further, the anisotropic fluctuation becomes
insupportable due to diffusive motion.

Balatsky and Trugman \cite{AVBalatsky:1997} considered only the DOS
at $E=0$ due to the suppression of superconductivity by paramagnetic
impurity potential. They needed a large volume fluctuation, $V\geq
\xi^d$, which is less probable and yields lower DOS than that of
Eq.~(\ref{anisAction}). \onlinecite{IVekhter:2003} checked whether
local suppression of the gap from $\DD$ to $E$ due to a large number
of impurities with {\it uncorrelated} spins (as opposed to a
ferromagnetic OF above) is advantageous. For $1-\epsilon\ll 1$ the
local pairbreaking rate, $\gamma$, needed to reduce the gap to $E$
is $\gamma\tau_s\approx 1+ (1-\epsilon)(\DD\tau_s)^{2/3}$, and the
volume of the region has to be at least equal to that of the
anisotropic OF to avoid high kinetic energy cost (this is an
underestimate since it ignores proximity coupling to bulk). In that
case the optimal action ${\mathrm S_{BT}}/{\mathrm S_0} \approx
(\DD\tau_s)^{1/3} (E_F/\DD) \bar c$, where $\bar
c=n_{imp}\lambda_F^d$ is the atomic concentration of impurity atoms.
As a result, for realistic values of $\bar c$ and clean samples
${\mathrm S_{BT}}\gg {\mathrm S_0}$, and the DOS given by the action
in Eq.~(\ref{anisAction}) is higher.

Therefore the ballistic limit of the action obtained by
\onlinecite{AVShytov:2003} is expected to be valid near the
crossover between the ballistic and the diffusive regimes.

\subsubsection{Ballistic regime, strong scattering}

As of today, we are not aware of any investigations of the structure
of the optimal fluctuation in the ballistic regime, when there exist
bound states on individual magnetic impurities. It is reasonable to
assume that the result differs from the standard Lifshitz formula
for the same reason as in the section above: the wave functions of
the states localized on magnetic impurities in superconductors
oscillate with the Fermi wavelength, see Sec.~\ref{sec:Shiba}. As a
result, in the dilute impurity limit, the shift of the energy level
localized on, for example, two impurities located at distance $R\gg
p_F^{-1}$, will be suppressed by the typical factor $\exp(-R/\xi_0)$
\cite{AIRusinov:1968}. Consequently, the states significantly below
the impurity band must be created by a large number of impurities or
impurities located on neighboring lattice sites. This problem still
awaits further investigation.

\subsubsection{Summary}

In $s$-wave superconductors with magnetic impurities the density of
states does not vanish {\em irrespective of the concentration and
nature of the impurity scattering}. The tails of the density of
states extend into the mean field gap. Therefore {\em all
superconductors with magnetic impurities are gapless.} This behavior
is qualitatively illustrated in Fig.~\ref{fig:OF-DOS}.

\begin{figure}
  \caption{Qualitative picture of the density of states in an
  $s$-wave superconductor with magnetic impurities. Blue shade
  denotes regions where the mean field density of states is
  finite. Red shading signifies the finite DOS induced by the
  deviations of the local impurity distribution from the average.
  The density of states in these tails is exponentially small but
  finite. If the impurities are weak, the well-defined impurity
  band is absent and the tail extend from the mean field gap
  edge.}  \label{fig:OF-DOS}
\end{figure}

\section{Summary and outlook}
\label{sec:Conclusion}

While considering the role of impurities in conventional and
unconventional superconductors, this review focused on theoretical
and experimental results that highlight the new physics beyond
standard Abrikosov-Gor'kov theory, Anderson theorem and average
lifetime effects.  The studies of disorder in $s$-wave
superconductors were carried out in detail in the 1960's. We
discussed more recent results in this field. Our main emphasis has
been on how individual impurities influence local electronic states
in their immediate vicinity, and on deviations from the standard
Abrikosov-Gorkov theory on mesoscopic scales. This focus is dictated
both by the advances in experimental techniques, which can now use
NMR methods and STS measurements to probe electronic states with
atomic spatial resolution, at the scales where impurities perturb
their surrounding, \cite{OFischer:2004}, and the concomitant
development of new theoretical approaches.

The stimulus for such extensive studies is that impurities are
markers that allow to reveal the nature of correlations and pairing
of the state where impurities are placed. Indeed particular pattern
of impurity-induced electronic states is closely connected to the
symmetry of the superconducting gap and helps us to understand the
nature of superconducting pairing. If strong electronic correlations
in the ground state  are present, they also are reflected in details
of impurity induced states. Therefore watching the waves created by
throwing a pebble in the pond of correlated electrons helps us
understand the properties of the underlying electronic liquid.

We kept the discussion general to allow applications to other
systems and materials. For instance, this was our rationale for
employing the BCS state to describe superconductivity. We believe
that it is a good approximation in heavy fermion, organic
superconductors and SrRuO$_4$, at very low energy. At the same time,
deviations from this mean field picture may provide additional
details on the underlying physics of the particular material.
Majority of the data at a moment are obtained in high-$T_c$
materials. It is clear that similar local effects are present around
impurities in other unconventional superconductors, e.g. in
Na$_x$CoO$_2\cdot y$H$_2$O superconductors~\cite{QHWang:2004},
although we are not aware of any data on single impurity states in
these materials. Given the importance of the impurity states, this
field will undoubtedly be extended to other systems by future
experiments.

{\em Outlook for the future}.  New ideas and directions continue to
emerge in the studies of electronic properties induced by
impurities. The suite of new experimental tools that address local
electronic effects, such as STM, will help to clarify the role of
interference between several impurities, and pave the way towards
connecting the microscopic local states with average properties.
Recent theoretical work addressed some aspects of this subject
\cite{DKMorr:2003b,DKMorr:2003c,LZhu:2003,WAAtkinson:2003,LZhu:2004,BMAndersen:2003b},
and is awaiting direct comparison with experiment.

Another promising avenue is combining the spatial resolution of
STM-STS with the time resolution. The subject is still in its
infancy, both theoretically and experimentally, but hold immense
promise for the future. Sec.~\ref{sec:DynamicalImp} reviewed some of
very recent work in this direction. Temporal and spatial
characterization of the states generated by dynamical impurities
allow  exploration of the correlations inside the electronic state
in which impurity is placed. One obvious example where such
characterization is crucial is the Kondo effect in a superconducting
state. It is desirable to have a time resolved measurements that
allow to visualize the Kondo effect in a superconductor. Another
interesting problem that needs further elaboration is a role of
collective modes in impurity-induced states. We are only starting to
investigate these questions, as discussed in
Sec.~\ref{sec:Interplay}.

Real progress on these problems will be made when we have real data.
As usual, one should expect that the data will have surprises that
were not anticipated in simple theoretical  models. This will
motivate further theoretical studies, stimulate more measurements,
and therefore will lead to a rapid further development of the field.
They can provide space (and time) resolved window into the intimate
workings of the correlated electron matter. We have every reason to
be enthusiastic and optimistic about the future the field of
impurity states in superconductors, and in other correlated electron
systems.

\section*{Acknowledgments}

We would like to thank many of our colleagues for useful
discussions. We are especially  grateful to our collaborators over
many years who were instrumental in our work on the subjects that
are reviewed here. Without their insights and knowledge this work
would be impossible. We thank Ar. Abanov, E. Abrahams, B. Altshuler,
A. R. Bishop, A. H. Castro Neto, J. C. Davis, D. Eigler, L. P.
Gor'kov, M. J. Graf, I. A. Gruzberg, P. J. Hirschfeld, W. Ho, C. R.
Hu, T. K. Lee, D. K. Morr, I. Martin, D. Pines, A. Rosengren, M. I.
Salkola, J. A. Sauls, D. J. Scalapino, J. R. Schrieffer, A. V.
Shytov, Q. Si, C. S. Ting, M. Vojta, Z. D. Wang, and J. Zaanen for
fruitful discussions. This work was supported by the U.S. Department
of Energy (A.V.B. and J.X.Z.) and by the Board of Regents of
Louisiana (I. V.).

\widetext
\section*{List of Symbols}
\begin{tabular}{ll}
Quantity & Explanation\\ \hline $a$ & Lattice parameter \\
$b(b^{\dagger})$ & Bosonic annihilation (creation) operators \\
$c(c^{\dagger})$ & Fermionic ahhihilation (creation) operators \\
$d$              & Spatial dimension \\
$D$              & Half energy bandwidth \\
$\Delta_0$         & Superconducting energy gap \\
$\Delta_{k}$       & Momentum-dependent superconducting energy gap\\
$\phi_{n}(\mathbf{r})$ & Electron eigenfunction \\
$E_{F}$ & Electron Fermi energy \\

$G(\tau,\tau^{\prime}), \ \ G(\tau, \br)$ & Electron Green's function in coordinate space\\
$G(\omega_n, \bk), \ \ G(\bk, \omega_n)$ & Electron Green's function in Matsubara frequency and momentum space\\
$H,\mathcal{H},H_{int}$      & Hamiltonian \\
$J,J_0,J_c$ & Exchange coupling\\
$L$ & Linear dimension of a system \\
$\mu$ & Chemical potential \\
$N(\epsilon)$ & Electron density of states \\
$N(\epsilon,\mathbf{r}), N(E,i)$ & Electron local density of
states \\
$\psi(\mathbf{r}(\psi^{\dagger}(\mathbf{r}))$ & Fermionic field
                              operators in continuum space \\
$\vert \Psi \rangle,  \ \ \vert \Psi_0 \rangle $  & BCS variational wavefunction\\
$\vert \Psi_{-1} \rangle,  \ \ \vert \Phi_{-1} \rangle $  &  Excited variational wavefunction with single particle excitation present\\
$\mathbf{r}$    & Spatial coordinates\\
$\bm{\sigma}$ & Pauli matrices in spin space \\
$\bm{\tau}$ & Pauli matrices in Nambu space \\
 $V_{\alpha\beta\gamma\delta},\tilde{V}_{\alpha\beta}$ &
                                                Superconducting
                                               pairing
                                               interaction\\
$\mathbf{S}$ & Local spin operator\\
$t,t^{\prime}$  & Electron hopping integral \\
 $u$ & Electron-like
Bogoliubov quasiparticle wavefunction
amplititude\\
$T(\omega)$ & $T$-matrix \\
$T$ &  Temperature\\
 $v$ &  Hole-like Bogoliubov
quasiparticle wavefunction
amplititude\\
$U$ & Hubbard on-site electron-electron interaction \\
$U_0$ & Impurity scattering potential \\
$W$ & Half energy bandwidth \\
$W_{\bk}$ &  $D$-density-wave order parameter \\
$\xi_0$  & BCS superconducting coherence length at low temperatures\\
$\xi(T)$ & BCS temperature dependent coherence length\\
\end{tabular}
\endwidetext

\bibliographystyle{apsrmp}
\bibliography{imp}

\end{document}